%%%%%%%%%%%%%%%%%%%%%%%%%%%%%%%%%%%%%%%%%%%%%%%%%%%%%%%%%%%%%%%%%%%%%%%%%%%%%%
%%%%%%%%%%%%%%%%%%%%%%%%%%%%%%%charm.tex%%%%%%%%%%%%%%%%%%%%%%%%%%%%%%%%%%%%%%
%%%%%%%%%%%%%%%%%%%%%%%%%%%%%%%%%%%%%%%%%%%%%%%%%%%%%%%%%%%%%%%%%%%%%%%%%%%%%%

%% site dependent options: 
%% \unredoffs and \redoffs define horizontal and vertical offsets 
%% respectively for unreduced and reduced modes. \speclscape defines
%% the \special{} call that sets printer to landscape (sideways) mode.
%% from standard set below, leave uncommented as appropriate or redefine
%
%%% next 400dpi
%\def\unredoffs{} \def\redoffs{\voffset=-.31truein\hoffset=-.48truein}
%\def\speclscape{\special{landscape}}
%
%%% apple lw
%\def\unredoffs{} \def\redoffs{\voffset=-.31truein\hoffset=-.59truein}
%\def\speclscape{\special{ps: landscape}}
%
%%% qms lasergrafix:
%\def\unredoffs{} \def\redoffs{\voffset=-.4truein\hoffset=.125truein}
%\def\speclscape{\special{qms: landscape}}
%
%%% saclay A4 paper:
\def\unredoffs{\hoffset-.14truein\voffset-.2truein} 
 
%\def\speclscape{\special{landscape}}
%
%---------------------------------------------------------------------%
%
\newbox\leftpage \newdimen\fullhsize \newdimen\hstitle \newdimen\hsbody
\tolerance=1000\hfuzz=2pt
\catcode`\@=11 % This allows us to modify PLAIN macros.
%\def\bigans{b }
%\message{ big or little (b/l)? }\read-1 to\answ
%
%\ifx\answ\bigans\message{(This will come out unreduced.}
\magnification=1095\unredoffs\baselineskip=16pt plus 2pt minus 1pt
\hsbody=\hsize \hstitle=\hsize %take default values for unreduced format
%
%\else\message{(This will be reduced.} \let\l@r=L
%\magnification=800\baselineskip=16pt plus 1pt minus 0.5pt \vsize=7truein
%\redoffs \hstitle=8truein\hsbody=4.75truein\fullhsize=10truein\hsize=\hsbody
%
%\output={\ifnum\pageno=0 %%% This is the HUTP version
%  \shipout\vbox{\speclscape{\hsize\fullhsize\makeheadline}
%    \hbox to \fullhsize{\hfill\pagebody\hfill}}\advancepageno}
%  \else
% \almostshipout{\leftline{\vbox{\pagebody\makefootline}}}\advancepageno 
%  \fi}
%\def\almostshipout#1{%\if L\l@r \count1=1 \message{[\the\count0.\the\count1]}
%      \global\setbox\leftpage=#1 \global\let\l@r=R
% \else 
%\count1=2
%  \shipout\vbox{\speclscape{\hsize\fullhsize\makeheadline}
%      \hbox to\fullhsize{\box\leftpage\hfil#1}}  \global\let\l@r=L\fi}
%\fi
%---------------------------------------------------------------------
%
\newcount\yearltd\yearltd=\year\advance\yearltd by -1900

%
% 	restores pagenumbers
%
%       use following instead of \Date on the preliminary draft, 
%       puts date/time on each page in big mode, writes labels in margins

\def\draftmode{\message{ DRAFTMODE }\def\draftdate{{\rm preliminary draft:
\number\month/\number\day/\number\yearltd\ \ \hourmin}}%
\headline={\hfil\draftdate}\writelabels\baselineskip=16pt plus 2pt minus 2pt
 {\count255=\time\divide\count255 by 60 \xdef\hourmin{\number\count255}
  \multiply\count255 by-60\advance\count255 by\time
  \xdef\hourmin{\hourmin:\ifnum\count255<10 0\fi\the\count255}}}
%       use \nolabels to get rid of eqn, ref, and fig labels in draft mode
\def\nolabels{\def\wrlabeL##1{}\def\eqlabeL##1{}\def\reflabeL##1{}}
\def\writelabels{\def\wrlabeL##1{\leavevmode\vadjust{\rlap{\smash%
{\line{{\escapechar=` \hfill\rlap{\sevenrm\hskip.03in\string##1}}}}}}}%
\def\eqlabeL##1{{\escapechar-1\rlap{\sevenrm\hskip.05in\string##1}}}%
\def\reflabeL##1{\noexpand\llap{\noexpand\sevenrm\string\string\string##1}}}
\nolabels
%
% tagged sec numbers
\global\newcount\secno \global\secno=0
\global\newcount\meqno \global\meqno=1
\def\newsec#1{\global\advance\secno by1\message{(\the\secno. #1)}
%\ifx\answ\bigans \vfill\eject \else \bigbreak\bigskip \fi  %if desired
\global\subsecno=0\eqnres@t\noindent{\bf\the\secno. #1}
\writetoca{{\secsym} {#1}}\par\nobreak\medskip\nobreak}
\def\eqnres@t{\xdef\secsym{\the\secno.}\global\meqno=1\bigbreak\bigskip}
\def\sequentialequations{\def\eqnres@t{\bigbreak}}\xdef\secsym{}
\global\newcount\subsecno \global\subsecno=0
\def\subsec#1{\global\advance\subsecno by1\message{(\secsym\the\subsecno. #1)}
\ifnum\lastpenalty>9000\else\bigbreak\fi
\noindent{\bf\secsym\the\subsecno. #1}\writetoca{\string\quad 
{\secsym\the\subsecno.} {#1}}\par\nobreak\medskip\nobreak}
\def\appendix#1#2{\global\meqno=1\global\subsecno=0\xdef\secsym{\hbox{#1.}}
\bigbreak\bigskip\noindent{\bf Appendix #1. #2}\message{(#1. #2)}
\writetoca{Appendix {#1.} {#2}}\par\nobreak\medskip\nobreak}
%
%       \eqn\label{a+b=c}	gives displayed equation, numbered
%				consecutively within sections.
%     \eqnn and \eqna define labels in advance (of eqalign?)
%
\def\eqnn#1{\xdef #1{(\secsym\the\meqno)}\writedef{#1\leftbracket#1}%
\global\advance\meqno by1\wrlabeL#1}
\def\eqna#1{\xdef #1##1{\hbox{$(\secsym\the\meqno##1)$}}
\writedef{#1\numbersign1\leftbracket#1{\numbersign1}}%
\global\advance\meqno by1\wrlabeL{#1$\{\}$}}
\def\eqn#1#2{\xdef #1{(\secsym\the\meqno)}\writedef{#1\leftbracket#1}%
\global\advance\meqno by1$$#2\eqno#1\eqlabeL#1$$}
%
%			 footnotes
\newskip\footskip\footskip14pt plus 1pt minus 1pt %sets footnote baselineskip
\def\footnotefont{\ninepoint}\def\f@t#1{\footnotefont #1\@foot}
\def\f@@t{\baselineskip\footskip\bgroup\footnotefont\aftergroup\@foot\let\next}
\setbox\strutbox=\hbox{\vrule height9.5pt depth4.5pt width0pt}
\global\newcount\ftno \global\ftno=0
\def\foot{\global\advance\ftno by1\footnote{$^{\the\ftno}$}}
%
%say \footend to put footnotes at end
%will cause problems if \ref used inside \foot, instead use \nref before
\newwrite\ftfile   
\def\footend{\def\foot{\global\advance\ftno by1\chardef\wfile=\ftfile
$^{\the\ftno}$\ifnum\ftno=1\immediate\openout\ftfile=foots.tmp\fi%
\immediate\write\ftfile{\noexpand\smallskip%
\noexpand\item{f\the\ftno:\ }\pctsign}\findarg}%
\def\footatend{\vfill\eject\immediate\closeout\ftfile{\parindent=20pt
\centerline{\bf Footnotes}\nobreak\bigskip\input foots.tmp }}}
\def\footatend{}
%
%     \ref\label{text}
% generates a number, assigns it to \label, generates an entry.
% To list the refs on a separate page,  \listrefs
%
\global\newcount\refno \global\refno=1
\newwrite\rfile
\def\ref{[\the\refno]\nref}
\def\nref#1{\xdef#1{[\the\refno]}\writedef{#1\leftbracket#1}%
\ifnum\refno=1\immediate\openout\rfile=refs.tmp\fi
\global\advance\refno by1\chardef\wfile=\rfile\immediate
\write\rfile{\noexpand\item{#1\ }\reflabeL{#1\hskip.31in}\pctsign}\findarg}
%	horrible hack to sidestep tex \write limitation
\def\findarg#1#{\begingroup\obeylines\newlinechar=`\^^M\pass@rg}
{\obeylines\gdef\pass@rg#1{\writ@line\relax #1^^M\hbox{}^^M}%
\gdef\writ@line#1^^M{\expandafter\toks0\expandafter{\striprel@x #1}%
\edef\next{\the\toks0}\ifx\next\em@rk\let\next=\endgroup\else\ifx\next\empty%
\else\immediate\write\wfile{\the\toks0}\fi\let\next=\writ@line\fi\next\relax}}
\def\striprel@x#1{} \def\em@rk{\hbox{}} 
\def\lref{\begingroup\obeylines\lr@f}
\def\lr@f#1#2{\gdef#1{\ref#1{#2}}\endgroup\unskip}
\def\semi{;\hfil\break}
\def\addref#1{\immediate\write\rfile{\noexpand\item{}#1}} %now unnecessary
\def\startrefs#1{\immediate\openout\rfile=refs.tmp\refno=#1}
\def\xref{\expandafter\xr@f}\def\xr@f[#1]{#1}
\def\refs#1{\count255=1[\r@fs #1{\hbox{}}]}
\def\r@fs#1{\ifx\und@fined#1\message{reflabel \string#1 is undefined.}%
\nref#1{need to supply reference \string#1.}\fi%
\vphantom{\hphantom{#1}}\edef\next{#1}\ifx\next\em@rk\def\next{}%
\else\ifx\next#1\ifodd\count255\relax\xref#1\count255=0\fi%
\else#1\count255=1\fi\let\next=\r@fs\fi\next}
%

%
% this is ugly, but moore insists
\newwrite\ffile\global\newcount\figno \global\figno=1
\def\fig{fig.~\the\figno\nfig}
\def\nfig#1{\xdef#1{fig.~\the\figno}%
\writedef{#1\leftbracket fig.\noexpand~\the\figno}%
\ifnum\figno=1\immediate\openout\ffile=figs.tmp\fi\chardef\wfile=\ffile%
\immediate\write\ffile{\noexpand\medskip\noexpand\item{Fig.\ \the\figno. }
\reflabeL{#1\hskip.55in}\pctsign}\global\advance\figno by1\findarg}
\def\xfig{\expandafter\xf@g}\def\xf@g fig.\penalty\@M\ {}
\def\figs#1{figs.~\f@gs #1{\hbox{}}}
\def\f@gs#1{\edef\next{#1}\ifx\next\em@rk\def\next{}\else
\ifx\next#1\xfig #1\else#1\fi\let\next=\f@gs\fi\next}
\newwrite\lfile
{\escapechar-1\xdef\pctsign{\string\%}\xdef\leftbracket{\string\{}
\xdef\rightbracket{\string\}}\xdef\numbersign{\string\#}}

\def\writestop{\def\writestoppt{\immediate\write\lfile{\string\pageno%
\the\pageno\string\startrefs\leftbracket\the\refno\rightbracket%
\string\def\string\secsym\leftbracket\secsym\rightbracket%
\string\secno\the\secno\string\meqno\the\meqno}\immediate\closeout\lfile}}
\def\writestoppt{}\def\writedef#1{}
\def\seclab#1{\xdef #1{\the\secno}\writedef{#1\leftbracket#1}\wrlabeL{#1=#1}}
\def\subseclab#1{\xdef #1{\secsym\the\subsecno}%
\writedef{#1\leftbracket#1}\wrlabeL{#1=#1}}
\newwrite\tfile \def\writetoca#1{}
\def\leaderfill{\leaders\hbox to 1em{\hss.\hss}\hfill}
%	use this to write file with table of contents
\def\writetoc{\immediate\openout\tfile=toc.tmp 
   \def\writetoca##1{{\edef\next{\write\tfile{\noindent ##1 
   \string\leaderfill {\noexpand\number\pageno} \par}}\next}}}
%       and this lists table of contents on second pass

%
\catcode`\@=12 % at signs are no longer letters
%
%	Unpleasantness in calling in abstract and title fonts
\edef\tfontsize{\ifx\answ\bigans scaled\magstep3\else scaled\magstep4\fi}
 \tfontsize  \tfontsize
 \tfontsize \font\titlei=cmmi10 \tfontsize
\font\titleis=cmmi7 \tfontsize \font\titleiss=cmmi5 \tfontsize
\font\titlesy=cmsy10 \tfontsize \font\titlesys=cmsy7 \tfontsize
\font\titlesyss=cmsy5 \tfontsize  \tfontsize
\skewchar\titlei='177 \skewchar\titleis='177 \skewchar\titleiss='177
\skewchar\titlesy='60 \skewchar\titlesys='60 \skewchar\titlesyss='60
 \ifx\answ\bigans\else scaled\magstep1\fi
\ifx\answ\bigans\else

 \font\absi=cmmi10 scaled\magstep1
\font\absis=cmmi7 scaled\magstep1 \font\absiss=cmmi5 scaled\magstep1
\font\abssy=cmsy10 scaled\magstep1 \font\abssys=cmsy7 scaled\magstep1
\font\abssyss=cmsy5 scaled\magstep1 
\skewchar\absi='177 \skewchar\absis='177 \skewchar\absiss='177
\skewchar\abssy='60 \skewchar\abssys='60 \skewchar\abssyss='60
\fi
\font\ninerm=cmr9 \font\sixrm=cmr6 \font\ninei=cmmi9 \font\sixi=cmmi6 
\font\ninesy=cmsy9 \font\sixsy=cmsy6 \font\ninebf=cmbx9 
\font\nineit=cmti9 \font\ninesl=cmsl9 \skewchar\ninei='177
\skewchar\sixi='177 \skewchar\ninesy='60 \skewchar\sixsy='60 
\def\ninepoint{\def\rm{\fam0\ninerm}% switch to footnote font
\textfont0=\ninerm \scriptfont0=\sixrm \scriptscriptfont0=\fiverm
\textfont1=\ninei \scriptfont1=\sixi \scriptscriptfont1=\fivei
\textfont2=\ninesy \scriptfont2=\sixsy \scriptscriptfont2=\fivesy
\textfont\itfam=\ninei \def\it{\fam\itfam\nineit}\def\sl{\fam\slfam\ninesl}%
\textfont\bffam=\ninebf \def\bf{\fam\bffam\ninebf}\rm} 
%
%---------------------------------------------------------------------
%

\hyphenation{anom-aly anom-alies coun-ter-term coun-ter-terms}
\def\inv{^{\raise.15ex\hbox{${\scriptscriptstyle -}$}\kern-.05em 1}}

\def\Dsl{\,\raise.15ex\hbox{/}\mkern-13.5mu D} %this one can be subscripted
\def\dsl{\raise.15ex\hbox{/}\kern-.57em\partial}

 %pound sterling
\def\lspace{\ifx\answ\bigans{}\else\qquad\fi}
\def\lbspace{\ifx\answ\bigans{}\else\hskip-.2in\fi} % $$\lbspace...$$
\def\boxeqn#1{\vcenter{\vbox{\hrule\hbox{\vrule\kern3pt\vbox{\kern3pt
	\hbox{${\displaystyle #1}$}\kern3pt}\kern3pt\vrule}\hrule}}}
\def\mbox#1#2{\vcenter{\hrule \hbox{\vrule height#2in
		\kern#1in \vrule} \hrule}}  %e.g. \mbox{.1}{.1}
%	matters of taste
%\def\tilde{\widetilde} \def\bar{\overline} \def\hat{\widehat}
%
% some sample definitions
  %     curly letters

\def\darr#1{\raise1.5ex\hbox{$\leftrightarrow$}\mkern-16.5mu #1}
 %pound sterling

\def\half{{\textstyle{1\over2}}} %puts a small half in a displayed eqn
 %ditto 4/9
 %ditto 3/2
\def\roughly#1{\raise.3ex\hbox{$#1$\kern-.75em\lower1ex\hbox{$\sim$}}}
\def\msbar{\overline{\rm MS}}

\def\p2inf{\mathrel{\mathop{\sim}\limits_{\scriptscriptstyle
{p^2 \rightarrow \infty }}}}
\def\kap2inf{\mathrel{\mathop{\sim}\limits_{\scriptscriptstyle
{\kappa \rightarrow \infty }}}}
\def\x2inf{\mathrel{\mathop{\sim}\limits_{\scriptscriptstyle
{x \rightarrow \infty }}}}
\def\Lam2inf{\mathrel{\mathop{\sim}\limits_{\scriptscriptstyle
{\Lambda \rightarrow \infty }}}}
\def\frac#1#2{{{#1}\over {#2}}}
\def\half{\hbox{${1\over 2}$}}

\def\Gev{{\rm GeV}}

\def\lsim{\mathrel{mathpalette\@v1000ersim<}}
\def\gsim{\mathrel{mathpalette\@versim>}}

\catcode`@=11 %This allows us to modify plain macros
\def\slash#1{\mathord{\mathpalette\c@ncel#1}}
 \def\c@ncel#1#2{\ooalign{$\hfil#1\mkern1mu/\hfil$\crcr$#1#2$}}
\def\lsim{\mathrel{\mathpalette\@versim<}}
\def\gsim{\mathrel{\mathpalette\@versim>}}
 \def\@versim#1#2{\lower0.2ex\vbox{\baselineskip\z@skip\lineskip\z@skip
       \lineskiplimit\z@\ialign{$\m@th#1\hfil##$\crcr#2\crcr\sim\crcr}}}
\catcode`@=12 %at signs are no longer letters

\def\PR{{\it Phys.~Rev.~}}

\def\NP{{\it Nucl.~Phys.~}}
\def\PL{{\it Phys.~Lett.~}}

\def\AP{{\it Ann.~Phys.~}}

\def\ZP{{\it Zeit.~Phys.~}}

\def\vol#1{{\bf #1}}
\def\vyp#1#2#3{\vol{#1} (#2) #3}

\def\Asl{\raise.15ex\hbox{/}\mkern-11.5mu A}
\def\psl{\lower.12ex\hbox{/}\mkern-9.5mu p}
\def\qsl{\lower.12ex\hbox{/}\mkern-9.5mu q}
\def\rsl{\lower.03ex\hbox{/}\mkern-9.5mu r}
\def\ksl{\raise.06ex\hbox{/}\mkern-9.5mu k}

%%%%%%%%%%%%%%%%%%%%%%%%%%%%%%%%%%%%%%%%%%%%%%%%%%%%%%%%%%%%%%%%%%%%%%%%%%%%%%%

\pageno=0\nopagenumbers\tolerance=10000\hfuzz=5pt
\line{\hfill RAL-TR-97-049}
\vskip 36pt
\centerline{\bf An Ordered Analysis of Heavy Flavour}
\vskip 8pt
\centerline{\bf Production In Deep Inelastic Scattering.}
\vskip 36pt
\centerline{R.~S.~Thorne and R.~G.~Roberts}
\vskip 12pt
\centerline{\it Rutherford Appleton Laboratory,}
\centerline{\it Chilton, Didcot, Oxon., OX11 0QX, U.K.}
\vskip 0.9in
{\narrower\baselineskip 10pt
\centerline{\bf Abstract}
\medskip
At low $Q^2$, charm production in deep--inelastic scattering is adequately 
described by assuming generation in electroweak boson--light parton
scattering (dominantly boson--gluon fusion) which naturally incorporates the 
correct threshold behaviour. At high $Q^2$ this description is inadequate, 
since it does not sum logs in $Q^2/m_c^2$, and is replaced by the treatment of 
the charm quark as a light parton. We show how the problem of going from one 
description to the other can be solved in a satisfactory manner to all orders.
The key ingredient is the constraint of matching the evolution of the physical
structure function $F_2$ order by order in $\alpha_s(Q^2)$ in addition to the 
matching of the value of $F_2$ itself. This leads to new expressions for the 
coefficient functions associated with the charm parton which are unique in
incorporating both the correct threshold and asymptotic behaviours at each
order in perturbation theory. The use of these improved coefficients lead to 
an improvement in global fits and an excellent description of the observed 
$F_{2,charm}$.}
   
\vskip 0.7in
\line{RAL-TR-97-049\hfill}
\line{September 1997\hfill}
\vfill\eject
\footline={\hss\tenrm\folio\hss}

%%%%%%%%%%%%%%%%%%%%%%%%%%%%%%%%%%%%%%%%%%%%%%%%%%%%%%%%%%%%%%%%%%%%%%%%%%%%%%%

%%%%%%%%%%%%%%%%%%%%%%%%%%%%%%%%%%%%%%%%%%%%%%%%%%%%%%%%%%%%%%%%%%%%%%%%%%%%%%%

\newsec{Introduction.}

The factorization theory in QCD \ref\factthe{J.C. Collins, D.E. Soper and 
G. Sterman, in: Perturbative Quantum Chromodynamics, ed. A.H. Mueller 
(World Scientific, Singapore, 1989), and references therein.}    
has existed for many years, and has been one 
of the triumphs of quantum field theory. However, in its original form it 
does not take account of a number of possible complications, i.e. it 
exists only for massless particles, and its ordering does not take account 
of possible enhancements at high orders in $\alpha_s$ at small $x$. 
Until the past few years both of these complications were
not of any real phenomenological importance. The lowest values of $x$ probed 
were large enough that it was unimportant to consider small
$x$ enhancement. Also, the up, down and strange quarks were considered light  
enough to be treated as massless whenever one was within the realm of 
perturbative QCD. Furthermore,
there was little data on the charm contribution to 
the structure function and it was generally such a small component of the 
total structure function that it could be treated very approximately.  

Both of the above complications have recently become a great deal more 
important 
due to the advent of HERA. This now probes structure functions at far lower 
values of $x$ than any previous experiments, as low as $x\sim 10^{-5}$
\ref\hone{H1 collaboration: \NP \vyp{B470}{1996}{3};
\NP \vyp{B497}{1997}{3}.}\ref\zeus{ZEUS
collaboration: M. Derrick {it et al}., \ZP \vyp{C69}{1996}{607};
\ZP \vyp{C72}{1996}{399}.}, and
the treatment of structure functions should really take proper account of any 
small $x$ complications. Also, the small $x$ structure functions now have a 
contribution due to the charm structure function which is far from 
insignificant, i.e. it can be more than $20\%$ of the total structure 
function and, moreover, in the past couple of years direct measurement of the 
charm structure function has also become possible 
\ref\honecharm{H1 collaboration: C. Adloff {\it et al}., \ZP 
\vyp{C72}{1996}{593}.}\ref\zeuscharm{ZEUS collaboration: J. Breitweg 
{\it et al}., preprint DESY-97-089.}. This has made it essential
to treat the contribution to the structure function due to massive
quarks in a correct manner.     

In this paper we propose a new method for the treatment of 
heavy quarks in structure functions. We begin by describing
the features a correct treatment must exhibit at both high and low $Q^2$,
and the techniques used in either of these limits. We then give a 
discussion of the correct way to take account of heavy quarks 
in a well--ordered manner over the full range of $Q^2$, 
showing how this relates to present treatments, and in particular 
demonstrating that one may choose to evolve the partons according to the 
massless evolution equations. However, we shall see that the detailed 
construction of the coefficient functions required is extremely
difficult if not impossible. Therefore, we provide instead 
a prescription for calculating structure functions including heavy quark 
effects which is somewhat simpler than the strictly correct treatment, and 
which is directly analogous to the normal manner in which one calculates 
order by order for massless partons, but which is  in practice essentially 
identical to the strictly correct treatment. Finally, we will present the 
results of a comparison of our method to data: both that for full structure 
functions and for the charm component to the structure functions. These
comparisons turn out to be very good. We also make predictions for 
the charm component of the longitudinal structure function. 
Complications due to leading $\ln(1/x)$ terms 
at all orders in $\alpha_s$ are ignored, and while a 
correct treatment of structure functions
should of course deal with this problem, we feel that this would 
overcomplicate our presentation, and besides we wish to compare directly 
with normal NLO in $\alpha_s$ approaches. A paper which takes account of both 
small $x$ complications and massive partons is in preparation.   
 
\newsec{Structure Functions With Massive Quarks.}

We consider the case of $n_f$ massless quarks and one heavy quark.
One of the simplest ways to deal with heavy flavour production in 
deep--inelastic scattering is to treat the mass of the heavy quark, $M$,
as a hard scale \ref\witten{E. Witten, \NP \vyp{B104}{1976}{445}.}. In 
this case the $n_{f}$ light quarks are always treated as partons, but all 
other quarks are never treated as partons at any scale: the cross--section 
for production of heavy quarks is expressed entirely in terms of 
coefficient functions depending on the heavy quark mass convoluted with 
parton distributions which only depend on light partons, i.e.
\eqn\heavycs{\sigma_i(x,Q^2,M^2)=\sum_{a} C^{FF}_{ia}
(Q^2/\mu^2, Q^2/M^2)\otimes f^{n_f}_a(\mu^2)+ {\cal O}(\Lambda^2/M^2),}
where $\sigma_i(x,Q^2,M^2)$ is the cross-section for scattering off a   
particular quark, either heavy or light,
 and $a$ runs over the light partons, i.e. the gluon and the 
$n_f$ light quarks. This approach is very
well--defined in theoretical terms, essentially being a simple 
generalization of the usual factorization theorem, 
with \heavycs\ being valid to all orders up to the higher twist corrections of 
${\cal O}(\Lambda^2/M^2)$. 

This approach is adopted by a number of groups \ref\grs{M. Gl\"uck, E. Reya 
and M. Stratmann, \NP \vyp{B422}{1994}{37}\semi
M. Gl\"uck, E. Reya and A. Vogt, \ZP \vyp{C67}{1995}{433}.},
and is usually known as the 
fixed flavour number scheme (FFNS). It is normally used in the particular
renormalization scheme where all diagrams with no heavy quark lines are 
renormalized in the $\msbar$ scheme, while those with heavy quarks are
renormalized at zero momentum. This is particularly convenient because 
the effect of the heavy particle decouples from the light sector, 
in particular the coupling is the 3 flavour $\msbar$ coupling and the 
light parton distributions evolve as in the 3 flavour $\msbar$ scheme.
The $C^{FF}_{ia}(z,Q^2/\mu^2, Q^2/M^2)$ have all been calculated to 
${\cal O}(\alpha_s)$ \witten\ref\shifman{M.A. Shifman, A.I. Vainshtein and 
V.I. Zakharov, \NP \vyp{B136}{1978}{157}.}\ref\gr{M. Gl\"uck and E. Reya, \PL 
\vyp{B83}{1979}{98}.} and ${\cal O}(\alpha_s^2)$ \ref\nlocalc{E. Laenen, S. 
Riemersma, J. Smith and W.L. van Neerven, \NP \vyp{B392}{1993}{162}\semi
B.W. Harris and J. Smith, \NP \vyp{B452}{1995}{109}.}
in this scheme, though analytic expressions 
only exist at ${\cal O}(\alpha_s)$. 

In principle this approach is a very good way to calculate the effects of
heavy quarks in structure functions. At each order it incorporates the 
kinematical threshold in the light parton--photon centre of mass energy
$\hat W^2\equiv Q^2(z^{-1}-1)\geq 4M^2$ in a smooth manner 
(which then guarantees the same smooth threshold in the invariant mass of the 
hadronic remnant $W^2$, up to proton mass corrections)
and the coefficient functions are calculated order by order in precisely 
the same manner as the light particle coefficient functions (though the 
actual calculations are rather more difficult). However, it does have one 
major shortcoming. As one calculates to higher orders in $\alpha_s$ 
one encounters higher powers of $\ln (Q^2/M^2)$ and $\ln(\mu^2/M^2)$. 
Letting $\mu^2=Q^2$, and thus eliminating all logs in $Q^2/\mu^2$, 
then for $Q^2\to \infty$ the coefficients at $m_{th}$--order in 
$(\alpha_s(Q^2)/2\pi)^m$ have the series expansion
\eqn\heavcoeff{ C^{FF,m}_{ia}(z,Q^2/M^2)=
\sum_{n=0}^m f_n(z)
\ln^n(Q^2/M^2).}
Thus, working order--by--order in $\alpha_s$ in this approach one is failing
to take account of these large logs. This is not only a practical concern
in the sense that these large logs in $Q^2/M^2$ at higher orders in 
$\alpha_s$ can potentially be phenomenologically important\foot{They are
not important for $Q^2<<M^2$ because the large logs are killed by 
factors coming from the kinematical threshold.}, but is also a 
theoretical concern insofar as at each order in $\alpha_s$ the 
leading power of $\ln(Q^2/M^2)$ is the leading in $\alpha_s$ part of the
overall coefficient function with this $\ln(Q^2/M^2)$ behaviour, and is really
part of the leading order expression for the structure function as a whole. 
The same reasoning applies for the next--to--leading power of 
$\ln(Q^2/M^2)$ {\it etc.}.
This is similar in principle to the problem of increasing powers of 
$\ln(1/x)$ with increasing powers of $\alpha_s$. It is more difficult in 
one sense, in so much that in the expressions for the coefficient functions 
these large logs in $(Q^2/M^2)$ are hidden within very complicated expressions.
However, it is far simpler in the particular limit $Q^2>>M^2$ because we 
know exactly how to sum the logarithms in $Q^2$, i.e. we solve the 
renormalization group equation for fixed order in $\alpha_s$.   

Thus, in order to sum these large logs in $Q^2/M^2$ it is convenient to 
consider the heavy quark to be a parton and for its distribution function 
to satisfy the renormalization group (DGLAP) equations as do the light 
partons.  An extremely simple approach which incorporates this idea is 
the zero--mass variable flavour number scheme (ZM-VFNS). 
This treats the massive
parton as being infinitely massive below some threshold in $\mu^2$,
and totally massless
above the threshold, i.e. all coefficient functions coupling directly to the 
charm quark turn on at the threshold, the evolution of the charm quark 
begins at this threshold, and the number of flavours in
the coefficient functions, anomalous dimensions and the 
running coupling constant 
increases by one to $n_f+1$ discontinuously at the threshold. 
Despite the simplicity of 
the approach this procedure must in principle be done with care if 
the correct results are to be obtained in the asymptotic limits 
\ref\colltung{J.C. Collins and W.K. Tung, \NP \vyp{278}{1986}{934}.}
(see below for details). In particular,
the decoupling theorem tells one how the coupling constant must change in 
order to get the correct results well below threshold. Also,
the parton distributions just above the chosen threshold must be carefully 
defined in terms of those below threshold in order to
guarantee that the correct result is obtained as $Q^2 \to \infty$. 
In practice at low orders 
the situation is relatively simple, e.g., if the threshold 
is chosen to be precisely $\mu^2=M^2$,
then at next to leading order, the light parton distributions are continuous 
across the threshold (in $\msbar$ scheme) and the evolution of the charm
parton distribution begins 
from zero. At higher orders the parton distributions must change 
discontinuously across the threshold and in particular the charm 
evolution must begin from a non--zero value. 

For many years the above approach was that most commonly used in global fits. 
The CTEQ collaboration used the approach at next to leading order, as
explained above \ref\cteq{H.L. Lai {\it et al}., \PR \vyp{D55}{1997}{1280}.}, 
while the MRS collaboration motivated their choice of 
threshold by phenomenological considerations rather than the strict 
theoretical ones \ref\mrs{A.D. Martin , R.G. Roberts and W.J. Stirling,
\PR \vyp{D50}{1994}{6734}.}, but in practice this resulted in a very 
similar choice of 
threshold (i.e. $2.7 \Gev^2$ for MRS compared to $2.56\Gev^2$ for CTEQ).
While the charm contribution to the structure functions near the region of 
threshold was not too important this simple treatment was perfectly adequate. 
However, it is clear from its construction that it will not suffice  
as a good description of charm production in the region of the charm 
threshold. In particular charm production has a sharp threshold at a 
chosen $\mu^2$, rather than a smooth threshold in $W^2$. 

\medskip

Hence, some approach which extrapolates smoothly from the FFNS at low $Q^2$ 
to the ZM-VFNS at high $Q^2$ is required in order to produce a good 
description of the effect of heavy quarks on structure functions over the
whole range of $Q^2$. Let us discuss how this may be achieved. In order to 
do this we first put the ZM-VFNS on a more solid theoretical footing. If we 
regard the quark mass $M^2$ as being a soft scale then the factorization 
theorem tells us that
\eqn\facttheh{\sigma_i(x,Q^2,M^2)=\sum_{b}C^{n_f+1}_{ib}(Q^2/\mu^2)\otimes 
f^{n_f+1}_b(\mu^2, M^2/\mu^2)+ {\cal O}(M^2/\mu^2),}
where $b$ runs over the light partons and the massive quark.
We are able to remove the large logs in $Q^2/M^2$ from the coefficient 
functions, and hence obtain the normal massless coefficient functions, 
and absorb them into the definition of the parton distributions
at the expense of having potential ``higher twist'' corrections of 
${\cal O}(M^2/\mu^2)$. While the parton distributions depend on $M^2$,
if the operators defining the partons undergo 
ultraviolet operator regularization in the $\msbar$ scheme
then their evolution depends only on the anomalous dimensions obtained
from this ultraviolet regularization. These are independent of the mass
of the heavy parton, and the evolution is as if for $n_f+1$ massless
quarks in the $\msbar$ scheme. Hence we have the formal definition of the 
ZM--VFNS which will become exact for $Q^2>>M^2$.  

However, we have one more degree of freedom in \facttheh\ than in \heavycs,
i.e. we have the heavy parton distribution to parameterize at some 
arbitrary starting scale for evolution,
and also no apparent reference to the mass--scale $M^2$ in the definition of 
the parton distributions. This is not in fact true since it can be shown that
\eqn\partdef{f^{n_f+1}_b(z,\mu^2,\mu^2/M^2)=A^{ba}(\mu^2/M^2)\otimes 
f^{n_f}_a(\mu^2),}
where the operator matrix elements $A^{ba}(z,\mu^2/M^2)$ contain logs 
in $(\mu^2/M^2)$, and are calculable 
order by order in perturbation theory \ref\vanneerone{M. Buza {\it et al}.,
\NP \vyp{B472}{1996}{611}.}\ref\vanneertwo{M. Buza {\it et al}., 
preprint DESY 96-278, {\tt hep-ph/9612398}, to be published in \ZP C.}. Hence,
the partons in the ZM-VFNS can in fact be generated from those in the 
FFNS at all $\mu^2$ by using the leading logarithmic expressions for the 
operator matrix elements and the expression \partdef,
rather than using the four--flavour evolution equations at all. 
Indeed, if the starting
scale is chosen as $\tilde \mu^2 \not= M^2$ then strictly speaking all the 
leading logs in $(\tilde \mu^2/M^2)$ should be included in the matching 
condition which is just as complicated as using \partdef\ at all scales. 
However, if the scale at which
evolution begins is precisely $\mu^2=M^2$, then the matching condition 
for the partons in the two schemes is a power series in $\alpha_s$
with no logs. Therefore, 
it simplest to use \partdef\ only to define the order--by--order parton 
distributions at the starting scale, 
and then to calculate the parton distributions at other scales by
evolving using $n_f+1$ massless flavours.
This procedure guarantees the 
correctness of the ZM-VFNS calculation in the limit $Q^2>>M^2$.

By comparing the expressions \facttheh\ and \heavycs\ at $Q^2>>M^2$, and
using the relationship \partdef, one can calculate the FFNS coefficient
functions, up to ${\cal O}(M^2/Q^2)$ corrections in terms of the massless
$\msbar$ coefficient functions for $n_f+1$ flavours \vanneerone\vanneertwo, 
i.e.  
\eqn\vnetalone{C^{FF}_{ia}(z,Q^2/\mu^2,Q^2/M^2)= C^{n_f+1}_{ib}(Q^2/\mu^2)
\otimes A^{ba}(\mu^2/M^2) +{\cal O}(M^2/\mu^2).}
The detailed expressions of this form can be found in 
\vanneertwo, where they
are used to calculate the $Q^2\to \infty$ limit of the heavy quark coefficient
functions in terms of the known light quark coefficient functions and 
calculated operator matrix elements. These authors
then define $F^{ASYMP}$ as the structure function obtained from these
asymptotic expressions for the coefficient functions and the parton 
distribution in \heavycs. They then, through purely phenomenological
motivation, define a variable flavour number scheme 
\vanneertwo\ref\vanneerthree{M. Buza {\it et al}., preprint DESY 97-124,
{\tt hep-ph/9707263}.}
by the formal definition
\eqn\vnvfns{F^{VFNS}=F^{ZM-VFNS}-F^{ASYMP}+F^{FFNS}.}
This then extrapolates smoothly from one limit to the other, being
guaranteed to reduce to the correct limit order by order in $\alpha_s(Q^2)$
at high $Q^2$, though only approximately to $F^{FFNS}$ order by order 
at low $Q^2$.  

\newsec{ A Complete Treatment of Charm Mass Corrections.}

Although we agree with \vnetalone\ and hence with the 
results at high $Q^2$ regarding coefficient functions
in \vanneerone\vanneertwo, we believe one may be more 
ambitious. Rather than simply accepting the uncertainty of 
${\cal O}(M^2/\mu^2)$ in \vnetalone\ we can be more systematic and
demand that there is a scheme which uses the definition of the parton 
distributions in \facttheh\ and \partdef\ but which is correct up to 
${\cal O}(\Lambda^2/M^2)$. Inserting \partdef\ into \facttheh\ and subtracting
from \heavycs\ it is clear that the difference is
\eqn\diff{ c_i^a(M^2/\mu^2)\otimes f^{n_f}_a(\mu^2)=\bigl[
C^{FF}_{ia}(Q^2/\mu^2, Q^2/M^2)-C^{n_f+1}_{ib}(Q^2/\mu^2)\otimes 
A^{ba}(\mu^2/M^2)\bigr]\otimes f^{n_f}_a(\mu^2),}
where $c_i^a(z,M^2/\mu^2)$ is ${\cal O}(M^2/\mu^2)$.
Also making use of \partdef\ this difference can be written as 
\eqn\diffa{c_i^a(M^2/\mu^2)\otimes (A^{ba}(M^2/\mu^2))^{-1}\otimes
f^{n_f+1}_b(\mu^2,\mu^2/M^2), }
and so $c_i^a(M^2/\mu^2)\otimes (A^{ba}(M^2/\mu^2))^{-1}$ plays the 
role of the correction to the massless coefficient functions which
accounts for the ${\cal O}(M^2/\mu^2)$ corrections. Defining a corrected
coefficient function as
\eqn\corrcoeff{\eqalign{C^{VF}_{ib}(z,Q^2/\mu^2, Q^2/M^2) &= 
C^{n_f+1}_{ib}(z,Q^2/\mu^2)
+c_i^a(M^2/\mu^2)\otimes (A^{ba}(M^2/\mu^2))^{-1}\cr
&=C^{FF}_{ia}(Q^2/\mu^2, Q^2/M^2)\otimes
(A^{ba}(\mu^2/M^2))^{-1},\cr} }
then we have the factorization theorem 
\eqn\facttheha{\sigma_i(x,Q^2,M^2)=\sum_{a}
C^{VF}_{ib}(Q^2/\mu^2, M^2/\mu^2)\otimes 
f^{n_f+1}_b(\mu^2 ,M^2/\mu^2)+ {\cal O}(\Lambda^2/M^2).}

Thus, \facttheha\ gives us a method for defining the structure function 
including a heavy quark up to errors of ${\cal O}(\Lambda^2/M^2)$ 
but where all partons evolve according to the massless evolution equations.
It seems sensible that the best way to proceed for the calculation of 
structure functions in the presence of a heavy quark is to use the 
FFNS up to some scale of ${\cal O}(M^2)$ and then switch to the scheme 
defined by \facttheha\ above this scale. (Changes of renormalization scheme 
across threshold applying to situations of this general type were 
first proposed in 
\ref\cwz{J.C. Collins, F. Wilczek and A. Zee, \PR \vyp{D18}{242}{1978}.}.) 
We shall call this a variable flavour number scheme (VFNS).
Moreover, we believe that it is sensible to choose the renormalization
and factorization scale $\mu^2=Q^2$ in both schemes, for all scales
and for both light and heavy quark structure functions.\foot{Of course, if
we reach sufficiently low $Q^2$ then we must introduce some finite 
renormalization scale in order to have a finite expression for heavy quark
photoproduction. Since we only consider $Q^2>1\Gev^2$ we do not consider this
problem in this paper.}   
This very simple choice automatically avoids having different scales for 
different components of the complete structure function, and means that all
mass effects are contained entirely within the coefficient functions.
It also agrees with the 
normal asymptotic choice of $\mu^2=Q^2$ and removes all problems of logs of 
$Q^2/\mu^2$ (the solution of the evolution equations summing such terms) and 
$\mu^2/M^2$, and we are left just with the problems of 
$\ln(Q^2/M^2)$.\foot{In the asymptotic expressions for the FFNS coefficient 
functions in \vanneerone, this choice leads to significant 
simplification.} This choice is expressed explicitly in all our equations 
from now on. Finally, as already mentioned, 
if we choose the transition scale as precisely $\mu^2\equiv Q^2=M^2$ then all
the logs in $Q^2/M^2$ disappear, and the matching conditions between 
the partons in the two schemes in \partdef\ are a simple power series in 
$\alpha_s(M^2)$. Thus, performing the matching at $M^2$, and
solving order by order in $\alpha_s(Q^2)$ as in the strictly massless case 
we are guaranteed to sum the logs in $Q^2/M^2$ correctly at zeroth order
in $M^2/Q^2$. Combining with the mass corrected coefficient functions to the
appropriate order, we should then get the mass corrected structure 
functions correctly order by order. Unfortunately, the procedure is not 
quite as simple as this.    

We see that the defining expression for $C^{VF}_{ia}(z,Q^2/M^2)$ is 
in fact of exactly the same form as \vnetalone, 
except that it is now exact at all $Q^2$, rather than having 
corrections of ${\cal O}(M^2/Q^2)$, and that this time it is the 
$n_f+1$ flavour coefficient functions which are the unknowns to be solved 
in terms of the FFNS coefficient functions and the operator matrix 
elements, rather than the asymptotic form of the FFNS coefficient functions. 
However, this leaves us with an ambiguity. The index $a$ runs over the 
gluon and the light quarks while $b$ also includes the heavy quark. Hence, 
while the asymptotic FFNS coefficient functions in \vnetalone\ were defined 
uniquely in terms of the light $n_f+1$ coefficient functions, solving
\corrcoeff\ for the $C^{VF}_{ia}(z,Q^2/M^2)$ in terms of 
the FFNS coefficient functions does not lead to a unique solution. 

In order to demonstrate this let us write out our equations for the 
VFNS in full. For the case where the photon couples directly to the heavy 
quark, $H$,  we have two equations: 
\eqn\bosglufus{C^{FF,S}_{Hg} = A^S_{gg,H}\otimes 
C^{VF,S}_{Hg} + n_f A^S_{qg,H} \otimes C^{VF,PS}_{Hq}
+A^S_{Hg}\otimes\bigl[C^{VF,NS}_{HH}+C^{VF,PS}_{HH}\bigr]}
and
\eqn\bosglufusi{C^{FF,S}_{Hq} = A^{PS}_{Hq}\otimes 
\bigl[C^{VF,NS}_{HH} + C^{VF,PS}_{HH}\bigr] +\bigl[A^{NS}_{qq,H} 
+n_f A^{PS}_{qq,H}\bigr]\otimes C^{VF,PS}_{Hq}+
A^S_{gq,H}\otimes C^{VF,S}_{Hg},}
where $S$, $NS$ and $PS$ refer to the flavour singlet, non--singlet and 
pure--singlet (singlet minus non--singlet) respectively. 
In the case where the photon couples directly to a light quark we have three
equations. Denoting the massless $\msbar$ coefficient functions with $n_f$
light flavours by $C_{ia}(n_f)$ and the contributions to the light flavour
coefficient functions in the FFNS due to heavy quark generation by 
$C^{FF}_{ia}$ we have 
\eqn\bosglufusii{C^{NS}_{qq} +C^{FF,NS}_{qq} = A^{NS}_{qq,H}\otimes 
C^{VF,NS}_{qq},}
\eqn\bosglufusiii{C^S_{qg}+ C^{FF,S}_{qg} = A^{S}_{qg,H}
\otimes C^{VF,NS}_{qq} + A^{S}_{gg,H} \otimes C^{VF,S}_{qg}
+n_f A^S_{qg,H}\otimes C^{VF,PS}_{qq}+ A^S_{Hg}\otimes C^{VF,PS}_{qH}}
and
\eqn\bosglufusiv{ C^{PS}_{qq}+ C^{FF,PS}_{qq} = \bigl[
A^{NS}_{qq,H}+n_f A^{PS}_{qq,H}\bigr]
\otimes C^{VF,PS}_{qq} + A^{PS}_{Hq} \otimes C^{VF,PS}_{qH}
+A^{PS}_{qq,H}\otimes C^{VF,NS}_{qq}+ A^S_{gq,H}\otimes C^{VF,S}_{qg}.}
These are very similar to the equations (2.31)--(2.35) in 
\vanneertwo\ and, as in those equations, we have implicitly divided all pure
singlet quantities coupling to quarks and all singlet quantities coupling to  
gluons by $n_f$. Also, as in these previous equations, it is implicit 
that all quantities on the left--hand side are expanded in the $n_f$--flavour
$\msbar$ coupling constant while those on the right--hand side are 
expanded in terms of the $(n_f+1)$--flavour $\msbar$ coupling. 
The relationship between the two couplings 
was calculated in \ref\coupchangeone{W. 
Bernreuther and W. Wetzel, \NP \vyp{197}{1982}{228}\semi
W. Bernreuther, \AP \vyp{151}{1983}{127}.} and corrected in 
\ref\coupchangetwo{S.A. Larin, T. van Ritbergen and J.A.M. Vermaseren,
\NP \vyp{B438}{1995}{278}.}. It is   
\eqn\couprel{\eqalign{\alpha_{s,n_f+1}(Q^2)= &\alpha_{s,n_f}(Q^2)
+ \alpha^2_{s,n_f}(Q^2){1\over 3\pi}T_f\ln(Q^2/M^2)\cr
&\hskip -0.8in 
+\alpha^3_{s,n_f}(Q^2){1\over \pi^2}\Bigl[{1\over 9}T^2_f\ln^2(Q^2/M^2)+
{1\over 12}(5C_A T_f+4C_f T_f)\ln(Q^2/M^2) +{13\over 48}T_f C_f-
{2\over 9}T_f C_A\Bigr]+\cdots,\cr}}
where the coefficient of the leading log at each order in 
$\alpha_{s,n_f}(Q^2)$ is the same in all schemes, but other coefficients 
depend on details of renormalization, in particular whether the mass $M$ is
the fixed or running mass. The particular choice above corresponds to a fixed
heavy quark mass at NLO. 
 
The difference between our expressions for the coefficient functions and
those in \vanneertwo\ is that the coefficient functions on the right--hand 
side are 
the VFNS coefficient functions. Not only does this mean that the equations 
are meant to hold including terms of ${\cal O}(M^2/Q^2)$, and that we solve 
for the coefficient functions on the right--hand side, but also that there is 
a difference between the coefficient functions which couple to the heavy quark
distribution  and those coupling to the light quark distributions. For
example, 
while $C^{VF,NS}_{HH}$ and $C^{VF,NS}_{qq}$ must be identical in the limit 
$Q^2\to \infty$ they certainly do not have to be identical at moderate
$Q^2$, and physical intuition suggests they should not be. This means that
unlike \vanneertwo\ we do not have five equations for five unknowns, but
we have five equations for eight unknowns. In order to reduce to the correct
ZM--VFNS at very high $Q^2$ we must choose definitions for the 
mass--corrected coefficient functions which reduce to the $n_f+1$ light
parton coefficient functions as $Q^2\to \infty$, but this constraint still
leaves a great deal of freedom.     

As an 
example let us consider what is in practice the most important case, 
the equation for the boson--gluon fusion coefficient function
for the heavy quark structure function $F_{2,H}(x,Q^2)$, \bosglufus.
The expansion of 
$C^{FF,S}_{2,Hg}$ begins at ${\cal O}(\alpha_s(Q^2))$ as does $
C^{VF,S}_{2,Hg}$ and $A^S_{Hg}$, while $A^S_{gg,H}$ and 
$C^{VF,NS}_{2,HH}$ begin at zeroth order. Using the known expressions
for the operator matrix elements we obtain the lowest order equation 
relating the FFNS coefficient functions and the VFNS coefficient functions
\eqn\bosglufusex{C^{FF,S,1}_{2,Hg}(z,Q^2/M^2) =  C^{VF,S,1}_{2,Hg}(z,Q^2/M^2)
+ (\ln(Q^2/M^2)+c_{rs})
P^0_{qg}\otimes C^{VF,NS,0}_{2,HH}(Q^2/M^2),}
where $P^0_{qg}(z)$ is the lowest order splitting function, and $c_{rs}$ is 
renormalization scheme dependent, but $c_{rs}=0$ in $\msbar$ scheme. Hence, we 
have freedom in how we choose our zeroth order heavy quark non--singlet 
coefficient function, and this then determines our first--order 
mass--corrected gluon coefficient function. More generally, we have 
freedom in how we define each of the three coefficient functions coupling to 
the heavy quark $C^{VF,NS}_{HH}$, $C^{VF,PS}_{HH}$ and $C^{VF,PS}_{qH}$
at each order in perturbation theory, being constrained only by the 
requirement that they are of the correct form as $Q^2\to \infty$. 

Of course, there cannot truly be an ambiguity in the order--by--order
definition of the structure functions. In order to illustrate this
consider the structure function $F_2(x,Q^2)$.
We also come back to the point concerning
renormalization scheme dependence. In order to maintain renormalization scheme 
consistency we must be very careful about the way in which we order the 
expressions. Doing this correctly does not remove the ambiguity in
our definitions of the coefficient functions, but it does render this 
ambiguity physically meaningless, even order by order. Let us consider 
specifically the heavy quark contribution to the structure function 
$F_2(x,Q^2)$ in the general VFNS. In fact we will discuss its 
$\ln Q^2$--derivative since it is the evolution of $F_{2,H}(x,Q^2)$ which is 
a more natural quantity. 
Taking the $\ln Q^2$ derivative of $F_{2,H}(x,Q^2)$ and keeping all terms up 
to ${\cal O}(\alpha_s(Q^2))$ 
multiplying the VFNS parton distributions we obtain
\eqn\derivone{\eqalign{{d F_{2,H}(x,Q^2)\over d\ln(Q^2)}=&
{d C^{VF,NS,0}_{2,HH}(Q^2/M^2)\over d\ln(Q^2)}\otimes(H(Q^2)+\bar H(Q^2))_0 \cr
&\hskip -0.4in +{\alpha_{s,n_f+1}(Q^2) \over 2\pi}C^{VF,NS,0}_{2,HH}(Q^2/M^2)
\otimes\Bigl(P^0_{qg}\otimes g^{n_f+1}_0(Q^2)+P^0_{qq}\otimes
(H(Q^2) +\bar H(Q^2))_0\Bigr)\cr
&\hskip -0.4in +{\alpha_{s,n_f+1}(Q^2) \over 2\pi}
\biggl[{d C^{VF,1}_{2,Hg}(Q^2/M^2)
\over d\ln(Q^2)}\otimes
g^{n_f+1}_0(Q^2)\cr
&\hskip 0.5in+{d C^{VF,NS,1}_{2,HH}(Q^2/M^2)\over d\ln(Q^2)}\otimes
(H(Q^2)+\bar H(Q^2))_0\biggr].\cr}}
Asymptotically the second and third term in this expression reduce to the
required form for the leading order expression in the ZM-VFNS. All other terms
fall off to zero in this limit so we are guaranteed the correct asymptotic 
expression 
using this prescription. However, at low $Q^2$ the exact form of the 
expression is highly sensitive to our particular choice of 
coefficient functions. 
This clearly means that we do not have a truly well--ordered solution
and this is because the true ordering of the 
coefficient functions $C^{VF}_{2,Hb}(z,Q^2/M^2)$
is not as simple as just order by order in $\alpha_{s,n_f+1}(Q^2)$ 
due to their 
dependence on the quark mass. Indeed, their ordering is 
the crux of the problem, and we will explore this below. 

In order to examine the true ordering of our expression we will express it 
in terms of unambiguously defined quantities, and also in terms of those 
where the ordering is relatively straightforward. 
Hence we will we will express it in terms of the FFNS parton distributions,
the mass--dependent coefficient functions $C^{VF}_{2,Hb}(z,Q^2/M^2)$,
the operator matrix elements and the coupling $\alpha_{s,n_f}(Q^2)$.
The FFNS parton distributions are 
correctly ordered simply by solving their evolution equations to a 
given order. The operator matrix elements are ordered according to the 
power of $\alpha_{s,n_f+1}(Q^2)$ minus the power of $\ln(Q^2/M^2)$, i.e. the 
leading--order term is of the form 
\eqn\loome{A^{0}_{ab}(z,Q^2/M^2)=\delta_{ab}\delta(1-z)+\sum_{n=1}^{\infty}
\biggl({\alpha_{s,n_f+1}(Q^2)\over 2\pi}\biggr)^n
\ln^n(Q^2/M^2) a_n(z).}
The $n_f+1$--flavour coupling constant is defined in terms of the 
$n_f$--flavour
coupling in an analogous manner, i.e. the leading order relation is
\eqn\locoup{\alpha_{s,n_f+1}(Q^2)=\alpha_{s,n_f}+\sum_{n=1}^{\infty}
\alpha_{s,n_f}^{n+1}(Q^2)\biggl({T_f\over 3\pi}\biggr)^n \ln^n(Q^2/M^2) .}

First using the expression for $C^{VF,1}_{2,Hg}(z,Q^2/M^2)$ \bosglufusex, 
but only 
keeping the leading order part of the operator matrix element, i.e. 
leaving out the $c_{rs}$, and substituting into \derivone\ we obtain
\eqn\derivtwo{\eqalign{{d F_{2,H}(x,Q^2)\over d\ln(Q^2)}=&
{d C^{VF,NS,0}_{2,HH}(Q^2/M^2)\over d\ln(Q^2)}\otimes(H(Q^2)+\bar H(Q^2))_0 \cr
&\hskip -0.5in +{\alpha_{s,n_f+1}(Q^2)\over 2\pi}
C^{VF,NS,0}_{2,HH}(Q^2/M^2)\otimes P^0_{qq}\otimes
(H(Q^2) +\bar H(Q^2))_0\cr
&\hskip -0.5in +{\alpha_{s,n_f+1}(Q^2)\over 2\pi}
\biggl[{d C^{FF,1}_{2,Hg}(Q^2/M^2)\over 
d\ln(Q^2)}
+{d C^{VF,NS,0}_{2,HH}(Q^2/M^2)\over d\ln(Q^2)}\otimes P^0_{qg}
\ln(Q^2/M^2)\biggr]\otimes
g^{n_f+1}_0(Q^2)\cr
&\hskip -0.5in +{\alpha_{s,n_f+1}(Q^2)\over 2\pi}
{d C^{VF,NS,1}_{2,HH}(Q^2/M^2)
\over d\ln(Q^2)}\otimes(H(Q^2)+\bar H(Q^2))_0.\cr}}

We can then be more detailed by using the
explicit expressions for $(H(z,Q^2)+\bar H(z,Q^2))_0$ 
and $g^{n_f+1}_0(z,Q^2)$ in terms of the FFNS parton distributions, i.e. 
\eqn\partdefcharm{(H(z,Q^2)+\bar H(z,Q^2))_0 ={\alpha_{s,n_f}(Q^2)\over 2\pi}
\ln(Q^2/M^2)P^0_{qg}\otimes
g^{n_f}_0(Q^2)+{\cal O}(\alpha_{s,n_f}^2(Q^2)\ln^2(Q^2/M^2)),} 
and
\eqn\partdefglu{g^{n_f+1}_0(z,Q^2) =g^{n_f}_0(z,Q^2)- 
{\alpha_{s,n_f}(Q^2)\over 6\pi} \ln(Q^2/M^2) 
g^{n_f}_0(z,Q^2)+{\cal O}(\alpha_{s,n_f}^2(Q^2)\ln^2(Q^2/M^2)),} 
and also the expression for the $n_f$--flavour coupling, \locoup.
Doing this and 
remembering that ${d C^{VF,NS,1}_{2,HH}(z,Q^2/M^2)\over d\ln(Q^2)}=
{\cal O}(M^2/Q^2)$ then we obtain
\eqn\derivthree{\eqalign{{d F_{2,H}(x,Q^2)\over d\ln(Q^2)}=
&{\alpha_{s,n_f}(Q^2)\over 2\pi}
\biggl[{d C^{FF,1}_{2,Hg}(Q^2/M^2)\over d\ln(Q^2)}
\otimes g^{n_f}_0(Q^2)\cr
&- {\alpha_{s,n_f}(Q^2)\over 6\pi} \ln(Q^2/M^2) 
{d C^{FF,1}_{2,Hg}(Q^2/M^2)\over d\ln(Q^2)}\otimes g^{n_f}_0(Q^2)\cr
&+{\alpha_{s,n_f}(Q^2)\over 2\pi}
\ln(Q^2/M^2)C^{VF,NS,0}_{2,HH}(Q^2/M^2)\otimes 
P^0_{qq}\otimes P^0_{qg} \otimes g^{n_f}_0(Q^2)\cr
&+{\cal O}(M^2/Q^2)\cdot {\cal O}(\alpha_{s,n_f}^2(Q^2)\ln^2(Q^2/M^2))\biggr].
\cr}}
Hence, as well as asymptotically reducing to the correct leading order 
expression, the prescription of keeping all terms up 
to ${\cal O}(\alpha_{s,n_f+1}(Q^2))$ which multiply the leading order VFNS 
parton 
distributions has resulted in a unique ${\cal O}(\alpha_{s,n_f}(Q^2))$
expression for the derivative of the heavy quark coefficient function 
which also (and necessarily) has the correct threshold behaviour. However,  
it is clear that the ${\cal O}(\alpha_{s,n_f}^2(Q^2)
\ln Q^2/m^2)$ expression, while
having the correct asymptotic limit, has  
behaviour for $Q^2 \sim M^2$ which is sensitive to our choice of 
coefficient functions. In particular, the behaviour of these 
terms will not generally respect the threshold in $\hat W^2$. 
It is clear that at higher 
orders in $\alpha_{s,n_f}(Q^2) \ln (Q^2/M^2)$ while we will obtain the correct 
asymptotic behaviour, our lowish $Q^2$ behaviour will be dependent on 
the choice of coefficient functions. 

If we were to use the expression for the  structure function itself,
rather than its derivative, in the VFNS by combining the lowest order in 
$\alpha_{s,n_f+1}(Q^2)$ coefficient function 
with the lowest order VFNS parton distributions, i.e. 
\eqn\struczero{F_{2,H}(x,Q^2)=C^{VF,NS,0}_{2,HH}(Q^2/M^2)\otimes 
(H(Q^2)+\bar H(Q^2))_0,}
then again we would be guaranteed the correct LO expression in the 
asymptotic limit. However, even the leading term in $\alpha_{s,n_f}(Q^2)
\ln(Q^2/M^2)$
(when expressed in terms of the FFNS parton distributions and operator 
matrix elements) is now completely dependent on the choice of coefficient
function, and there is no requirement to have the correct threshold 
behaviour at all. 

It should be no surprise that we have this problem. As mentioned 
earlier in the FFNS the coefficient functions to all orders contain 
renormalization--scheme--independent leading--order contributions.
By working in the VFNS we have managed to extract the asymptotic form of 
this leading order contribution in a relatively simple manner. However, 
in order to have the full leading order expression for the structure 
functions in the VFNS in the threshold region, we need to extract all the 
information from the leading order contribution to the FFNS coefficient
functions. In principle, the full LO FFNS expression should 
contain the leading parts of the coefficient functions
at all orders in $\alpha_{s,n_f}(Q^2)$, and the 
LO VFNS should include coefficient functions constructed from the full
LO FFNS coefficient functions and the full LO operator matrix elements.
Absolutely correct matching between the FFNS and the VFNS at $Q^2=M^2$
leads to the absolutely correct renormalization scheme consistent 
description of both of these schemes.  
Thus, in practice the strictly correct LO VFNS is no simpler than using the 
strictly correct LO FFNS coefficient functions. This is extremely difficult 
indeed, and in fact probably impossible, there being no clear 
unique way in which we
subtract out the leading order, renormalization scheme invariant part of the 
${\cal O}(\alpha_{s,n_f}^n(Q^2))$ FFNS coefficient function except in 
the asymptotic 
limit. Indeed, if we were to proceed further for our above example of
${d F_{2,H}(x,Q^2)\over d\ln(Q^2)}$ we would find that our definition of the 
LO contribution at ${\cal O}(\alpha_{s,n_f}^2(Q^2))$ would rely on being
able to extract an unambiguous LO, renormalization scheme independent part out
of ${d C^{FF,S,2}_{2,Hg}(z,Q^2/M^2)\over d\ln(Q^2)}$. 
Though this is simple in the limit $Q^2\to\infty$ \vanneerone\vanneertwo, 
there does not seem to be any good prescription for arbitrary $Q^2$.   
Therefore it appears as though the VFNS is only any advantage at 
all in so much that it gives a definition of the charm parton distribution. 
There does not seem to be any tractable way to produce a prescription 
for calculating heavy quark structure functions which both correctly sums
the leading logarithms and which has absolutely correct, unique 
threshold behaviour.  

\newsec{A practical VFNS.} 

Bearing in mind the difficulty, or indeed probable impossibility of 
producing the unambiguous well--ordered calculation of structure functions,
it is our aim to produce a prescription for calculating heavy quark structure 
functions order by order in $\alpha_s(Q^2)$ 
in such a way that we obtain relatively simple 
expressions yet maintain 
as much accuracy as possible over the whole range of $Q^2$.
Let us first consider the region of $Q^2=M^2$ and below.
In this case if we 
work order by order in $\alpha_{s,n_f}(Q^2)$ in the FFNS, i.e. define the 
$n_{th}$--order expression for the heavy quark structure function by 
\eqn\nffns{F^n_{2,H}(x,Q^2)=
\sum_{m=0}^n \sum_a \biggl({\alpha_{s,n_f}(Q^2)\over 2\pi}\biggr)^{n-m+1}
C^{FF,n-m+1}_{2,Ha}(M^2/Q^2)
\otimes f_m^{n_f}(Q^2),\quad n=1\to\infty,}
we know that the strictly
leading order terms we ignore are really an order of $\alpha_{s,n_f}(Q^2)$ 
down on those 
we keep, with no large $\ln (Q^2/M^2)$ enhancement, for these values of $Q^2$.
Adopting this procedure, when working to ${\cal O}(\alpha_{s,n_f}^n(Q^2))$ 
we have an 
error of ${\cal O}(\alpha_{s,n_f}^{n+1}(Q^2))$ compared to the (in principle) 
correct calculation, which is 
the same size as terms not yet calculated and the same size as the 
renormalization scheme uncertainty. This seems perfectly satisfactory for 
this region. 

Above $Q^2=M^2$ we want to order our 
calculation as in the massless case so that in the asymptotic limit of 
$Q^2>>M^2$ we will obtain correctly ordered expressions. Therefore,
we order the calculation by using up to ${\cal O}
(\alpha_{s,n_f+1}^n(Q^2))$ coefficient 
functions when solving the evolution equations using up to ${\cal O}
(\alpha_{s, n_f+1}^{n+1}(Q^2))$ anomalous dimensions, 
as required by renormalization
scheme consistency. e.g. the leading order expression is 
\eqn\logen{F^0_{2,i}(x,Q^2)=\sum_a C^{VF,0}_{2,ia}(M^2/Q^2)\otimes 
f_0^{n_f+1}(Q^2),}
the next--to--leading order expression is 
\eqn\nlogen{\eqalign{F^1_{2,i}(x,Q^2)=&F^0_{2,i}(x,Q^2)+\sum_a \bigl[
{\alpha_{s,n_f+1}(Q^2)\over 2\pi}C^{VF,1}_{2,ia}(M^2/Q^2)\otimes 
f_0^{n_f+1}(Q^2)\cr 
&\hskip 2.1in +C^{VF,0}_a(M^2/Q^2)\otimes f_1^{n_f+1}(Q^2)\bigr],\cr}}
{\it etc.}. We stress that this is not a choice, but a strict requirement of
obtaining ordered asymptotic expressions for the structure function itself 
or its $\ln(Q^2)$ derivative. 
Of course, in this region of $Q^2$ we now have the ambiguity in the 
definition of the coefficient functions. Thus, since we are not performing
the strictly correct ordering we have to make a choice for these coefficient 
functions. We do this by defining them order by order in 
$\alpha_{s,n_f+1}(Q^2)$ 
using the equations \bosglufus-\bosglufusiv\ which guarantee correctness  
to all orders, and also by using the freedom to choose some 
coefficient functions, i.e. the three coefficient functions coupling to the 
heavy quarks, to bring us as close to the really correct calculation as
possible. 

In perturbation theory it is not really the structure function 
at a particular value of $Q^2$ for which we solve but the evolution at 
all $Q^2$ in terms of the structure functions at some particular $Q^2$. 
Bearing 
this in mind it seems sensible to constrain our coefficient functions by
making the slope of the structure 
functions at a given order in $\alpha_s(Q^2)$
to be continuous across the transition point. In order to 
examine this constraint,
let us again consider the $\ln Q^2$--derivative  of $F_{2,H}(x,Q^2)$.
Approaching the transition point from below, our prescription gives
the lowest order $\alpha_{s,n_f}(Q^2)$ expression for the 
$\ln(Q^2)$--derivative as
\eqn\derivfour{{d F_{2,H}(x,Q^2)\over d\ln(Q^2)}=
{\alpha_{s,n_f}(Q^2)\over 2\pi}
{d C^{FF,1}_{2,Hg}(Q^2/M^2)\over d\ln(Q^2)}\otimes
g^{n_f}_0(Q^2).}
Just above $Q^2=M^2$ the $\ln(Q^2)$--derivative of the LO expression in the 
VFNS is
\eqn\derivfive{\eqalign{{d F_{2,H}(x,Q^2)\over d\ln(Q^2)}=&
{d C^{VF,NS,0}_{2,HH}(Q^2/M^2)\over d\ln(Q^2)}\otimes(H(Q^2)+\bar H(Q^2))_0 \cr
&\hskip -0.5in + 
{\alpha_{s,n_f+1}(Q^2)\over 2\pi}C^{VF,NS,0}_{2,HH}(Q^2/M^2)\otimes
\Bigl(P^0_{qg}\otimes g^{n_f+1}_0(Q^2)+P^0_{qq}\otimes
(H(Q^2) +\bar H(Q^2))_0\Bigr),\cr}}
where at the transition point the coupling at this order is continuous. 
Also we see that the artificial zeroth order term in \derivfive\
disappears at $Q^2=M^2$ (it is actually cancelled in the complete calculation 
as seen in \derivone-\derivthree), and it is indeed possible to demand the 
continuity of the derivative across the transition point. Using 
the constraint and our simple 
prescription for constructing the structure function in the two regions 
we now have a unique form for the previously ambiguous 
$C^{VF,NS,0}_{2,HH}(z,Q^2/M^2)$. Using the fact that $(H(z,Q^2)+
\bar H(z,Q^2))_0=0$ at $Q^2=M^2$, we immediately obtain  
\eqn\deflocf{C^{VF,NS,0}_{2,HH}(Q^2/M^2)\otimes P^0_{qg}
={d C^{FF,1}_{2,Hg}(z,Q^2/M^2)\over d\ln(Q^2)},}
at $Q^2=M^2$, and we define $C^{VF,NS,0}_{2,HH}(z,Q^2/M^2)$ by demanding 
that it satisfy this relationship at all $Q^2$. 
As well as guaranteeing the continuity of the evolution of the structure 
function this definition also reduces to the correct form for $Q^2>>M^2$ 
since in 
this limit ${d C^{FF,1}_{2,Hg}(z,Q^2/M^2)\over d\ln(Q^2)}\to P^0_{qg}(z)$
(as we shall see explicitly in \S 5).
This means that the evolution will clearly reduce to the correct asymptotic 
form of a delta function in the limit $Q^2 \to \infty$. 
Above $Q^2=M^2$ terms are not 
exactly as prescribed by the absolutely correct procedure explained in the 
last section but they do explicitly maintain the correct threshold behavior 
since ${d C^{FF,1}_{2,Hg}(z,Q^2/M^2)\over d\ln(Q^2)}$ is zero for 
$\hat W^2<4M^2$. At leading order we have in principle an error of 
${\cal O}(\alpha_s^2(Q^2))$ at 
the transition point due to the truncation of the FFNS expansion at 
${\cal O}(\alpha_{s,n_f}(Q^2))$ 
(where this error falls like $(M^2/Q^2)$ as we approach 
the correct asymptotic limit) and an error generated by the evolution
which is zero at the transition point, and grows like 
$\alpha_{s,n_f+1}(Q^2)\ln^n(Q^2/M^2)$, but falls like $(M^2/Q^2)$ as we 
evolve up from this point. These errors are quite minimal, 
always being small compared to the quantity being calculated. 
From \bosglufusex\ we see that we
have also completely defined $C^{VF,NS,1}_{2,Hg}(z,Q^2/M^2)$, i.e. in 
the $\msbar$ scheme it is 
\eqn\defnlogcf{C^{VF,1}_{2,Hg}(z,Q^2/M^2)= C^{FF,1}_{2,Hg}(z,Q^2/M^2)-
\ln (Q^2/M^2){d C^{FF,1}_{2,Hg}(z,Q^2/M^2)\over d\ln(Q^2)},}
though we have 
not yet made use of this coefficient function. However, we notice that each 
term in this coefficient function separately has the correct threshold 
behavior in $\hat W^2$.   

At leading order in this prescription the effect discussed above is 
the only real complication, i.e. the choice for $C^{VF,NS,0}_{2,HH}(z,Q^2/M^2)$
is the only one to make. Above the transition point the evolution equations
for the partons are now in terms of $n_f+1$ massless quarks, and the coupling 
constant becomes the $\msbar$ coupling for $n_f+1$ massless flavours. 
But all parton distributions and all other zeroth order coefficient functions 
are continuous across the transition. 

Of course, although we have determined the lowest order derivative of the 
coefficient functions on both sides of the boundary we must also discuss the 
value of the structure function itself at $Q^2=M^2$. 
Using the zeroth order expression \deflocf, the vanishing of the 
charm quark distribution at $Q^2=M^2$ leads to the charm structure function
being zero there. Likewise the fact that at zeroth order in the FFNS the
coefficient functions for charm production all vanish leads to the 
zeroth order value of $F_{2,H}(x,M^2)$ being zero also. Thus, 
the two expressions
are consistent. However, this is unsatisfactory for two reasons. Firstly,
the leading order (order $\alpha_s(Q^2)$) derivative of the charm structure 
function is non--zero both above and below $Q^2=M^2$, provided $x$ is low 
enough that we are above the threshold in $W^2$. Hence, starting with a 
value of $F_{2,H}(x,M^2)=0$ would lead to negative values for this structure 
function for $Q^2<M^2$. Also, one would naturally expect the LO expression 
for a quantity to be a reasonable approximation to the quantity itself. The 
value of $F_{2,H}(x,M^2)$ is not zero, and so the zeroth order expression is
not a good representation of the true value. These problems come about 
because of a peculiarity of $F_2(x,Q^2)$ already discussed in 
\ref\sx{R. Thorne, preprint RAL-96-065, {\tt hep-ph/9701241}.}. 
In general its value at a given 
$Q_0^2$ begins at zeroth order in $\alpha_s(Q_0^2)$,
but the ${\cal O}(\alpha_s(Q_0^2))$ term 
is also really part of the leading order 
expression since it is renormalization--scheme independent. In contrast the 
derivative begins at ${\cal O}(\alpha_s(Q^2))$, and all corrections are
renormalization scheme--dependent and genuinely higher order. Thus, as 
argued in \sx, the input should contain both the zeroth order 
term and the ${\cal O}(\alpha_s(Q_0^2))$ term, 
but the latter should play no part in the evolution. 

Adopting this procedure we can now specify our leading order 
expressions for the charm structure function as follows. Below $Q^2=M^2$ 
we take the LO expression to be
\eqn\loffns{F^{FF,0}_{2,H}(x,Q^2)={\alpha_{s,n_f}(Q^2)\over 2\pi}
C^{FF,1}_{2,Hg}(Q^2/M^2)\otimes g^{n_f}_0(Q^2),}
which is equal to the ${\cal O}(\alpha_s)$ value at $Q^2=M^2$, and
incorporates the LO evolution down from this scale (up to small
corrections). Above $Q^2=M^2$ the LO expression is     
\eqn\lovfns{ F^{VF,0}_{2,H}(x,Q^2)=F^{FF,0}_{2,H}(x,M^2)+
C^{VF,0}_{2,HH}(Q^2/M^2)\otimes (H(Q^2)+\bar H(Q^2))_0,}
which (up to the constant term) is of the standard form \logen, 
and incorporates the 
correct LO evolution. In practice the constant term becomes almost 
insignificant as soon as $Q^2>4M^2$. Now we should consider the NLO
expressions. 

\medskip

At next to leading order the situation is rather more complicated because
more terms come into play.
We now define FFNS expressions by including terms up to 
order $\alpha_{s,n_f}^2(Q^2)$ relative to the lowest order parton 
distributions. 
The NLO VFNS expression is defined as in \nlogen\ (up to a constant again). 
At this order the situation becomes more complicated because the pure singlet
FFNS coefficient function becomes non--zero as does the contribution due to 
coefficient functions where the photon couples to a light quark but where
heavy quarks are generated. Let us examine the NLO expressions for the 
derivative of the heavy quark structure function. First consider the 
${\cal O}(\alpha_{s,n_f}^2(Q^2)))$ 
expression for the derivative of the heavy quark 
structure function in the FFNS. This is 
\eqn\derivsix{\eqalign{{dF_{2,H}(x,Q^2)\over d\ln(Q^2)}=&
\biggl({\alpha_{s,n_f}(Q^2)\over 2\pi}\biggr)^2
\biggl[-\beta_{n_f}^0 C^{FF,1}_{2,Hg}(Q^2/M^2)\otimes g^{n_f}_0(Q^2)\cr
&+C^{FF,1}_{2,Hg}(Q^2/M^2)\otimes\bigl(P^{0,n_f}_{gg}\otimes g^{n_f}_0(Q^2)+
P^{0}_{gq}\otimes \Sigma^{n_f}_0(Q^2)\bigl)\cr
&+{d C^{FF,2}_{2,Hg}(Q^2/M^2)\over d \ln (Q^2)}\otimes g^{n_f}_0(Q^2)+
{d C^{FF,2}_{2,Hq}(Q^2/M^2)\over d \ln (Q^2)}\otimes \Sigma^{n_f}_0(Q^2)
\biggr]\cr
&+{\alpha_{s,n_f}(Q^2)\over 2\pi}
{d C^{FF,1}_{2,Hg}(Q^2/M^2)\over d \ln (Q^2)}\otimes 
g^{n_f}_1(Q^2),\cr}}
where $\Sigma^{n_f}(z,Q^2)$ is the singlet light quark distribution.\foot{We 
label $P^0_{gg}(z)$ by the number of flavours because it is the only 
leading--order splitting function which depends on this number. The decrease
of this splitting function above a threshold accounts for the fact that there 
is a new parton distribution, and guarantees overall conservation of momentum 
in the evolution.}  In the VFNS the situation is even more complicated. 
Taking the derivative of the 
NLO expression, and ignoring those terms already in \derivfive, we obtain
\eqn\derivseven{\eqalign{{dF_{2,H}(x,Q^2)\over d\ln(Q^2)}=& 
{d C^{VF,0}_{2,HH}(Q^2)\over d\ln(Q^2)}\otimes (H(Q^2)+\bar H(Q^2))_1\cr
&+{\alpha_{s,n_f+1}(Q^2)\over 2\pi}\biggl[{d C^{VF,1}_{2,HH}(Q^2)
\over d\ln(Q^2)}\otimes 
(H(Q^2)+\bar H(Q^2))_0+{d C^{VF,1}_{2,Hg}(Q^2)\over d\ln(Q^2)}\otimes 
g^{n_f+1}_0(Q^2)\cr
&+C^{VF,0}_{2,HH}(Q^2/M^2)\otimes(P^{0}_{qq}\otimes
(H(Q^2)+\bar H(Q^2))_1+ 
P^0_{qg}\otimes g^{n_f+1}_1(Q^2))\biggr]\cr
&+\biggl({\alpha_{s, n_f+1}(Q^2)\over 2\pi}\biggr)^2\biggl[ -\beta_{n_f}^0 
\bigl(C^{VF,1}_{2,Hg}(Q^2/M^2)
\otimes g^{n_f+1}_0(Q^2) \cr
&+C^{VF,1}_{2,HH}(Q^2/M^2)\otimes (H(Q^2)+\bar H(Q^2))_0
\bigr)\cr
&+C^{VF,1}_{2,HH}(Q^2/M^2)\otimes(P^{0}_{qq}
\otimes(H(Q^2)+\bar H(Q^2))_0
+ P^0_{qg}\otimes g^{n_f+1}_0(Q^2))\cr
&+C^{VF,1}_{2,Hg}(Q^2/M^2)\otimes
(P^{0}_{gq}\otimes\Sigma^{n_f+1}_0(Q^2)+ 
P^{0,n_f+1}_{gg}\otimes g^{n_f+1}_0(Q^2))\cr
&+C^{VF,0}_{2,HH}(Q^2/M^2)\otimes(P^{NS,1,n_f+1}_{qq}
\otimes(H(Q^2)+\bar H(Q^2))_0\cr
&+P^{PS,1,n_f+1}_{qq}\otimes \Sigma^{n_f+1}_0(Q^2) 
+P^{1,n_f+1}_{qg}\otimes g^{n_f+1}_0(Q^2))\biggr].\cr}}
These expressions are very difficult to compare in general. However, expressing
the four flavour quantities in terms of the three flavour quantities 
the two are identical at NLO in $\msbar$ scheme at $Q^2=M^2$
(the discontinuities in
both the parton distributions and the coupling begin at NNLO). Thus, 
the heavy parton distributions $(H(z,Q^2)+\bar H(z,Q^2))_0$ and 
$(H(z,Q^2)+\bar H(z,Q^2))_1$
vanish at this point, and so do many other terms in \derivseven. From the 
definition of $C^{VF,0}_{2, HH}(z,Q^2)$ we can see that the term depending on
$g_1(z,Q^2)$ is the same in both expressions, and using the equation 
\defnlogcf\ we can see that ${d C^{VF,1}_{2,Hg}(z,Q^2)\over d\ln(Q^2)}=0$
at $Q^2=M^2$. Also in the combination $-\beta_{n_f}^0f(z)+P^{0,n_f}_{gg}\otimes
f$ the flavour dependence cancels between the two terms, so this combination
is the same in both expressions at $Q^2=M^2$. 

Thus we have a great deal of simplification when comparing the two expressions 
at $Q^2=M^2$. As in the LO case we can equate the terms coupling to the 
gluon in the two expressions, i.e.  
\eqn\defnlocf{{d C^{FF,2}_{2,Hg}(z,Q^2/M^2)\over d\ln(Q^2)}=
C^{VF,1}_{2,HH}(Q^2/M^2)\otimes P^0_{qg}+C^{VF,0}_{2,HH}(Q^2/M^2)
\otimes P^{1,n_f+1}_{qg},}
at $Q^2=M^2$, and this serves as a definition of the coefficient function
$C^{VF,1}_{2,HH}(z,Q^2/M^2)$ at this $Q^2$. However, unlike the LO case we 
cannot
define the coefficient function at all $Q^2$ simply by extending this 
expression to all $Q^2$. This is because it will not result in the correct
asymptotic expression for $C^{VF,1}_{2,HH}(z,Q^2/M^2)$, i.e. 
${d C^{FF,2}_{2,Hg}(z,Q^2/M^2)\over d\ln(Q^2)}$ contains a $\ln(Q^2/M^2)$ term
which must be cancelled. It is quite easy to find the generalization of 
\defnlocf, however. If one differentiates both sides of \bosglufus, and keeps
those terms of ${\cal O}(\alpha_{s,n_f}^2(Q^2)))$ which survive as 
$Q^2\to \infty$
(all terms of the form ${d C^{VF}_{2,ba}(z,Q^2/M^2)\over d\ln(Q^2)}$ vanish in 
this limit since the VFNS coefficient functions tend to constants) then
one obtains 
\eqn\defnlocfi{\eqalign{ {d C^{FF,2}_{2,Hg}(z,Q^2/M^2)\over d\ln(Q^2)}=&
C^{VF,NS,1}_{2,HH}(Q^2/M^2)\otimes {d A^1_{Hg}(Q^2/M^2)\over d\ln (Q^2)}\cr
&\hskip -0.5in +C^{VF,NS,0}_{2,HH}(Q^2/M^2)
\otimes {d A^2_{Hg}(Q^2/M^2)\over d\ln (Q^2)}
+{1\over 3\pi}\ln (Q^2/M^2) C^{VF,0}_{2,HH}(Q^2/M^2)\otimes P^0_{qg},\cr}}
where the last term comes about from the difference in the derivatives of the
three and four flavour couplings. This expression guarantees the 
correct asymptotic expression for $C^{VF,1}_{2,HH}(z,Q^2/M^2)$, while
\defnlocf\ guarantees the continuity of the NLO derivative of $F_{2,H}(x,Q^2)$
in the gluon sector, and hence the definition of $C^{VF,1}_{2,HH}(z,Q^2/M^2)$
must satisfy \defnlocf\ at $Q^2=M^2$ and \defnlocfi\ as $Q^2\to \infty$.
In fact, at $Q^2=M^2$ the two expressions are identical, i.e. 
\eqn\reducei{{d A^1_{Hg}(z,Q^2/M^2)\over d\ln (Q^2)}=P^0_{qg}(z),}
and
\eqn\reduceii{{d A^2_{Hg}(z,Q^2/M^2)\over d\ln (Q^2)}=
(P^0_{qq}\otimes P^0_{qg}
+P^0_{qg}\otimes P^{0,n_f+1}_{gg}-\beta^{n_f+1}_0P^0_{qg})
\ln(Q^2/M^2)+P^{1,n_f+1}_{qg},}
and we have the 
very neat result that \defnlocfi\ is the generalization of \defnlocf\ for all
$Q^2$, and $C^{VF,1}_{2,HH}(z,Q^2/M^2)$ is defined by \defnlocfi.   

The above definition of $C^{VF,1}_{2,HH}(z,Q^2/M^2)$, when substituted into
\bosglufus, determines the expression for $C^{VF,2}_{2,Hg}(z,Q^2/M^2)$
which will be used at NNLO. However, we have now used up our single degree
of freedom involved with the heavy quark structure function at NLO. 
Looking at the terms coupling to the singlet quark distribution in the two
expressions \derivsix\ and \derivseven\ we find that the first contains
\eqn\nlosingff{C^{FF,1}_{2,Hg}(Q^2/M^2)\otimes P^0_{gq}+
{d C^{FF,2}_{2,Hq}(z,Q^2/M^2)\over d\ln(Q^2)},} 
while the second contains
\eqn\nlosingvf{C^{VF,1}_{2,Hg}(Q^2/M^2)\otimes P^0_{gq}+
C^{FF,0}_{2,HH}(Q^2/M^2)\otimes P^{1,PS,n_f+1}_{qq}.}
There is no degree of freedom in either of these equations, and no reason for 
them to be equal at $Q^2=M^2$, and they are not. Indeed there was no
further degree of freedom in the relationships \bosglufus\ and
\bosglufusi\ required of the heavy quark coefficient functions. Up to
this order the only one available was for $C^{VF}_{2,HH}(z,Q^2/M^2)$, and this
has been determined by imposing the continuity of the evolution of the 
structure function in the gluon sector. Indeed, looking at \bosglufusi\
at ${\cal O}(\alpha_s^2(Q^2))$ we see that we have already determined   
$C^{VF,PS,2}_{2,Hq}(z,Q^2/M^2)$, i.e.
\eqn\defsingnlo{C^{VF,PS,2}_{2,Hq}(z,Q^2/M^2)=C^{FF,S,2}_{2,Hq}(z,Q^2/M^2)-
C^{VF,0}_{2,HH}(Q^2/M^2)\otimes
A^{PS,2}_{Hq}(Q^2/M^2).}
Using the framework we have chosen to define the structure functions 
this discontinuity in the derivative of the heavy quark structure function 
in the singlet sector is unavoidable. There are simply not enough degrees of
freedom to avoid it. In practice, since the evolution of 
the heavy quark structure function is driven very largely by the gluon,
since this discontinuity begins only at NLO, and since \nlosingff\ and
\nlosingvf\ are not too different at $Q^2=M^2$ the effect is tiny.  
Of course, any discontinuity is only an artifact of the manner in 
which we are forced to do our fixed order calculations, and would 
disappear if we were to work all orders. In fact one can show that the 
discontinuity of the derivative in the singlet sector gets formally smaller 
as one works to higher orders.  

So now we have the definition of our NLO expressions for the heavy--quark
structure function both above and below threshold. In the FFNS the definition 
is the simple extension of \loffns, being just
\eqn\nloffns{\eqalign{F^{FF,1}_{2,H}(x,Q^2)&={\alpha_{s,n_f}(Q^2)\over 2\pi}
\bigl(C^{FF,1}_{2,Hg}
(Q^2/M^2)\otimes g^{n_f}_0(Q^2)+C^{FF,1}_{2,Hg}(Q^2/M^2)
\otimes g^{n_f}_1(Q^2)\bigr)\cr
&+\biggl({\alpha_{s,n_f}(Q^2)\over 2\pi}\biggr)^2
\bigl(C^{FF,2}_{2,Hg}(Q^2/M^2)
\otimes g^{n_f}_0(Q^2)+C^{FF,1}_{2,Hq}(Q^2/M^2)
\otimes \Sigma^{n_f}_0(Q^2)\bigr),\cr}}
which is equal to the ${\cal O}(\alpha_s^2)$ (i.e. NLO)
value for the structure function at $Q^2=M^2$ and incorporates the NLO
(i.e. ${\cal O}(\alpha_{s,n_f}^2(Q^2)$) evolution down from this scale (up to
small corrections). The VFNS NLO expression is  
\eqn\nlovfns{ \eqalign{F^{VF,1}_{2,H}(x,Q^2)=&\biggl({\alpha_{s}(M^2)
\over 2\pi}\biggr)^2
\bigl(C^{FF,2}_{2,Hg}(1)
\otimes g^{n_f}_0(M^2)+C^{FF,1}_{2,Hq}(1)
\otimes \Sigma^{n_f}_0(M^2)\bigr)\cr
&+C^{VF,0}_{2,HH}(Q^2/M^2)\otimes (H(Q^2)+\bar H(Q^2))_0\cr
&+C^{VF,0}_{2,HH}(Q^2/M^2)\otimes (H(Q^2)+\bar H(Q^2))_1\cr
&+{\alpha_{s,n_f+1}(Q^2)\over 2\pi}
C^{VF,NS,1}_{2,HH}(Q^2/M^2)\otimes (H(Q^2)+\bar H(Q^2))_0\cr
&+{\alpha_{s,n_f+1}(Q^2)\over 2\pi}C^{VF,1}_{2,Hg}(Q^2/M^2)
\otimes g^{n_f+1}_0(Q^2),\cr}}
which again, up to the constant term, which is the NLO input
(the LO part of the input now being included automatically), is of the 
standard form and incorporates the correct NLO evolution across the 
transition point.

At this order we have to make some decision about how we treat the light
quark sector. The lowest order contribution the heavy quark makes to a
light sector FFNS coefficient function is for the nonsinglet
coefficient function at ${\cal O}(\alpha_{s,n_f}^2(Q^2))$. 
Thus in the matching 
conditions between the FFNS coefficient functions and those in the VFNS 
in the light quark sector there are no mass-dependent corrections 
to the VFNS coefficient functions to ${\cal O}(\alpha_{s,n_f+1}(Q^2))$. 
Hence, the 
evolution of the light quark coefficient functions above $Q^2=M^2$ 
is exactly as in the massless $n_f+1$ flavour case. Nevertheless, we must
decide on the form of the structure function at $Q^2=M^2$ 
and below this transition point. For the heavy quark structure function
we have been keeping heavy quark coefficient functions to one order higher
in $\alpha_s(Q^2)$ in the FFNS than in the VFNS. This has been for the 
reason that the explicit $\ln(Q^2)$--dependence in the coefficient functions
means that they contribute to the $\ln(Q^2)$--derivative of the structure 
function at effectively one higher order in $\alpha_s(Q^2)$ than the VFNS
coefficient functions, and also because the lack of the usual
zeroth order coefficient function makes the ${\cal O}(\alpha_{s,n_f}(Q^2))$ 
coefficient function the LO one, the ${\cal O}(\alpha^2_{s,n_f}(Q^2))$ 
the NLO one, {\it etc.}. For the light structure functions there is a zeroth 
order coefficient function, so the second argument no longer holds. 
However, the former one still does, i.e. differentiating the expression for 
the light quark structure function below $Q^2=M^2$ and keeping terms of 
order $\alpha_{s,n_f}^2(Q^2)$ then ${d C^{FF,NS,2}_{2,qq}(z,Q^2/M^2)\over 
d \ln (Q^2)}$ appears in the expression. This contribution accounts for the 
effect of the heavy quark to the evolution turning on as $Q^2$ increases.
For this reason we continue to keep the coefficient functions containing 
heavy quarks to one higher order than those with only light quarks even
in the light sector. 

For the heavy quark structure function, because we had terms of higher order 
in $\alpha_s(Q^2)$ below $Q^2$ than above it, 
in order to impose continuity of the 
structure function at $Q^2=M^2$ we had to put a contribution 
to the VFNS expression which is constant, and one order in $\alpha_s$ higher 
than the rest of the expression (we also justified this from renormalization
scheme consistency). We now have to adopt a similar procedure for the 
light quark expressions. The NLO expression for the nonsinglet structure 
function for $Q^2<M^2$ is 
\eqn\nloffnsl{\eqalign{F^{FF,NS,1}_{2,q}(x,Q^2)=&C^{NS,n_f,0}_{2,qq}
\otimes f^{NS,n_f}_0(Q^2)+C^{NS,n_f,0}_{2,qq}
\otimes f^{NS,n_f}_1(Q^2)\cr
&\hskip -0.8in+{\alpha_{s,n_f}(Q^2)\over 2\pi}
C^{NS,n_f,1}_{2,qq}
\otimes f^{NS,n_f}_0(Q^2)
+\biggl({\alpha_{s,n_f}(Q^2)\over 2\pi}\biggr)^2 C^{FF,NS,2}_{2,qq}(Q^2/M^2)
\otimes f^{NS,n_f}_0(Q^2).\cr}}
That for $Q^2\geq M^2$ is equal to    
\eqn\nlovfnsl{\eqalign{F^{VF,NS,1}_{2,q}(x,Q^2)=&C^{NS,n_f+1,0}_{2,qq}
\otimes f^{NS,n_f+1}_0(Q^2)+C^{NS,n_f+1,0}_{2,qq}
\otimes f^{NS,n_f+1}_1(Q^2)\cr
&\hskip -0.8in +{\alpha_{s,n_f+1}(Q^2)\over 2\pi}C^{NS,n_f+1,1}_{2,qq}
\otimes f^{NS,n_f+1}_0(Q^2)
+\biggl({\alpha_s(M^2)\over 2\pi}\biggr)^2C^{FF,NS,2}_{2,qq}(1)
\otimes f^{NS,n_f}_0(M^2)\bigr).\cr}}
In principle both sides should also contain a term $\propto \alpha^2_s(Q_0^2)$
for the genuinely light NLO input, where $Q_0^2$ is the scale at which the 
inputs are chosen. Such a term is always ignored, and would be 
very small. In practice all the ${\cal O}(\alpha_s^2(Q^2))$ terms in the above 
expression are extremely small as well.
The ${\cal O}(\alpha_s^2(Q^2))$ evolution derived from the above equations is 
not precisely continuous at $Q^2=M^2$ due to terms of inverse powers of 
$Q^2/M^2$ present in  ${d C^{FF,NS,2}_{2,qq}(z,Q^2/M^2)\over d \ln (Q^2)}$.
This discontinuity will decrease as we go to higher orders, and these 
mass-dependent terms get absorbed by higher order mass--dependent
VFNS coefficient functions. We note that leaving the 
${\cal O}(\alpha_{s,n_f}^2(Q^2))$
term out of of \nloffnsl\ would also lead to a discontinuous evolution
(actually more so) since the evolution would take account of $n_f$ massless 
flavours below threshold but $n_f+1$ massless flavours above threshold. 

Finally at NLO the light quark pure singlet structure functions have 
no complications due to the heavy quarks at all. The first non--zero
FFNS coefficient functions do not appear until 
${\cal O}(\alpha_{s,n_f}^3(Q^2))$,
and so do not contribute to the evolution until NNLO. So at NLO we just use 
the $n_f$ massless flavour expressions below $Q^2=M^2$ and the    
$n_f+1$ massless flavour expressions above $Q^2=M^2$. Continuity of both the
structure function and its evolution are automatic. 

\medskip

One could in principle work to progressively higher orders, but of course in 
practice the NNLO 
splitting functions and the NNLO FFNS coefficient functions are all unknown
at present. Nevertheless, we outline the procedure to be adopted at all 
orders. For the heavy quark structure function there is essentially nothing 
new as we progress to higher orders. At $n_{th}$ nontrivial order we include
all FFNS coefficient functions up to order $\alpha_{s,n_f}^n(Q^2)$, and all 
VFNS coefficient functions up to order $\alpha_{s,n_f+1}^{n-1}(Q^2)$. 
In the VFNS expression
we always include the ${\cal O}(\alpha_s^n(M^2))$ term which ensures 
continuity of the structure function. We determine 
$C^{VF,n-1}_{2,HH}(z,Q^2/M^2)$ by demanding continuity of the derivative of the
structure function at ${\cal O}(\alpha_s^n)$
in the gluon sector, and this determination 
predetermines $C^{VF,n}_{2,Hg}(z,Q^2/M^2)$ and $C^{VF,n}_{2,Hq}(z,Q^2/M^2)$
by using \bosglufus\ and \bosglufusi\ to ${\cal O}(\alpha_{s,n_f}^n(Q^2))$. 
At ${\cal O}(\alpha_{s,n_f+1}^2(Q^2))$ the coefficient function 
$C^{VF,n}_{2,HH}(z,Q^2/M^2)$
becomes the sum of the nonsinglet and pure singlet coefficient functions. 
Neither the condition \bosglufus\ nor the continuity of the structure function
and its derivatives determine these two contributions separately, so we are 
free to separate them as we wish, using the condition that each tends to 
the correct asymptotic limit. It would also be desirable to choose each 
so that they respect the kinematic threshold.  

For the light quark structure function the procedure at higher orders 
is also straightforward. At $n_{th}$ nontrivial order we include
all pure light quark contributions to coefficient functions below $Q^2=M^2$
up to order $\alpha_{s,n_f}^{n-1}(Q^2)$; all mass dependent
FFNS coefficient functions up to order $\alpha_{s,n_f}^n(Q^2)$, and all 
VFNS coefficient functions up to order $\alpha_{s,n_f+1}^{n-1}(Q^2)$. 
In the VFNS expression
we always include the ${\cal O}(\alpha_s^n(M^2))$ term which ensures 
continuity of the structure function.
Starting with the ${\cal O}(\alpha_{s,n_f+1}^2(Q^2))$ 
coefficient function we determine 
$C^{VF,n-1}_{2,qH}(z,Q^2/M^2)$ by demanding continuity of the derivative of the
light structure function at ${\cal O}(\alpha_s^n(Q^2))$
in the gluon sector, analogously to the heavy quark sector.
With this one degree of freedom eliminated in this way
all other VFNS coefficient functions are determined uniquely order by order in 
$\alpha_s$ by \bosglufusii--\bosglufusiv, i.e.
this determination of $C^{VF,n-1}_{2,qH}(z,Q^2/M^2)$
predetermines $C^{VF,n}_{2,qg}(z,Q^2/M^2)$ and $C^{VF,PS,n}_{2,qq}(z,Q^2/M^2)$
by using \bosglufusiii\ and \bosglufusiv\ to ${\cal O}(\alpha_s^n(Q^2))$. 

Thus, we have completely defined our prescription for calculating order by
order for the structure function $F_2(x,Q^2)$. We can sum it up in the 
form of a table. This is shown in table 1.  
This method uniquely determines all VFNS coefficient functions, and while 
not
leading to absolutely correctly ordered expressions it is a relatively
simple prescription for obtaining order by order structure functions which 
are very similar to the strictly correctly ordered ones, which reduce to 
the correctly ordered expressions in the asymptotic limit
and which order by order are consistent with all physical requirements. 
All prescriptions which obey \bosglufus--\bosglufusiv\ will be correct 
when summed to all orders, but some ways of choosing the heavy quark
coefficient functions will clearly stay closer to the correct ordering
than others. We believe that our prescription is the best available at
present, and we see no easy way to improve upon it.   

\medskip

We will demonstrate the results using our prescription in the next 
section, and see that indeed they do seem to work very well.
However, first let us mention another currently available VFNS, the
ACOT scheme \ref\oltung{F.I. Olness and W.K. Tung, \NP 
\vyp{B308}{1988}{813}.}\ref\acotpapone{M. Aivazis, F. Olness and W.K. Tung,
\PR \vyp{D50}{1994}{3085}.}\ref\acotpaptwo{M. Aivazis, J.C. Collins, 
F. Olness and W.K. Tung, \PR \vyp{D50}{1994}{3102}.}. 
Although there is currently no all--orders \ref\allacot{J.C. Collins, 
(in preparation).}, or even NLO definition (for developments see 
\ref\refnloacot{P. Agrawal, F.I. Olness, S.T. Riemersma and W.K. Tung,
Proceedings of the XXXth Rencontres de Moriond, ``QCD and High Energy 
Hadronic interactions'', Les Arcs, France, March 1995\semi
C. Schmidt, to appear in Proceedings of the International Workshop on 
Deep Inelastic Scattering and QCD (DIS97), Chicago, IL, USA, April 1997,  
{\tt hep-ph/9706496.}}, of the ACOT
VFNS (which we will denote by ACOT) in print we believe that the definition 
of the coefficient functions in this scheme must be equivalent to that in 
\facttheha, i.e. the VFNS coefficient functions are related to those in the 
FFNS by the equations \bosglufus--\bosglufusiv.
Indeed, at what they call LO, the ACOT coefficient functions 
satisfy \bosglufusex. However, they determine the expression for 
$C^{VF,NS,0}_{2,HH}$ from the tree--level diagram for a massive quark 
scattering from a boson, and for a photon this gives
\eqn\acota{\hat C^{VF,NS,0}_{HH}(z,Q^2/M^2)=z\delta(\hat x_0-z)\biggl(1+
{4M^2\over Q^2}\biggr), \hskip 0.4in \hat x_0=\biggl(1+{M^2\over Q^2}
\biggr)^{-1},}
where the modified argument of the delta--function follows from demanding 
the on--shell condition for the massive quark, and the remaining factor
follows from the parton model for the longitudinal structure function, 
$F_L=4M^2/Q^2$, which is added to the transverse component to give $F_2$. 
Inserting into \bosglufusex\ for arbitrary $\mu$ then gives the expression for 
$\hat C^{VF,S,1}_{2,Hg}(z,Q^2/\mu^2, Q^2/M^2)$. Presumably the ACOT scheme 
works at higher orders in a similar manner, with the higher order 
heavy quark coefficient functions being calculated explicitly (but needing
explicit subtraction of divergences in $(Q^2/M^2)$ beyond leading order).
However, we note that ACOT do not usually 
use the scale choice $\mu^2=Q^2$ as we do. More common is $\mu^2=M^2+0.5
Q^2(1-M^2/Q^2)^2$ \acotpaptwo, which grows more slowly than our choice from 
the same value at $Q^2=M^2$ and is $\mu^2=0.5Q^2$ asymptotically. 

ACOT claim a smooth transition from the 
FFNS at low $Q^2$ order by order. Their ``LO'' expression for the 
structure function is 
\eqn\loacot{\eqalign{F^{LO}_{2,H}(x,Q^2)=  \hat C^{VF,NS,0}_{2,HH}(Q^2/M^2)
\otimes(&H(\mu^2)+\bar H(\mu^2))_0\cr
&+{\alpha_{s, n_f+1}(\mu^2)\over 2\pi}
\hat C^{VF,S,1}_{2,Hg}(Q^2/\mu^2,Q^2/M^2)\otimes
g^{n_f+1}_0(\mu^2).\cr}}
where from \bosglufusex\ and \acota,
\eqn\acotb{\hat C^{VF,S,1}_{2,Hg}(z,Q^2/\mu^2,Q^2/M^2)=
C^{FF,S,1}_{2,Hg}(z,Q^2/M^2)
-(\ln(\mu^2/M^2)+c_{rs})P^0_{qg}\otimes
z\delta(\hat x_0-z)\biggl(1+{4M^2\over Q^2}\biggr).}

There are a number of odd features associated with these expressions.
Firstly, the ``correct'' threshold behaviour comes about only from a 
conspiracy of cancellation. Neither term in \loacot\ respects
the physical threshold individually and 
$\hat C^{VF,S,1}_{2,Hg}(z,Q^2/\mu^2,Q^2/M^2)$
has a part with a threshold in $\hat W^2$ and a part going like \acota. 
In fact, since the first term in \loacot\ grows more quickly than the 
subtraction term in the second term in \loacot, there will be nonzero 
(albeit
very small) heavy quark structure function for $W^2<4M^2$. Once 
all the necessary cancellation has taken place, the result is very good. 
This can 
be seen in fig. 8 of \acotpaptwo, and also in \fig\acotone{Charm quark 
structure function, $F_{2,c}(x,Q^2)$ for $x=0.05$ and 
$x=0.005$ calculated using the ACOT ``LO'' prescription, our input parton 
distributions evolved at LO and renormalization scale $\mu^2=Q^2$. Shown are 
the total, the two contributions due to convolution of the coefficient 
function $\hat C^{VF,1}_{2,Hg}(z,Q^2/m_c^2)$ with the gluon
distribution (the subtraction term making a negative contribution), and the
contribution directly due to the charm quark.} 
which is calculated
using the ACOT ``LO'' prescription, our choice of renormalization scale,
and the parton distributions obtained from our best fit 
(see later for details). There is a smoother transition in fig. 8 of 
\acotpaptwo\
than in \acotone\ because their complicated choice of scale leads to 
$\mu^2$ departing slowly from $M^2$ and staying well below $Q^2$ and hence to 
the growth of the charm parton distribution being effectively much slower 
than for the simple $Q^2=\mu^2$ choice. The effect of the choice of 
renormalization scale on the speed of departure of the ACOT result from the 
LO FFNS result can be seen nicely in fig. 1 of \ref\smith{J. Smith,
to appear in Proceedings of New Trends in HERA Physics, Ringberg, May 1997
{\tt hep-ph/9708212}}.    

However, even though the cancellation of terms works well,  
\loacot\ is at odds with the usual way of defining a LO expression, 
which usually
only involves zeroth coefficient functions convoluted with the 
parton distributions obtained from the one--loop evolution equations. It
is clearly of mixed order, and indeed, part of the expression is in fact 
renormalization scheme dependent, which is certainly not correct for a
LO expression. If we go to $Q^2>>M^2$, \acotb\ does
not reduce to any fixed order expression in the ZM--VFNS. The first
term in \acotb, represented by the dotted line in \acotone,
becomes the LO expression in the ZM--VFNS, but the second
belongs to the NLO expression. One can see in \acotone\ that the total 
LO ACOT result is significantly different from the asymptotic ZM--VFNS result
even at $Q^2=1000\Gev^2$. Similarly the derivative of \loacot\ leads 
to terms both of ${\cal O}(\alpha_{s,n_f+1})$ and 
${\cal O}(\alpha_{s,n_f+1}^2)$, and will 
have a renormalization scheme dependent part. This mixing of orders is not 
acceptable.

Alternatively, with the choice of 
$\hat C^{VF,NS,0}_{2,HH}(z,Q^2/M^2)$ made, 
the usual way of ordering the expansion 
for a structure function leads to serious problems. 
Using  what one would normally consider the LO
expression, $F_{2,H}^{0}(x,Q^2)=  \hat C^{VF,NS,0}_{HH}(Q^2/M^2)
\otimes(H(\mu^2)+\bar H(\mu^2))$, has only a sharp threshold in $Q^2$
and the rate of growth of $F_{2,H}(x,Q^2)$ would be very discontinuous at  
$Q^2=M^2$ and a great deal too fast just above this. This can easily be 
seen 
on \acotone\ where this contribution is represented by the dotted line and 
labelled ``charm quark''. It deviates very quickly from both the 
continuation of the FFNS expression and from the total expression. 
Using the NLO expression ordered in the usual manner\foot{It is an expression 
of this general form that is used in the recent global fits to data
\ref\lai{H.L. Lai and W.K. Tung, \ZP \vyp{C74}{463}{1997}.}.} 
the effect would be lessened, but would still be significant.
The subtraction piece in $\hat C^{FF,S,1}_{2,Hg}
(z,Q^2/\mu^2,Q^2/M^2)$ 
would largely cancel the quick growth generated by the LO
evolution of the charm parton distribution, but the NLO evolution would 
still be uncancelled. This effect can be seen in fig. 9 in \acotpaptwo\
where NLO parton distributions are combined with what is called the LO 
coefficient functions and in \fig\acottwo{Same as \acotone, but with the 
partons evolved at NLO.}, where we do the same thing
using our parton distributions and $\mu^2=Q^2$. 
Here the subtraction term only partially 
cancels the charm quark contribution and the total quickly departs 
from the continuation 
of the FFNS structure function, and the effect increases at smaller $x$. 
The all--orders definition of the coefficient functions in  
\bosglufus--\bosglufusiv, if indeed it is the all orders definition
in the ACOT scheme, 
guarantees that the correct low $Q^2$ behaviour will be restored when 
working to all orders, but in this scheme this behaviour will come about 
only due to the mixing of effects at different orders. At low orders 
the discrepancy is still large. 
We note that the MRRS scheme \ref\mrrs{A.D. Martin, R.G. Roberts, M.G. Ryskin
and W.J. Stirling, preprint RAL-TR-96-103, {\tt hep-ph/9612449}, 
to be published in \ZP C.}, 
which incorporates mass effects into the 
evolution, but has a similar definition 
of coefficient functions to ACOT (though
with the usual ordering), suffers badly from this problem outlined above. 
At the transition point where the heavy quark starts contributing to
the heavy quark coefficient function directly there is a very distinct kink,
and the total rises very quickly above the continuation of the FFNS
expression, as seen in figs. 6 and 7 of their paper.  

We do not believe that the method used by ACOT (or MRRS) is a satisfactory 
way in which to define the 
coefficient functions in a VFNS, and we certainly do not believe that it is 
unique. It is a choice, as our prescription is a choice, and as we have 
discussed in \S 3, we do not believe that any are strictly ``correct''.
However, using the ACOT choice the calculation of the heavy quark 
coefficient functions proceeds as though the heavy quark parton distribution
is due to intrinsic presence of the heavy quark rather than it being 
generated from (at least mainly) the gluon. 
In particular the heavy quark coefficient function 
contains no reference to the kinematic threshold in $\hat W^2$. 
This necessitates 
a mixing of orders to get satisfactory results. 
We believe it is far more useful to choose the heavy quark coefficient 
functions so that they reflect the physics and all automatically contain
at least the correct form of low $Q^2$ behaviour, and our prescription 
guarantees this. 

\newsec{The VFNS in Practice.}

We now discuss how our procedure is implemented in practice. 
Of course, in practice the first heavy quark we encounter is the charm quark
with $m_c\approx 1.5\Gev$.
First we consider the LO expression. Denoting $\epsilon=m_c^2/Q^2$ the LO FFNS 
heavy quark to gluon coefficient function is
\eqn\lochgffns{\eqalign{C^{FF,1}_{2,cg}(z,\epsilon) = &
\biggl[(P^0_{qg}(z)
+ 4\epsilon z (1-3z) -8\epsilon^2 z^2)\ln \biggl({1+v\over 1-v}
\biggr)\cr
&+(8z(1-z)-1-4\epsilon z(1-z))v\biggr] \theta(\hat W^2-4m_c^2),\cr}}   
where $\hat W^2=Q^2(1/z-1)$, the gluon quark centre of mass energy, 
$v$ is the velocity of the heavy quark or antiquark in the 
photon--gluon centre--of--mass frame, defined by $v^2=1-4m_c^2/\hat W^2$,
and $P^0_{qg}(z) = (z^2 + (1-z)^2)$. These $v$--dependent terms ensure 
that the coefficient function tends to zero
smoothly as $\hat W^2=4m_c^2$ is approached from below,
and hence the structure function has a smooth threshold in $W^2$. 
Taking the $\ln Q^2$--derivative of this 
is a straightforward matter and results in 
\eqn\lodchgffns{\eqalign{{d C^{FF,1}_{2,cg}(z,\epsilon) \over d \ln Q^2} =& 
\biggl[(P^0_{qg}(z)
+ 2\epsilon {z(1-2z^2)\over 1-z} -16\epsilon^2 z^2){1\over v}\cr
&+(-4\epsilon z(1-3z) +16 \epsilon^2 z^2)\ln \biggl({1+v\over 1-v}
\biggr) +(4\epsilon z(1-z))v\biggr] \theta(\hat W^2-4m_c^2),\cr}}   
and it is easy to see that in the limit $Q^2\to \infty$, 
\eqn\limit{{d C^{FF,1}_{2,cg}(z,\epsilon) \over d \ln Q^2} \to P^0_{qg}(z).}
Hence, from \deflocf, we see that $C^{FF,NS,0}_{2,cc}(z,\epsilon)$ must indeed
tend to the simple form $z\delta (1-z)$ in this limit.  

Solving \deflocf\ for $C^{FF,NS,0}_{2,cc}(z,\epsilon)$ at arbitrary 
$\epsilon$ is not too complicated. Taking moments of both sides the 
Mellin transformation of $C^{FF,NS,0}_{2,cc}(z,\epsilon)$ is the product of 
the Mellin transformation
of ${d C^{FF,1}_{2,cg}(z,\epsilon) \over d \ln Q^2}$ and the inverse of the 
Mellin transformation of $(P^0_{qg}(z))$, which is 
\eqn\mell{ \int_0^1 z^{n-1} P^0_{qg}(z) dz ={n^2+n+2 \over
n(n+1)(n+2)}.}
This leads to the following expression for the convolution of 
$C^{FF,NS,0}_{2,cc}(z,\epsilon)$ with the heavy quark distribution,
\eqn\conv{\eqalign{C^{FF,NS,0}_{2,cc}(\epsilon)\otimes (c(Q^2)+
\bar c(Q^2))=& - \int_{x}^{x_0} dz \, {d C^{FF,1}_{2,cg}(z,\epsilon) 
\over d \ln Q^2}
\biggl({x\over z}\biggr)^2 {d (c(x/z,Q^2)+\bar c(x/z,Q^2))
\over d (x/z)}\cr
& \hskip -0.8in+ 3\int_{x}^{x_0} dx \, {d C^{FF,1}_{2,cg}(z,\epsilon) 
\over d \ln Q^2}
\biggl({x\over z}\biggr) (c(x/z,Q^2)+\bar c(x/z,Q^2))\cr
& \hskip -0.8in- 2\int_{x}^{x_0} dz \, {d C^{FF,1}_{2,cg}(z,\epsilon) 
\over d \ln Q^2}
\int_{{x/z}}^1 d z'\, r(z') {x\over zz'}
(c(x/zz',Q^2)+\bar c(x/zz',Q^2)),\cr}}
where $x_0=(1+4\epsilon)^{-1}$ and $r(z)$ is given by
\eqn\defr{r(z) = z^{\half}\biggl[ \cos\Bigl({\sqrt 7 \over 2}\ln 
{1\over z}\Bigr) + {3\over \sqrt 7} \sin \Bigl({\sqrt 7 \over 2}\ln 
{1\over z}\Bigr)\biggr].} 

Using this expression we are able to calculate the LO contribution to the  
heavy quark structure function using a particular set of parton 
distributions. In practice we use those obtained from a global fit to 
structure function data using the NLO formalism (details later). In order
to get the LO parton distributions we simply take the same input 
parameterizations for the partons but evolve them using the LO evolution 
equations. Our prescription for the LO $\alpha_s(Q^2)$ across the charm 
threshold is to define 
\eqn\coupdefzero{\alpha_s(Q^2, n)={\beta_0^n \over 4\pi 
\ln(Q^2/\Lambda_{QCD}^2)},}
and
\eqn\coupdefi{\alpha_{s,4}(Q^2)=\alpha_s(Q^2,4),} 
i.e. $\Lambda_{QCD}$ is defined for 4 flavours, and take for three 
flavours
\eqn\coupdefii{\alpha^{-1}_{s,3}(Q^2)=\alpha^{-1}_{s}(Q^2,3)+
\alpha^{-1}_{s}(m_c^2,4)-\alpha^{-1}_{s}(m_c^2,3).}
This prescription precisely reproduces the results of summing the leading 
logs in $(Q^2/m_c^2)$ in \locoup. 
The results of the LO contribution for the heavy quark 
coefficient function are shown in \fig\correctone{Charm quark 
structure function, $F_{2,c}(x,Q^2)$ for $x=0.05$ and 
$x=0.005$ calculated using our LO prescription, our input parton 
distributions evolved at LO and renormalization scale $\mu^2=Q^2$. Also shown 
are the continuation of the LO FFNS expression and the ZM--VFNS expression 
both calculated using the same parton distributions and same choice of scale.},
along with the 
continuation of the LO FFNS expression and also the LO ZM--VFNS expression. 
One can see that the LO VFNS expression departs very smoothly from the 
continuation of the LO FFNS expression, then rises above it, and in the limit 
of very high $Q^2$ becomes essentially identical to the LO ZM--VFNS 
result. This is precisely the behaviour we would expect. We also note that 
unlike other approaches the expression does not rely on any cancellation 
between terms. 
 
\medskip

We now consider also the NLO expression for the heavy quark structure function.
As well as the LO coefficient function just introduced we include the 
${\cal O}(\alpha_{s,n_f+1}(Q^2))$ coefficient functions. The expression for 
$C^{VF,1}_{2,cg}(z,\epsilon)$ is as in \defnlogcf, and is in terms of 
quantities we have written explicitly above in \lochgffns\ and 
\lodchgffns, i.e.  
\eqn\defnlogcf{C^{VF,1}_{2,cg}(z,\epsilon)= C^{FF,1}_{2,cg}(z,\epsilon)-
\ln (Q^2/m_c^2){d C^{FF,1}_{2,cg}(z,\epsilon)\over d\ln(Q^2)},}
Hence, there are no new problems in implementing 
$C^{VF,1}_{2,cg}(z,\epsilon)$. 
In the limit $Q^2\to \infty$ the first of these becomes
\eqn\limita{C^{FF,1}_{2,cg}(z,\epsilon)  \to P^0_{qg}(z)\ln\biggl({(1-z)
\epsilon\over z}\biggr) + (8z(1-z)-1).}
Using this, along with \limit\ and the definition \defnlogcf, we see that in 
the limit $\epsilon \to \infty$ 
\eqn\limita{\eqalign{C^{VF,1}_{2,cg}(z,\epsilon)  
&\to P^0_{qg}(z)\ln\biggl({1-z
\over z}\biggr) + (8z(1-z)-1)\cr
= C^{n_f+1,1}_{2,qg}(z),\cr}}
in the $\msbar$ renormalization scheme.  

From the arguments leading up to \defnlocf\ it is clear that 
$C^{VF,NS,1}_{2,cc}(z,\epsilon)$ also tends to the correct asymptotic limit,
and indeed, all coefficient function are constructed so that this will be
true. However, it is not possible to exhibit this in such an explicit manner
since the expression for $C^{VF,NS,1}_{2,cc}(z,\epsilon)$ depends on 
$C^{FF,2}_{2,cg}(z,\epsilon)$ for which there is no analytic 
expression.\foot{We are grateful to Jack Smith and Steve Riemersma for 
providing the extensive program to compute the
${\cal O}(\alpha_s^2)$ FFNS coefficient functions \ref\grid{S. Riemersma, 
J. Smith and W.L. van Neerven, \PL \vyp{B347}{1995}{143}.}.} 
Likewise, it would be extremely difficult to implement 
$C^{VF,NS,1}_{2,cc}(z,\epsilon)$ into the calculation precisely. In practice we
find that the contribution to the total heavy quark structure function of 
this term convoluted with the heavy quark parton distribution is only a very 
small fraction of the total. Hence, we include this contribution 
to the total heavy quark structure function in an approximate manner, 
being confident that it is very far from being significant
at the present level of accuracy required. 

Using our NLO prescription we use our NLO partons to calculate the heavy 
quark structure function. Our prescription for the NLO $\alpha_s(Q^2)$ is to 
define $\alpha_s(Q^2,n)$ by the standard two loop extension of 
\coupdefzero, and then to use \coupdefi\ and \coupdefii\ once again. 
\coupdefii\ does not sum all leading and next--to--leading logs in 
$(Q^2/m_c^2)$ absolutely correctly, but is an extremely accurate 
representation of the precise expression.   
The NLO charm structure function is shown in \fig\correcttwo{Same as 
\correctone, but with NLO prescriptions and NLO parton distributions.} along
with the continuation of the NLO FFNS expression and the NLO ZM--VFNS
result. As at LO the VFNS departs very smoothly from the continuation of the
FFNS expression. Although at this order we have not been able to 
demand absolute continuity of the derivative of the structure function across 
$Q^2=m_c^2$ we see that there is no visible evidence of discontinuity at all. 
In fact the transition from one scheme to the other is smoother than at LO. 
Also the VFNS stays closer to the continuation of the FFNS at higher $Q^2$
at this order. This is as we would expect, since as one works to higher 
orders in $\alpha_s$ in the FFNS one automatically includes more 
$\ln(Q^2/m_c^2)$ terms which are completely 
summed in the VFNS. At all orders the two schemes become identical 
insofar as physical quantities are concerned. At very high $Q^2$ then our
expression tends towards the NLO ZM--VFNS exactly as required. 

Of course, at higher $Q^2$ we also have effects due to the bottom quark
which has $m_b\approx 5\Gev$. Below $Q^2=m_b^2$ there is no bottom quark
distribution and we take account of the bottom 
quark effects by using treating all diagrams including bottom quarks in 
the fixed flavour scheme, and all other effects decouple. At $Q^2=m_b^2$
we switch to a VFNS for inclusion of the bottom quark effects. Analogously
to the charm quark this involves switching to a 5 flavour coupling constant, 
defined by
\eqn\coupdefiii{\alpha^{-1}_{s,5}(Q^2)=\alpha^{-1}_{s}(Q^2,5)+
\alpha^{-1}_{s}(m_b^2,4)-\alpha^{-1}_{s}(m_b^2,5),}
and beginning the evolution of the bottom quark density. The VFNS coefficient
functions are defined using a generalization of \bosglufus--\bosglufusiv.
There are now two heavy quarks in the definition of the fixed flavour 
number scheme, so we have two extra equations for $C_{2,bg}^{FF,S}$ and
$C_{2,bq}^{FF,S}$, and there are now extra VFNS coefficients such as 
$C_{2,bg}^{VF,S}$ or $C_{2,bc}^{FF,PS}$.
Also, the finite operator matrix elements will depend on 
both the charm and bottom mass effects. However, exactly the same principles
as outlined in the last section apply for determining the VFNS coefficient 
functions. At low orders in $\alpha_s(Q^2)$ 
there is no mixing between the charm mass effects and the bottom mass 
effects. Hence, the VFNS charm coefficients functions we have mentioned 
explicitly above remain the same above $Q^2=m_b^2$ (except for a 
completely negligible change in $C_{2,cc}^{FF,NS,1}$)
and the bottom quark coefficient 
functions are obtained simply by replacing $m_c$ with $m_b$ and 
$n_f+1$--flavour splitting functions with $n_f+2$--flavour splitting 
functions. At higher
orders the VFNS charm coefficient functions change above $Q^2=m_b^2$, acquiring
$m_b$ dependent corrections (in particular $C_{2,cb}^{FF,PS}$ comes into
existence), and bottom coefficient functions acquire charm mass corrections. 
 
\medskip

Thus, we have described how one may implement our prescription for the 
VFNS in practice, showing that there is no real difficulty. We have also
demonstrated that the results
have precisely the properties that our theoretical arguments in the 
previous section lead us to expect. In order to make even more concrete
statements regarding the suitability of our VFNS for the calculation 
of structure functions we will now discuss a comparison with data. 

\newsec{Phenomenological Results.}

Using the prescription for heavy partons discussed above we can calculate
the full structure functions in terms of input parton densities for the 
light quarks and gluon. The input scale is chosen as $Q_0^2=1\Gev^2$, and
the input parton distributions are then determined by performing a best fit 
to a wide variety of structure function and related data. Hence we
repeat the type of procedure adopted by MRS and CTEQ (and others) in 
their global determination of parton distributions. We note that apart from
the masses $m_c$ and $m_b$, which we fix at $1.35\Gev$ and $4.3\Gev$ 
respectively, the heavy quark contributions to structure functions 
have no free parameters. The overall description of the data is shown in 
table 2.\foot{We note that we do not alter the values of $F_2(x,Q^2)$
for the HERA data to take account of our predictions for $F_L(x,Q^2)$, as
should really be done. The $F_L(x,Q^2)$ values used in \hone\ and \zeus\
are obtained using a NLO--in--$\alpha_s(Q^2)$ calculation, and so are not
very different from ours in general, and the number of points affected is 
relatively small. Hence the quality of the overall fit is very insensitive 
to the neglect of this small 
correction.} There is a clear improvement when compared to 
the MRS(R$_2$) fit \ref\mrsr{A.D. Martin, R.G. Roberts and W.J. Stirling, 
\PL \vyp{B387}{1996}{419}.}, which uses the ZM--VFNS prescription with a 
phenomenologically motivated smoothing function, and 
a small but definite improvement when compared to the MRRS fit.\foot{The fit 
is not as good as the LORSC fit \ref\sxlet{R.S. Thorne, \PL 
\vyp{B392}{1997}{463}.}\sx, 
which includes $\ln(1/x)$ corrections but not yet charm mass corrections.}    

Full details of a global analysis using this procedure will be presented in
a future publication. However, here we will concentrate on those data
which describe the charm contribution to the structure function only. 
The component $F_{2,c}$ has been measured at intermediate $x$ values by 
EMC \ref\emc{EMC collaboration: J.J. Aubert {\it et al.}, \NP
\vyp{B213}{1983}{31.}} (via the detection of inclusive muons) and at small 
$x$ by the H1 \honecharm\ and ZEUS \zeuscharm\   
collaborations at HERA (through measuring $D$ and $D^{*}$ cross--sections).   
The latter indicate that as much as $20-25\%$ of the total $F_2(x,Q^2)$ is 
due to $F_{2,c}(x,Q^2)$. 
While these data on $F_{2,c}(x,Q^2)$ are not included as 
part of the general fitting procedure we can compare them with our 
theoretical predictions. The results are shown in \fig\charmdat{Our 
prediction for $F_{2,c}(x,Q^2)$ using our NLO prescription, the NLO
partons obtained from our global fit and three different values of 
$m_c$ compared with the EMC and HERA data.}. 
A very good description of both the small and medium $x$ data is achieved for
a charm mass of $m_c=1.35\Gev$,\foot{There is also a single EMC data point at 
$x=0.422$ and $Q^2=78\Gev^2$ not shown in \charmdat, which has 
$F_{2,c}(x,Q^2) = 0.00274 \pm 0.00152$ compared to a prediction of $0.0003$.}
although there is a strong suggestion that a 
slightly higher mass would be preferred, i.e. the curves for $m_c=1.5\Gev$
give a rather better description. In fact it is the data for 
$Q^2\sim 2\Gev^2$ which strongly favour this higher value of $m_c$. 
Since in this 
region of $Q^2$ the theoretical approach is unambiguous, i.e. the true 
result must be very similar to the FFNS calculation, 
these points may be thought 
of as determining the value of $m_c$ at about $1.5\Gev$, and a 
value very similar to this that should be used over the whole range.
Also, in \fig\percent{The ratios $F_{2,c}/F_2$ and $F_{2,b}/F_2$ at fixed
values of $Q^2$ resulting from our NLO parton distributions and taking
$m_c=1.35\Gev$ and $m_b=4.3\Gev$. The experimental data point shows the
estimate from ref. \honecharm\ for $F_{2,c}/F_2$ in the kinematic range
$10\Gev^2<Q^2<100\Gev^2$.} we show the 
relative importance of the charm and bottom components to the total 
structure function, and note the large fraction which is due to charm in 
the HERA low $x$ region. The bottom contributes no more than $\sim 4\%$
in any currently accessible range of $x$ and $Q^2$.

\newsec{The Longitudinal Structure Function.}

Finally we discuss our prediction for the heavy quark contribution to the
longitudinal structure function. Although there is no data directly 
available on this quantity, we feel that it is an important issue. This is 
because the implementation is a little different from the case of 
$F_2(x,Q^2)$, also because the 
charm contribution has a very important bearing on the total longitudinal 
structure function, contributing up to about $35\%$ in the ZM-VFNS for example,
and finally because our results are very different from those in other 
ways of implementing a variable flavour number scheme. 

As for $F_2(x,Q^2)$, and for the same reasons, there is no way we can obtain 
the (hypothetical) absolutely correctly ordered expression. Therefore,
as in this previous case we want some relatively simple prescription
which will reflect the physics involved correctly. 
There is a lot of similarity between our order--by--order prescription
for the heavy quark contribution to $F_L(x,Q^2)$ and that for $F_2(x,Q^2)$,
and the equations that the VFNS coefficient functions must satisfy in terms 
of the operator matrix elements and the FFNS coefficient functions are 
once again \bosglufus--\bosglufusiv. One large difference between the two is 
the fact that in a zero--mass formulation there are no zeroth order in 
$\alpha_s(Q^2)$ coefficient functions for the longitudinal structure 
function, and
hence the ${\cal O}(\alpha_s(Q^2))$ coefficient functions are leading order
and renormalization scheme independent. All previous implementations  
of a VFNS \acotpapone\acotpaptwo\mrrs\ have included a zeroth 
order heavy quark longitudinal coefficient
function, i.e. the term in \acota\ which $\propto 4M^2/Q^2$. 
This procedure means that there is a coefficient function at lower
order than the one which becomes leading order in the ZM--VFNS limit, and
hence in order to reach this limit with the LO VFNS expression
coefficient functions at both zeroth and  first order in 
$\alpha_{s,n_f+1}(Q^2)$ would need to be included. Also, if one
includes any zeroth order coefficient function, using 
the expression \bosglufus\ for the ${\cal O}(\alpha_{s,n_f+1}(Q^2))$ 
gluon coefficient 
function results in $C^{VF,1}_{L,Hg}(z,Q^2/M^2)$ having a component which is 
renormalization scheme dependent. 
 
Hence, we choose not to have any zeroth order contribution to the 
longitudinal coefficient functions. As with $F_2(x,Q^2)$ our VFNS
coefficient functions are then determined entirely by the requirements of 
reduction to the ZM-VFNS order by order as $Q^2 \to \infty$ and continuity 
with the FFNS across the boundary $Q^2=M^2$. Therefore, the prescription 
for the VFNS longitudinal structure function is very similar to that for 
$F_2(x,Q^2)$, except that the relative order of heavy quark coefficient 
functions above $M^2=Q^2$ and light quark coefficients at all $Q^2$ 
is one higher, i.e. ${\cal O}(\alpha_{s,n_f+1}(Q^2))$ is leading order 
${\it etc.}$. 
The prescription for the FFNS structure function at fixed order is then very
straightforward, i.e.   
\eqn\nffnslong{F^n_{L,i}(x,Q^2)=
\sum_{m=0}^n \sum_a \biggl({\alpha_{s,n_f}(Q^2)\over 2\pi}\biggr)^{n-m+1}
C^{FF,n-m+1}_{L,ia}(M^2/Q^2)\otimes f_m^{n_f}(Q^2),\quad n=1\to\infty,}
for both the heavy and light quark structure functions. 
The general form of the expression above $Q^2=M^2$ is the same as this, i.e.
\eqn\nvfnslong{F^n_{L,i}(x,Q^2)=
\sum_{m=0}^n \sum_a \biggl({\alpha_{s,n_f+1}(Q^2)\over 2\pi}\biggr)^{n-m+1}
C^{VF,n-m+1}_{L,ia}(M^2/Q^2)\otimes f_m^{n_f+1}(Q^2),\quad n=1\to\infty.}
Since the expressions are now of an identical form both sides of the
transition point (which was impossible for $F_2(x,Q^2)$ because of the
requirement of zeroth order heavy quark coefficient functions above 
$Q^2=M^2$), and are identical to all orders, continuity of the 
structure functions themselves is guaranteed order by order in $\alpha_s(Q^2)$ 
across the transition point. However,   
as in the case of $F_2(x,Q^2)$ the heavy quark coefficient functions at each 
order have to be determined using some prescription. 
(This ambiguity has no effect on the continuity of the structure function
since at $n_{th}$ order in the expression for the structure function the 
$n_{th}$--order heavy quark coefficient functions only appear coupling to 
the zeroth order heavy quark distribution, which vanishes at $Q^2=M^2$.)  

As with $F_2(x,Q^2)$ it would be nice to demand both continuity of the 
structure function and its $\ln Q^2$--derivative across the transition 
point. Since the expressions for the structure function are of exactly the 
same form both above and below $Q^2=M^2$ in this case (essentially because 
there are no zeroth order terms in the longitudinal structure function),
we can now attempt to equate the $\ln Q^2$ derivatives of the $n_{th}$
order terms in both expressions rather than making the derivative of the 
$n_{th}$--order VFNS expression match on to the $n_{th}$ order derivative 
in the FFNS as was necessary for $F_2(x,Q^2)$. 
As in the previous case we have enough choice to demand only that this is true
in the gluon sector, but again by far the dominant contribution to this 
derivative comes from this sector. However, now we have an additional 
problem compared to the previous case. This can be seen by
examining the lowest order expressions. 

In the FFNS the lowest order 
expression for the heavy quark structure function is
\eqn\lolongffns{F^0_{L,H}(x,Q^2)={\alpha_{s,n_f}(Q^2)\over 2\pi}
C^{FF,1}_{L,Hg}(M^2/Q^2)\otimes 
g_0^{n_f}(Q^2),} 
while in the VFNS it is 
\eqn\lolongvfns{F^0_{L,H}(x,Q^2)={\alpha_{s,n_f+1}(Q^2)\over 2\pi}
C^{VF,1}_{L,Hg}(M^2/Q^2)\otimes 
g_0^{n_f+1}(Q^2)+C^{VF,1}_{L,HH}(M^2/Q^2)\otimes 
(H(Q^2)+\bar H(Q^2))_0^{n_f+1},} 
and from \bosglufus\ we have 
\eqn\locoflong{C^{VF,1}_{L,Hg}(z,M^2/Q^2)=C^{FF,1}_{L,Hg}(z,M^2/Q^2)
=\biggl[4z(1-z)v-8\epsilon z^2 \ln\biggl({1+v \over 1-v}\biggr)
\biggr],}
for the charm quark, where again $\epsilon=m_c^2/Q^2$ and $\theta(\hat W^2 -
4m_c^2)$ is implied whenever $v$ appears. 
Clearly the structure functions are the same at $Q^2=M^2$. It is also clear 
that the ${\cal O}(\alpha_s(Q^2))$ expression for the $\ln Q^2$--derivative is 
the same on both sides, i.e. 
\eqn\lolongderiv{{dF_{L,H}(x,Q^2)\over d\ln Q^2}={\alpha_s(Q^2)\over 2\pi}
{dC^{FF,1}_{L,Hg}(Q^2/M^2)\over d\ln Q^2}\otimes g_0(Q^2).}
However, this expression, which vanishes as $Q^2/M^2\to \infty$,
is lower order than the leading order asymptotic expansion, which is 
${\cal O}(\alpha_s^2(Q^2))$. It is this, 
rather than a zeroth order coefficient function, 
which truly reflects the fact that 
the heavy quark longitudinal structure function has behaviour which begins 
at lower order than the massless expression.
This ${\cal O}(\alpha_s(Q^2))$ derivative means that while the asymptotic 
${\cal O}(\alpha_s^2(Q^2))$ expression for the total derivative of the 
longitudinal structure function is renormalization scheme independent, it 
contains a part which vanishes as $Q^2/M^2 \to\infty$ which is renormalization
scheme dependent. This is different to the case for $F_{2,H}(x,Q^2)$ where
the leading asymptotic and ${\cal O}(M^2/Q^2)$ parts of the derivative   
are both of the same order, i.e. ${\cal O}(\alpha_s(Q^2))$. 

If we treat the ${\cal O}(\alpha_s(Q^2))$ component of ${d F_{L,H}(x,Q^2)\over 
d\ln Q^2}$ as a superleading part which is trivially continuous across
$Q^2=M^2$, and then 
examine the form of the ${\cal O}(\alpha_s^2(Q^2))$ terms coming from the 
derivatives of \lolongffns\ and \lolongvfns, then since each of the 
leading--order expressions is renormalization scheme 
independent then so are the 
contributions obtained in the expressions for the derivatives. Explicitly we
obtain in the FFNS
\eqn\loderlongffns{\eqalign{{dF_{L,H}(x,Q^2)\over d\ln Q^2}=
&\biggl({\alpha_{s,n_f}(Q^2)\over 2\pi}\biggr)^2
C^{FF,1}_{L,Hg}(Q^2/M^2)\otimes \Bigl(-\beta_0^{n_f} g_0^{n_f}(Q^2)
+P^{0,n_f}_{gg}\otimes g_0^{n_f}(Q^2)\cr
&+P^0_{gq}\otimes \Sigma^{n_f}_0(Q^2)\Bigr),\cr}}
and in the VFNS
\eqn\loderlongvfns{\eqalign{{dF_{L,H}(x,Q^2)\over d\ln Q^2}=
&\biggl({\alpha_{s,n_f+1}(Q^2)\over 2 \pi}\biggr)^2
\biggl[C^{VF,1}_{L,Hg}(Q^2/M^2)\otimes \Bigl(-\beta_0^{n_f+1} 
g_0^{n_f}(Q^2)
+P^{0,n_f+1}_{gg}\otimes g_0^{n_f}(Q^2)\cr
&+P^0_{gq}\otimes \Sigma^{n_f}_0(Q^2)\Bigr)
+C^{VF,1}_{L,HH}(Q^2/M^2)\otimes 
\Bigl(P^0_{qg}\otimes g_0^{n_f+1}(Q^2)\cr
&+P^0_{qq}\otimes 
(H(Q^2)+\bar H(Q^2))_0\bigr)-\beta_0^{n_f+1}(H(Q^2)+\bar H(Q^2))_0
\biggr].\cr}}
From previous arguments it is clear that the terms 
$\propto C^{VF,1}_{L,Hg}(z,Q^2/M^2)$ in each equation are equal at $Q^2=M^2$.
The vanishing of the heavy quark distribution at this scale, lead to the 
single condition 
\eqn\locondlong{C^{VF,1}_{L,HH}(z,1)=0,}
in order to match these ${\cal O}(\alpha_s^2(Q^2))$ contributions to the 
derivative. Thus, we have this condition, along with the fact that 
$C^{VF,1}_{L,HH}(z,Q^2/M^2)$ 
must reduce to the correct asymptotic form, in order
to determine $C^{VF,1}_{L,HH}(z,Q^2/M^2)$. It is clearly possible to choose
forms for $C^{VF,1}_{L,HH}(z,Q^2/M^2)$ which satisfy these conditions, but
there is rather less guidance as to the precise form required than for
$F_{2,H}(x,Q^2)$, where the condition at $Q^2=M^2$ contained a component
which was clearly identifiable as the asymptotic expression. 

This indeterminacy is due to the fact that the ${\cal O}(M^2/Q^2)$ 
contributions to the derivative begin at one lower order than the asymptotic 
form, rather than our chosen manner of imposing the matching. If we had 
chosen to match \loderlongvfns\ to the total ${\cal O}(\alpha_s^2(Q^2))$ 
expression for ${d F_{L,H}(x,Q^2)\over d\ln Q^2}$ in the FFNS, 
rather than just the part 
coming from \loderlongffns, i.e. analogously to $F_{2,H}(x,Q^2)$, we would have
encountered a different problem. In this case the determined value of 
$C^{VF,1}_{L,HH}(z,1)$ would have contained a part $\propto {d C^{FF,2}_{L,Hg}
(z,Q^2/M^2)\over d \ln Q^2}$, which contains $C^{VF,n_f+1,1}_{L,qq}(z)$,
and the asymptotic limit would therefore appear more naturally.
However, the full expression for 
${d C^{FF,2}_{L,Hg}(z,Q^2/M^2)\over d \ln Q^2}$,
and consequently the full expression for $C^{VF,1}_{L,HH}(z,Q^2/M^2)$ implied,
contains ${\cal O}(M^2/Q^2)$ parts which are renormalization scheme dependent
(since they are subleading to the ${\cal O} (\alpha_s(Q^2))$ expression). 
This is not satisfactory in the definition of the leading order VFNS
coefficient function, and the renormalization scheme dependent part of 
the expression should be removed. However, there is no unique way to do this,
and hence the definition of $C^{VF,1}_{L,HH}(z,M^2/Q^2)$ would be just as
ambiguous as when using our chosen matching condition.    

Hence, we have to live with the fact that there is no completely satisfactory
way to 
determine $C^{VF,1}_{L,HH}(z,Q^2/M^2)$ from physical arguments. We choose to
impose \locondlong, as well as the fact that $C^{VF,1}_{L,HH}(z,Q^2/M^2)$
must reduce to the correct asymptotic form, and also choose the coefficient
function so that a smooth threshold in $\hat W^2$ is automatically 
incorporated. A simple choice satisfying all these requirements is 
\eqn\locoflongq{C^{VF,1}_{L,HH}(z,Q^2/M^2)={8\over 3}v(1-M^2/Q^2)z.}
In practice this ambiguity has little effect phenomenologically since the vast
majority of the LO expression for $F^{0}_{L,H}(x,Q^2)$ comes from the 
gluon contribution which is determined uniquely. Using \bosglufus\ we 
have now also defined $C^{VF,2}_{L,Hg}(z,Q^2/M^2)$, i.e.  
\eqn\nlocoflongg{C^{VF,2}_{L,Hg}(z,Q^2/M^2)=C^{FF,2}_{L,Hg}(z,Q^2/M^2)
-\ln(Q^2/M^2)P^0_{qg}\otimes C^{VF,1}_{L,HH}(Q^2/M^2),}
in $\msbar$ scheme, although we do not have to make use of this yet. The fact 
that $C^{VF,1}_{L,HH}(z,Q^2/M^2)$ reduces to the correct asymptotic limit
guarantees that $C^{VF,2}_{L,Hg}(z,Q^2/M^2)$ does.

As far as the light quark 
contribution is concerned the coefficient functions are identical 
above and below $Q^2=M^2$, and the only effect is the change of the evolution
of the parton distributions and in the running of the coupling. 
The $\ln Q^2$--derivatives of these LO light quark distributions which are
entirely of ${\cal O}(\alpha_s^2(Q^2))$, are not quite continuous across the 
transition point because of the flavour dependence of $\beta_0$ and of the
lowest order splitting functions, i.e. of $P^0_{gg}(z)$. As in the   
${\cal O}(\alpha_s^2(Q^2))$ derivative for 
$F_{2,i}(x,Q^2)$ there is continuity in 
the gluon sector, but not in the quark sector. Phenomenologically the 
discontinuity is very small, and becomes formally smaller as we work to 
higher orders. 

The result of our leading order calculation of $F_{L,c}(x,Q^2)$,
using the same LO parton distributions as before, is shown in 
\fig\lolongfig{Charm quark 
structure function, $F_{L,c}(x,Q^2)$ for $x=0.05$ and 
$x=0.005$ calculated using our LO prescription, our input parton 
distributions evolved at LO and renormalization scale $\mu^2=Q^2$. 
Also shown 
are the continuation of the LO FFNS expression and the ZM--VFNS expression 
both calculated using the same parton distributions and same choice of scale.},
along with the LO FFNS and the LO ZM--VFNS results. As in the case of 
$F_{2,c}(x,Q^2)$ one can see that the transition from the FFNS result 
is extremely smooth, and of course, the the correct asymptotic limit is 
reached. We note that at low $Q^2$ the VFNS result for $F_{L,c}(x,Q^2)$ is 
very different indeed from that in the ZM--VFNS. This leads to a very 
significant
difference between the results for the total $F_{L}(x,Q^2)$ in the two 
different schemes, and important phenomenological implications. 
We also show explicitly the contribution made by the  
charm quark distribution. At high $Q^2$ this is
unambiguously defined, and at low $Q^2$ it is very small indeed. Therefore,
the ambiguity in the low $Q^2$ heavy quark contribution is not really
significant. 

A very important difference should be noted between this approach and 
previous VFNS approaches at this point. As already mentioned, all previous
approaches have used a zeroth order charm quark coefficient function of the 
form
\eqn\zerocoefflong{\hat C^{VF,0}_{L,cc}(z,\epsilon)=4\epsilon z 
\delta(\hat x_0-z).}
If one were to regard the LO expression for $F_{L,c}(x,Q^2)$ as just this 
coefficient function convoluted with the heavy quark distribution function
then the behaviour would be rather strange, having a sharp
threshold at $Q^2=m_c^2$, growing quickly, then turning over and going to zero 
as $Q^2/m_c^2 \to \infty$. If, as is more likely, the LO expression is taken 
to include both the zeroth order and ${\cal O}(\alpha_s(Q^2))$ 
coefficient functions,
so that the correct asymptotic LO limit is reached, then 
$\hat C^{VF,1}_{L,cg}(z,\epsilon)$ is defined by \bosglufusex, i.e. 
\eqn\wronglocoeff{\hat C^{VF,1}_{L,cg}(z,\epsilon)=
\hat C^{FF,1}_{L,cg}(z,\epsilon)-
{4 m_c^2\over Q^2}P^0_{qg}\otimes z\delta(\hat x_0-z)(\ln (Q^2/m_c^2)
+c_{rs}).} 
As well as this introducing incorrect renormalization scheme dependence into 
a leading order expression (via $c_{rs}$), it has unfortunate
phenomenological consequences. The VFNS differs from the FFNS expression 
by 
\eqn\diff{4\epsilon z \delta(\hat x_0-z)\otimes \bigl[(c(Q^2)+\bar c(Q^2))_0 - 
\alpha_{s,4}(Q^2)
P^0_{qg}\ln (Q^2/m_c^2)\otimes g^{4}_0(Q^2)\bigr],}
where we have used $\msbar$ scheme. These two terms are intended to
largely cancel at and just 
above $Q^2=m_c^2$, ensuring a relatively smooth transition as in the ACOT 
prescription for the LO expression for $F_{2,c}(x,Q^2)$. The procedure 
works well in the case of $F_{2,c}(x,Q^2)$, and the transition is quite 
smooth, as we have seen. However, the cancellation is not exact (otherwise
we would just have the FFNS), $(c(z,Q^2)+\bar c(z,Q^2))_0\approx 
\alpha_{s,4}(Q^2)
P^0_{qg}\ln (Q^2/m_c^2)\otimes g^{4}_0(Q^2)$ for $Q^2$ just above $m_c^2$, 
but the resummation of the logs in the evolution of the charm quark 
distribution leads to differences appearing. To a rough approximation, 
\eqn\diffpart{\eqalign{(c(z,Q^2)+\bar c(z,Q^2))_0 -\alpha_{s,4}(Q^2)
P^0_{qg}&\ln (Q^2/m_c^2)\otimes g^{4}_0(Q^2)\cr
&\approx (\alpha_{s,4}(Q^2)\ln (Q^2/m_c^2))^2 P^0_{qg}
\otimes P^{0,4}_{gg}\otimes g^{4}_0(Q^2),\cr}}
at moderate $Q^2$. Inserting this into \diff\ leads to
\eqn\diffex{4\epsilon (\alpha_{s,4}(Q^2)\ln (Q^2/m_c^2))^2 
z\delta(\hat x_0-z)\otimes P^0_{qg}
\otimes P^{0,4}_{gg}\otimes g^{4}_0(Q^2).}
For $Q^2\approx 5-10\Gev^2$ this expression is comparable in size to the 
FFNS component of the full expression for $F_{L,c}(x,Q^2)$, which 
$\sim \alpha_s g^{4}_0(Q^2)$ with
damping due to kinematic factors (and which is more than
10 times smaller than the LO FFNS component for $F_{2,c}(x,Q^2)$).
However, it falls away quickly at larger
$Q^2$. This leads to the LO VFNS expression for $F_{L,c}(x,Q^2)$
increasing very quickly above the FFNS expression above the transition point,
dramatically slowing, or perhaps even falling at $Q^2\sim 5m_c^2$, 
and then smoothly 
approaching the correct asymptotic limit. i.e. there is a very pronounced 
unphysical bulge in the value of $F_{L,c}(x,Q^2)$ calculated in this way. 
When one calculates $R_c=F_{L,c}/(F_{2,c}-F_{L,c})$, which exhibits the 
relative rate of growth of $F_{L,c}$ and $F_{2,c}$, the effect is 
demonstrated much more clearly as a distinct hump peaking at about 
$Q^2=3m_c^2$. 
This can be seen very clearly in fig. 9 of \mrrs, where the effect is
particularly dramatic since the evolution of the heavy quark distribution
there is even quicker than in $\msbar$, and is at NLO. 
However, the treatment of
coefficient functions follows the same general principles as ACOT, 
and the same type of effect, if somewhat smaller 
(the reduction depending very strongly on the particular choice of 
renormalization scale -- one which departs extremely slowly from $m_c^2$ as 
$Q^2$ increases could remove the effect)
will be clearly seen in their expressions.\foot{In fact, since as we see in 
\acotone\ at $x=0.005$ the subtraction term is larger than the heavy
parton distribution, the effect will be negative.}    
Even in the absence of detailed data this type of effect seems sufficient
to rule out this approach as a suitable way to order a VFNS expression. 

\medskip

We now consider the NLO expressions for the longitudinal structure functions. 
For both heavy and light quark structure functions 
both above and below $Q^2=M^2$ we add to the LO expressions
the ${\cal O}(\alpha_s^2(Q^2))$ coefficient functions convoluted with the LO 
parton distributions and the ${\cal O}(\alpha_s(Q^2))$ coefficient functions 
convoluted with the NLO parton distributions. Let us first consider the  
heavy quark coefficient function. It is guaranteed by satisfying
\bosglufus--\bosglufusiv\ order by order in $\alpha_s(Q^2)$,
while also satisfying the correct relations between parton distributions and 
the coupling, that this procedure will lead to structure functions which are 
continuous across $Q^2=M^2$. This is straightforward, if a little tedious to 
check. Continuity of the derivative of the heavy quark structure function 
across the threshold is not guaranteed, but depends on the particular choice
of the heavy quark coefficient functions. We can compare the derivatives 
of the full NLO expressions in both the FFNS and VFNS up to 
${\cal O}(\alpha_s^3(Q^2))$. From the conditions we have already imposed 
using \bosglufus--\bosglufusiv\ it is
guaranteed that all new terms we introduce which behave like 
$\alpha_s^2(Q^2)$, 
i.e. those depending on ${d C^{F(V)F,2}_{L,Hg(q)}\over d \ln Q^2}$, will 
be continuous across the transition point. Again this is straightforward to 
check. If we examine the ${\cal O}(\alpha_s^3(Q^2))$ contributions to the 
expressions then both in the FFNS and VFNS these are very involved, i.e. 
containing rather more terms than \derivsix\ and \derivseven. However, as 
with \derivsix\ and \derivseven\ many of these terms vanish at $Q^2=M^2$, 
because the heavy parton distribution vanishes here, also because 
in this case $C^{VF,NS,1}_{L,HH}(z,1)=0$, and also because many other terms 
are the same in 
both expressions. A long, but entirely straightforward calculation reveals
that if we require continuity of the derivative in the gluon sector we
have the requirement that
\eqn\requirelong{\eqalign{C^{VF,S,2}_{L,HH}(Q^2/M^2)\otimes P^0_{qg}=&
C^{FF,2}_{L,Hg}(z,Q^2/M^2)\bigl(\beta_0^{n_f+1}-\beta_0^{n_f}\bigr)+
C^{FF,1}_{L,Hg}(z,Q^2/M^2)\bigl(\beta_1^{n_f+1}-\beta_1^{n_f}\bigr)\cr
&-C^{FF,1}_{L,Hg}(Q^2/M^2)\otimes \bigl(P^{1,n_f+1}_{gg}-
P^{1,n_f}_{gg}\bigr),}}
at $Q^2=M^2$. So at ${\cal O}(\alpha_s^2(Q^2))$, as at 
${\cal O}(\alpha_s(Q^2))$, 
there is no implication of the asymptotic form required of the 
heavy quark coefficient function in the condition at $Q^2=M^2$, but the 
condition is no longer that the coefficient function is zero at this value 
of $Q^2$. We can understand where the nonzero terms come from quite easily. 
If we had used the whole of the ${\cal O}(\alpha_{s,n_f}^3(Q^2))$ 
expression for the 
derivative of the heavy quark structure function in the FFNS, and 
equated this to our VFNS expression then the asymptotic form of 
$C^{FF,2}_{L,HH}(z,Q^2/M^2)$ would have appeared naturally in the expression 
for ${d C^{FF,3}_{L,Hg}\over d \ln Q^2}$. However, by examination of the 
expression for ${d C^{FF,3}_{L,Hg}\over d \ln Q^2}$ contained within 
\bosglufus\ we would find that the definition of $C^{VF,NS,2}_{L,HH}
(z,Q^2/M^2)$
would also need to contain terms of the sort in \requirelong, as well as 
others which vanish at $Q^2=M^2$, in order to reduce to the correct
asymptotic limit. However, in an analogous fashion to our previous 
discussion at leading order, we do not use this technique since parts of the
${\cal O}(M^2/Q^2)$ corrections to ${d C^{FF,3}_{L,Hg}\over d \ln Q^2}$
are properly of NNLO, i.e. are renormalization scheme dependent in such a 
way as to compensate for the renormalization scheme variation of the NLO
terms. This would require an ambiguous subtraction procedure for these
terms, and we would have no more real information than that contained in 
\requirelong\ and the asymptotic condition. 

Hence, as for the ${\cal O}(\alpha_{s,n_f+1}(Q^2))$ coefficient 
function we make a simple
choice for the coefficient function which satisfies \requirelong,
which reduces to the correct asymptotic limit, and which explicitly contains
the correct threshold behaviour. Once again we multiply the asymptotic limit,
which makes no appearance at $Q^2=M^2$, by $(1-M^2/Q^2)v$. We multiply 
the terms appearing in \requirelong, but which must disappear asymptotically,
by $M^2/Q^2$ (in this case the threshold behaviour is automatically contained
in the expressions). Hence we obtain  
\eqn\nlocoflongq{\eqalign{C^{VF,S,2}_{L,HH}(z,Q^2/M^2)=&\biggl(1-
{M^2\over Q^2}\biggr)v
C^{n_f+1,2}_{L,qq}(z)+{M^2\over Q^2}(P^0_{qg})^{-1}\otimes\biggl[
C^{FF,2}_{L,Hg}(Q^2/M^2)\bigl(\beta_0^{n_f+1}-\beta_0^{n_f}\bigr)\cr
&\hskip -0.4in+C^{FF,1}_{L,Hg}(Q^2/M^2)\bigl(\beta_1^{n_f+1}-
\beta_1^{n_f}\bigr)
-C^{FF,1}_{L,Hg}(Q^2/M^2)\otimes \bigl(P^{1,n_f+1}_{gg}-
P^{1,n_f}_{gg}\bigr)\biggr].}}
This definition is ambiguous at low $Q^2$, but as at leading order the 
total heavy quark structure function at NLO is totally dominated by the 
gluon contribution. We also note that the ambiguity introduced at LO
from the definition of the heavy quark coefficient function is very largely
negated at NLO by the inclusion of $C^{VF,NS,1}_{L,HH}(z,Q^2/M^2)$ in the 
expression for $C^{VF,2}_{Hg}(z,Q^2/M^2)$ \nlocoflongg. As we work to higher 
orders the ambiguity formally disappears. We also note that the coefficient 
function $C^{VF,S,2}_{L,HH}(z,Q^2/M^2)$
is the sum of the nonsinglet and pure singlet coefficient functions. 
We are free to separate them as we wish, using the condition that each tends 
to the correct asymptotic limit. It would also be desirable to choose each 
so that they respect the kinematic threshold. The choice has no bearing on 
the expression for the structure function but a simple choice is to let the 
nonsinglet part contain all parts $\propto v$, and to split the other
part simply in terms of the asymptotic form. 

Comparing the ${\cal O}(\alpha_s^3(Q^2))$ expressions for the derivative of the
heavy quark structure function which are proportional to the singlet quark
distribution, then as for the NLO derivative for $F_{2,H}(x,Q^2)$ we 
see that continuity is not achieved. The difference between the VFNS
expression and the FFNS expression is 
\eqn\disconlong{2(\beta_0^{n_f}-\beta_0^{n_f+1})C^{FF,PS,2}_{L,Hq}(Q^2/M^2)
\otimes \Sigma^{n_f}_0(Q^2),}
where $C^{FF,PS,2}_{Hq}(z,Q^2/M^2)=C^{VF,PS,2}_{L,Hq}(z,Q^2/M^2)$.
This NLO effect is very small, and as for $F_{2,H}(x,Q^2)$ the effect
disappears as we work to higher orders.  

For the light quark structure functions there is one choice to make. 
There is a mass dependent contribution to the nonsinglet coefficient 
function at ${\cal O}(\alpha_s^2(Q^2))$, but the form of the VFNS 
coefficient function is determined entirely by \bosglufusii. In essence 
the mass dependent correction to $C^{NS,n_f,2}(z)$ contain a piece which 
becomes constant asymptotically which represents $C^{NS,n_f+1,2}(z)-
C^{NS,n_f,2}(z)$, and a piece which grows like $\ln(Q^2/M^2)$ which takes 
account of the difference between the $n_f+1$ and $n_f$ flavour couplings. 
The ${\cal O}(\alpha_s^3(Q^2))$ $\ln Q^2$ derivative is slightly discontinuous
at $Q^2=M^2$, but this is corrected by inclusion of the 
${\cal O}(\alpha_s^3(Q^2))$ coefficient functions. 
For the pure singlet and gluon coefficient functions coupling to light partons
there are no mass dependent corrections in the FFNS at ${\cal O}
(\alpha_{s,n_f}^2(Q^2))$, and we
simply use the same coefficient functions above and below $Q^2=M^2$. 
Continuity of the NLO structure functions is then automatic. 
However, the pure singlet coefficient function $C^{VF,PS,2}_{L,qH}(z,Q^2/M^2)$
becomes nonzero at this order. It can be determined by 
demanding continuity of the $\ln Q^2$--derivative of the structure function 
in the gluon sector. This results in a similar procedure as for the heavy to 
heavy coefficient function: the asymptotic form is put in by hand and 
multiplied by $(1-M^2/Q^2)v$ while there are nonzero terms at $Q^2=M^2$ 
which are multiplied by $v$ to ensure that they vanish as 
$Q^2/M^2\to \infty$. The calculation is straightforward, and we do not 
present details here. As for the heavy quark structure function 
the ${\cal O}(\alpha_s^3(Q^2))$ $\ln Q^2$ derivative in the singlet quark 
sector is slightly discontinuous at $Q^2=M^2$, but again this is corrected at
next order by inclusion of the ${\cal O}(\alpha_s^3(Q^2))$ coefficient 
functions.   

Now that the NLO prescription for the longitudinal structure function is 
completely defined we can examine the results. Using our NLO coefficient 
functions \nlocoflongq\ and \nlocoflongg\ and the NLO partons obtained from 
the best fit we calculate the NLO charm quark longitudinal structure function. 
This is shown in \fig\nlolongfig{Same as 
\lolongfig, but with NLO prescriptions and NLO parton distributions.} along 
with the continuation of the 
NLO FFNS expression and the NLO ZM--VFNS result. Once again the VFNS 
increases above the FFNS result very smoothly despite the discontinuity in the 
$\ln Q^2$ derivative in the singlet quark sector which is now demonstrably
minute. At very high $Q^2$ our expression tends towards the ZM--VFNS as 
required, but as at LO the two become very different at lower $Q^2$. 
As in the case of $F_2(x,Q^2)$ at NLO the difference between the 
VFNS and the continuation of the FFNS is reduced compared to the
difference at LO for the same reasons. Hence we have every reason to consider 
our prescription for the longitudinal structure functions quite satisfactory. 
 
In fact we can compare to some data. At $x<0.1$ the VFNS, ZM-VFNS and FFNS 
values for the total $F_L(x,Q^2)$ are very similar. However, the NMC
collaboration have produced data for $0.11>x>0.0045$ and 
$1.3\Gev^2<Q^2<20.6\Gev^2$ \ref\nmc{NMC collaboration: \NP
\vyp{483}{1997}{3}.}, $Q^2$ increasing as $x$ increases. These data are 
in the region where our VFNS prescription produces very different results 
to the ZM--VFNS (but almost identical to the FFNS) for $F_{L,c}(x,Q^2)$,
and hence significantly different results for the total longitudinal structure 
function.  Using the parton distributions obtained from our best global fit
we produce predictions for $R(x,Q^2)=F_L(x,Q^2)/(F_2(x,Q^2)-F_L(x,Q^2))$
using the ZM--VFNS and the VFNS, and compare to the data. The results are shown
in \fig\nmccomp{Our 
prediction for $R(x,Q^2)$ using our NLO prescription, the NLO
partons obtained from our global fit and $m_c=1.35\Gev$ compared with the NMC 
data \nmc. Also shown is the prediction obtained using the same parton 
distributions but for the NLO ZM--VFNS prescription. The curves are computed 
using $Q^2=1.3\Gev^2$ for $x\leq 0.0077$ and $Q^2=262 x^{1.09}$ for 
$x\geq 0.0077$.}. The kink in both curves at the 
lowest  $x$ values
comes about because for all data points other than that at $x=0.0045$, as
$x$ decreases $Q^2$ also decreases smoothly, while for this point the 
extraction of $R(x,Q^2)$ relies on an extrapolation and the $Q^2$ is 
actually almost
identical to that for the $x=0.008$ point. The kink in the ZM--VFNS 
expression at $Q^2=m_c^2=1.8\Gev^2$ is due to all charm coefficient functions
turning on discontinuously at this point. Comparing to the data it is clear
that the ZM--VFNS curve is much too large for most of the $x$--range,
while down to
$x=0.01$ the VFNS curve matches the data quite well. 
Thus, there is strong evidence 
for including charm mass effects in the longitudinal structure function, 
and our prescription seems reasonably successful. 
Other VFNS prescriptions would lead to 
$R(x,Q^2)$ somewhere between the two curves. 
The theory is clearly below the 
data for the lowest $x$ and $Q^2$ points where the charm 
contribution to $R(x,Q^2)$ is extremely small, i.e. the VFNS and ZM--VFNS 
curves are almost identical. The smallness of $R(x,Q^2)$, and the decrease 
with decreasing $x$ at constant $Q^2$ in this region are largely due to 
a negative small $x$ contribution from $C^{n_f,2}_{L,ig}(z)$ which becomes 
increasingly important as $x$ and $Q^2$ fall. Thus, the difference between 
the theory and data for the two lowest $x$ points is perhaps a sign of 
the failure of the NLO--in--$\alpha_s(Q^2)$ calculation of structure functions 
at small $x$.\foot{The curve labelled $R_{QCD}$ in fig. 10 of the NMC paper
\nmc\ contains little information. For $F_L(x,Q^2)$
it uses a LO formula \ref\altmart{G. Altarelli and G. Martinelli, \PL
\vyp{B76}{1978}{89}.}(and hence does not contain the 
important NLO small $x$ effect) which assumes 4 massless quarks at all
$Q^2$, along with a gluon which has been extracted using a NLO fit in the 
FFNS \hone. Moreover, this gluon is not constrained at large $x$ and is highly
inconsistent with large $x$ data. From the momentum sum rule this means its 
form at small $x$ is also much different to a well constrained gluon.}      

\medskip

As with $F_2(x,Q^2)$ the NLO calculation is the best that can be done 
explicitly with the present knowledge of structure functions. However, 
as in this previous case, we outline the procedure for all orders. 
The general form of the expressions is presented in \nffnslong\ and
\nvfnslong, and for the heavy quark structure function nothing essentially 
new compared to the LO and NLO prescriptions occurs. At $n_{th}$ 
nontrivial order we determine $C^{VF,n}_{L,HH}(z,Q^2/M^2)$ by demanding
continuity of the $\ln Q^2$ derivative of the structure function   
in the gluon sector, and by demanding the correct asymptotic form. 
At each order the correct asymptotic form will not appear in the continuity
conditions and need to be introduced by hand. Each time we multiply by 
$(1-M^2/Q^2)v$. At each order there will also be terms introduced by the 
continuity demand which must vanish as $Q^2\to \infty$, and we multiply
these by $M^2/Q^2$. At every order this determination of
$C^{VF,n}_{L,HH}(z,Q^2/M^2)$
predetermines $C^{VF,n+1}_{L,Hg}(z,Q^2/M^2)$ and $C^{VF,n+1}_{L,Hq}(z,Q^2/M^2)$
by using \bosglufus\ and \bosglufusi\ to ${\cal O}(\alpha_{s}^{n+1}(Q^2))$.
The comments concerning the separation of $C^{VF,n}_{L,HH}(z,Q^2/M^2)$ 
into nonsinglet and pure singlet parts in \S 4 apply again.   

For the light quark structure function the procedure at higher orders 
is also straightforward. 
As with the ${\cal O}(\alpha_{s,n_f+1}^2(Q^2))$ 
coefficient function we determine 
$C^{VF,n}_{L,qH}(z,Q^2/M^2)$ by demanding continuity of the derivative of the
light structure function in the gluon sector, analogously to the heavy quark 
sector. With this one degree of freedom eliminated in this way
all other VFNS coefficient functions are determined uniquely order by order in 
$\alpha_{s,n_f+1}(Q^2)$ by \bosglufusii--\bosglufusiv, i.e.
this determination of $C^{VF,n}_{L,qH}(z,Q^2/M^2)$
predetermines $C^{VF,n+1}_{L,qg}(z,Q^2/M^2)$ and 
$C^{VF,PS,n+1}_{L,qq}(z,Q^2/M^2)$
by using \bosglufusiii\ and \bosglufusiv\ to ${\cal O}(\alpha_s^{n+1}(Q^2))$. 

Thus, we have completely defined our prescription for calculating 
the structure function $F_L(x,Q^2)$ order by order. As for $F_2(x,Q^2)$ we 
can sum it up in a simple diagram, shown in table 3.  
The generalization to the case of two heavy quarks follows the same lines as 
for the case of $F_2(x,Q^2)$ which was discussed at the end of \S 5. 
For $Q^2<m_b^2$ the bottom quark effects are all treated via FFNS 
coefficient functions while in the region above $Q^2=m_b^2$ we have a 
variable flavour scheme for both the charm and bottom quark. For high
orders in $\alpha_s(Q^2)$ there will be mixing of the effects of the
two quarks, but for the 
orders currently available in practice the mixing is extremely small
indeed, as with $F_2(x,Q^2)$, and the bottom coefficient functions are 
essentially the same as those for charm with $m_c\to m_b$ and with 
5 flavours rather than 4.     

Our prescription uniquely determines all VFNS coefficient functions, and 
as for $F_2(x,Q^2)$, while not leading to absolute correctly ordered 
expressions it is a 
relatively simple prescription for obtaining order by order structure 
functions which are very similar to the hypothetical strictly correct ones, 
which reduce to 
the correct asymptotic form order by order in $\alpha_{s,n_f+1}(Q^2)$, 
and which are 
consistent with physical requirements order by order. All ways of satisfying
both \bosglufus--\bosglufusiv\ and the correct asymptotic limits will be 
correct in a certain sense (provided they are consistent with
ordering within a given renormalization scheme), but many 
will have behaviour which is 
unsatisfactory for $Q^2$ not much larger than $M^2$, and we have seen an 
example of this. As with $F_2(x,Q^2)$ we believe our prescription to be 
very suitable. 

\newsec{Summary and Conclusion.}

In this paper we have constructed an order by order in $\alpha_s$ prescription
for calculating the neutral current structure function including the 
effects of a massive quark. For the region $Q^2 < M^2$ this has essentially 
just been the normal FFNS, where the heavy quark is not treated as a 
constituent of the hadron, but all heavy quarks in the final state are 
generated via the electroweak boson interacting with light partons. 
For $Q^2>M^2$ we have to solve the problem of summing large logs in $Q^2/M^2$ 
and $\mu^2/M^2$ which appear at all orders in $\alpha_s(\mu^2)$. 
The easiest way to do this is to 
to treat the heavy quark as a parton, in which case the logs will be summed
automatically when one solves the evolution equations for the partons. 
If one chooses the parton 
distributions above $\mu^2=M^2$ to evolve as though massless
and in the $\msbar$ scheme, then the new $n_f+1$ flavour parton distributions 
are determined in terms of the FFNS parton distributions at all $\mu^2$ by
well--defined, calculable matrix elements which contain logs in $\mu^2/M^2$.
In particular the heavy quark distribution is determined entirely in terms 
of the light parton distributions.   
The matrix elements can then be used to define the $n_f+1$ flavour
parton distributions in terms of the $n_f$ flavour distributions at some scale
(in practice $\mu^2=M^2$ is by far the most convenient), and the evolution 
upwards can take place in terms of $n_f+1$ massless flavours with the 
correct asymptotic limits being guaranteed. If the massless $n_f+1$
flavour coefficients functions are used then the correct asymptotic limit   
for the structure functions is also reached. 

The main problem lies in obtaining the correct description in the region 
not too far above $Q^2=M^2$. We have demonstrated that this is achieved to all
orders by defining the mass--dependent coefficient functions above 
$Q^2=M^2$ in terms of the operator matrix elements and the FFNS coefficient 
functions as in \bosglufus--\bosglufusiv. However, we have also demonstrated 
that since there are more degrees of freedom on the right hand side of these 
equations than on the left, the additional ones all being coefficient 
functions coupling to heavy quarks, there is freedom in precisely how the 
coefficient functions may be chosen. Although in a true well--ordered 
calculation this ambiguity disappears, this manner of ordering is at the 
very least extremely complicated, involving parts of the FFNS from all
orders in $\alpha_s(\mu^2)$ at each order in the calculation, and in practice 
is probably impossible, there being no unique prescription for ordering the 
${\cal O}(M^2/Q^2)$ terms. 
Hence we choose to order our calculation as in the normal
order--by--order in $\alpha_s(\mu^2)$ manner, choosing the very simple 
natural scale $\mu^2=Q^2$,
which puts all of the mass effects into the coefficient functions
and guarantees the 
correct asymptotic limit order by order in $\alpha_s(Q^2)$. We then determine 
the precise form of
our heavy quark coefficient functions by demanding continuity not only of the 
structure functions at $Q^2=M^2$ (which is automatic), but also the 
continuity of the $\ln Q^2$--derivative of the structure function. In 
practice this exact continuity 
is only possible for those terms proportional to the gluon, 
but this is by far the dominant contribution. Our constraint then determines 
our prescription for dealing with heavy quarks completely, and 
incorporates the correct qualitative threshold behaviour into
every coefficient function at each order of $\alpha_s(Q^2)$, not relying on 
cancellations between terms with incorrect behavior and of different orders 
to obtain satisfactory results. In practice the most important of our 
results are 
the zeroth order coefficient function for $F_{2,c}(x,Q^2)$, \deflocf, 
which exhibits the correct threshold behaviour in $\hat W^2$ as well as 
reducing to
the correct asymptotic form, and the absence of a zeroth order coefficient 
function for $F_{L,c}(x,Q^2)$, the ${\cal O}(\alpha_s(Q^2))$ coefficient
functions being \locoflong\ and \locoflongq, which again exhibit the 
correct threshold behaviour and asymptotic limits. 

We display the results obtained using our prescription for neutral current
structure functions in figs. 3, 4, 7 and 8, 
finding that they exhibit exactly the type of behaviour 
we would expect, i.e. smoothly deviating from the FFNS at low $Q^2$, and 
tending towards the $n_f+1$ massless results at high $Q^2$, in all cases.
In particular we notice that the bump in the charm quark longitudinal
structure function at $Q^2\approx 10 \Gev^2$ which occurs in other
variable flavour number schemes is absent here. We also see that our 
predictions agree very well with the current data on the charm structure 
function which 
exists from $1.5\Gev^2<Q^2<100\Gev^2$, implying a charm quark mass of 
$\sim 1.45\Gev$. We note that comparisons of theoretical 
predictions with the complete
range of data on the charm structure function appear very rarely
(in particular, detailed comparison with EMC data is frequently
omitted), and we strongly encourage this as the best constraint on 
any theory.   

The general technique can be applied to all other quantities in 
perturbative QCD which require the convolution of coefficient functions 
with parton distributions. We can always choose the parton distributions to 
evolve as though there are $n_f+1$ massless flavours in the $\msbar$--scheme, 
factor these into the mass dependent operator matrix elements and the 
FFNS parton distributions
and then obtain the coefficient functions in the variable flavour scheme in 
terms of those in the fixed flavour scheme by equating the parts proportional
to each FFNS parton distribution. Indeed, the expressions 
\bosglufus--\bosglufusiv\ are not exclusive to neutral current structure 
functions, but apply to all quantities which can be written as the 
sum of convolutions of coefficient functions with single parton distributions. 
In the appendix we discuss the case of the charged current structure 
functions as an example. For expressions involving more
than one parton distribution the generalization is clear, e.g. for 
proton--proton scattering the FFNS and VFNS coefficients are related by
equations of the form $C^{FF}_{iab}=C^{VF}_{icd}A^{ca}A^{db}$. In all
cases there will be ambiguity in definitions of the heavy parton coefficient
functions, but these can always be eliminated by demanding as much 
continuity of the $\ln Q^2$--derivative order--by--order in $\alpha_s(Q^2)$ as 
possible. 
 
Let us briefly discuss problems which arise in other approaches to heavy quark
structure functions.
Buza {\it et al.} do not provide a detailed prescription for the region
of $Q^2$ just above $M^2$. They have a means of extrapolating 
the structure function from the FFNS result at $Q^2<M^2$ to the 
ZM--VFNS result at $Q^2/M^2\to \infty$ in a way which guarantees 
smoothness \vanneerone\vanneertwo, 
but it seems phenomenologically motivated, with no
strict definition of the ordering and certainly no expressions for parton 
distributions and coefficient functions in the intermediate region. 
The ACOT group have a prescription which involves switching from $n_f$ to 
$n_f+1$ massless flavours in the evolution, and a way of determining the 
VFNS coefficient functions \acotpapone\acotpaptwo\ which at low orders 
appears to be the same as 
prescribed in \bosglufus-\bosglufusiv. However, their way of eliminating the 
free choices in the heavy quark coefficient functions involves assuming that
the behaviour is as if there is intrinsic charm in the proton at all
scales above the transition point, rather than charm being generated 
almost entirely from the gluon. This leads to coefficient functions 
having thresholds in $Q^2=M^2$ rather than $\hat W^2=4M^2$, and a mixing 
of orders
being required (and a complicated renormalization scale being 
advantageous) in order to ensure cancellations and that smooth 
behaviour occurs, e.g. the 
${\cal O}(\alpha_s(\mu^2))$ gluon coefficient function must appear at the same 
time as the zeroth order quark coefficient function. This mixing of orders
is incorrect, being at odds with well--ordered asymptotic expressions, 
but removing it results in a lack of smoothness in the structure functions.
Even when this mixing is retained the behaviour of the  
longitudinal structure function is still not smooth. 
The MRRS procedure \mrrs\ is based on the leading log limit of Feynman 
diagrams rather than the renormalization group and as such incorporates
mass--dependent effects in the evolution, but seems more difficult to define 
formally to all 
orders in $\alpha_s$. The definition of the heavy quark coefficient
functions uses similar reasoning to ACOT, but
in this case with ordering such that it reduces to the 
correct well--ordered form asymptotically. These coefficient functions 
along with the imposition of this correct ordering
leads to an unphysical lack of smoothness in the structure functions (which is
made slightly worse by the mass--dependent contributions to the evolution), 
particularly for the longitudinal structure function.
Our prescription has none of the above problems. It is well--defined to all
orders, reduces to correct well-ordered expressions at both low and high
$Q^2$, and exhibits precisely the behaviour one would expect. Hence,  
we believe that our prescription is the best currently available to 
describe the heavy quark contribution to structure functions.   

Before finishing let us mention a couple of points in which our treatment is
incomplete. Firstly, we have assumed that there is no intrinsic charm in 
the nucleon. Eqn. \heavycs\ is formally correct up to the quoted error, but 
this error has an unknown numerical factor and may be enhanced by functions of 
$x$. It appears that for intrinsic charm the numerical factor of this 
``higher twist'' correction is rather large and that the contribution is 
enhanced by a factor of $(1-x)^{-1}$. Therefore, at large $x$ , where the 
leading twist contribution to the charm structure function is not large 
anyway, it seems as though the ``higher twist'' intrinsic charm may 
constitute an important part of the total charm structure 
function \ref\brod{S.J. Brodsky, P. Hoyer, A.H. Mueller and W.K. Tang, 
\NP \vyp{B369}{1992}{519}, and references therein.}. The
treatment of this correction is outside the scope of this paper, which
deals only with the ``leading twist'' contribution to the structure 
function, and we believe it is not naturally dealt with in any other VFNS. 
However, it seems very unlikely that in most of the region where there
is current data on the charm structure function, or where the 
charm contribution is a 
sizable fraction of the total structure function, that this ``higher
twist'' contribution plays any significant role at all. Using the type of 
values expected for this intrinsic charm (see e.g. \ref\intrin{B.W. Harris, 
J. Smith and R. Vogt, \NP \vyp{B461}{1996}{181}.}) then adding to our 
values does bring the $x=0.422$ prediction in line with the EMC data point,
raises the $x=0.237$ predictions quite significantly (but neither really 
helps or hinders the comparison to the three data points), 
raises the $x=0.133$ predictions a little (tending to make the comparison 
a little worse), and has negligible effect for lower $x$. Hence, the 
$x=0.422$, $Q^2=78\Gev^2$ EMC data point may be seen as some evidence for 
this ``higher twist'' intrinsic charm.

Finally we note that throughout this paper we have completely ignored the 
problem of enhancement of higher orders in $\alpha_s$ by 
$\ln (1/x)$ terms. These terms certainly do have the potential to 
quantitatively alter the results of this paper. Correctly including 
the leading $\ln(1/x)$
terms within the context of only massless quarks is a complicated 
procedure, though it does appear to improve the description of small $x$ 
data \sx. Some results on heavy quark
coefficient functions which include leading $\ln(1/x)$ terms already 
exist \ref\catciaf{S. Catani, M. Ciafaloni and F. Hautmann, \PL 
\vyp{B242}{1990}{97}; \NP \vyp{B366}{1991}{135}.}\ref\cat{S. Catani, 
\ZP \vyp{C75}{1997}{665}; Proceedings of the
International Workshop on Deep Inelastic Scattering, Rome, April 1996, 
p. 165.}. It would clearly be desirable to extend 
this work and to include both 
the correct treatment of leading $\ln(1/x)$ terms and a correct description 
of heavy quark results within a single framework. Work along these lines 
is currently in progress. 

\appendix{A}{Charged Current Structure Functions.}

The treatment of the charged current structure function follows exactly the 
same reasoning as for the neutral current case. Let us consider $F_2(x,Q^2)$.
Equations \bosglufus--\bosglufusiv\ are derived in exactly the same way, 
but now take a different form because there are no nonsinglet coefficient
functions. For the case where a heavy quark is produced directly 
by the interaction with the $W$ boson, 
which we call the heavy quark structure function, we have 
\eqn\bosglufuscc{C^{FF,S}_{Hg} = A^S_{gg,H}\otimes 
C^{VF,S}_{Hg} + n_f A^S_{qg,H} \otimes C^{VF,PS}_{Hq}
+A^S_{Hg}\otimes C^{VF,PS}_{HH},}
and
\eqn\bosglufusicc{C^{FF,S}_{Hq} = A^{PS}_{Hq}\otimes 
C^{VF,PS}_{HH} +\bigl[A^{NS}_{qq,H} 
+n_f A^{PS}_{qq,H}\bigr]\otimes C^{VF,PS}_{Hq}+
A^S_{gq,H}\otimes C^{VF,S}_{Hg}.}
We note that what we have denoted the charm quark structure function here
may be interpreted physically as the unlike sign dimuon contribution. 
In the case where the $W$ boson directly produces a light quark, 
which we call the light quark structure function, we have
\eqn\bosglufusiiicc{C^S_{qg}+ C^{FF,S}_{qg} =  A^{S}_{gg,H} 
\otimes C^{VF,S}_{qg}
+n_f A^S_{qg,H}\otimes C^{VF,PS}_{qq}+ A^S_{Hg}\otimes C^{VF,PS}_{qH},}
and
\eqn\bosglufusivcc{ C^{PS}_{qq}+ C^{FF,PS}_{qq} = n_f A^{PS}_{qq,H}
\otimes C^{VF,PS}_{qq} + A^{PS}_{Hq} \otimes C^{VF,PS}_{qH}
+ A^S_{gq,H}\otimes C^{VF,S}_{qg}.}
It is not only the absence of the nonsinglet coefficient functions which is 
different, the ordering of the other coefficient functions also changes, in
particular $C^{PS}_{2,q_iq_j}$, $i\not= j$, begins at zeroth order. This 
changes the form of the relationship between the FFNS and the VFNS 
coefficient functions. For example, examination of \bosglufuscc\ reveals that 
we have the trivial equality
\eqn\trivequ{C^{FF,1}_{2,Hg}(z,Q^2/M^2)\equiv C^{FF,1}_{2,Hg}(z,Q^2/M^2),}   
whereas now we have the nontrivial relationship
\eqn\nontrivequ{C^{FF,1}_{2,qg}(z,Q^2/M^2)= C^{VF,1}_{2,qg}(z,Q^2/M^2)
-(\ln (Q^2/M^2) +c_{rs})P^0_{qg}\otimes C^{VF,PS,0}_{2,qH},}   
e.g. the zeroth order coefficient function for a charm quark to interact with 
a $W^{-}$ to produce a strange quark is undetermined. As in the previous
case we determine this zeroth order heavy quark coefficient function by 
demanding continuity of the $\ln Q^2$--derivative of the structure function, 
in the gluon sector (again at lowest order we have complete continuity),
along with demanding the correct asymptotic result. 
Unlike the neutral current case this time it is the strange quark 
(or down quark) structure function on which the condition is imposed,
rather than the charm quark structure function. This is because at lowest 
order the charm quark structure function is completely independent of the 
charm quark distribution, whereas the light quark structure functions 
do depend on it. However, in complete
analogy with the neutral current case our constraint results in
\eqn\cclocof{C^{VF,PS,0}_{2,qH}(Q^2/M^2)\otimes P^0_{qg}=
{d C^{FF,1}_{2,qg}(z,Q^2/M^2)\over \ln (Q^2)},}
where the left hand side automatically has the correct threshold behaviour
and the right--hand side $\to P^0_{qg}(z)$ as $Q^2/M^2 \to \infty$.     
Using this explicitly in \nontrivequ\ then results in the 
$C^{VF,1}_{2,qg}(z,Q^2/M^2)$ reducing to the correct massless $\msbar$ limit
as $Q^2/M^2\to \infty$, as it must by construction.  

This procedure can be repeated 
at all orders in exactly the same way as for the neutral 
current structure function. 
This time there are only two coefficient functions to be 
determined, $C^{VF,PS}_{2,Hq}(z,Q^2/M^2)$ as we have just seen, and
which exists at all orders, and
$C^{VF,PS}_{2,HH}(z,Q^2/M^2)$. The later begins at ${\cal O}(\alpha_s^2(Q^2))$
and will be determined by demanding continuity of the $\ln Q^2$-derivative
of the structure function where a heavy quark is produced directly at
the interaction vertex of the $W$ boson at ${\cal O}(\alpha_s^3(Q^2))$. The 
extension to the longitudinal charged current structure function is also
easily achieved using the above results and the discussion of the 
longitudinal neutral current structure function in \S 7.

\vfill 
\eject

\noindent{\bf Table 1.}

\bigskip

Prescription for the order by order in $\alpha_s(Q^2)$ determination of the 
VFNS coefficient functions for $F_2(x,Q^2)$.

\medskip

\halign{ #\hfil & \quad\hfil#& \qquad #\hfil \cr
{\bf Order of} & {\bf Eqn.} & {\bf Coefficient functions determined}\cr
{\bf equality}\hfil & &\cr
\noalign{\medskip}
$\alpha_s^0(Q^2)$ & \bosglufusii\ & $C^{VF,NS,0}_{2,qq}$\hfil\cr
\noalign{\smallskip}
$\alpha_s(Q^2)$ & \bosglufus\ & $C^{VF,NS,0}_{2,HH}$ (by continuity of 
$({dF_{2,H}\over d\ln Q^2})_{M^2}$ at ${\cal O}(\alpha_s(Q^2))$), 
$C^{VF,1}_{2,Hg}$\hfil\cr
&\bosglufusii\ & $C^{VF,NS,1}_{2,qq}$\hfil\cr
&\bosglufusiii\ &$C^{VF,1}_{2,qg}$\hfil\cr
\noalign{\smallskip}
$\alpha_s^2(Q^2)$&\bosglufus\ & $C^{VF,S,1}_{2,HH}$ (by continuity of 
$({dF_{2,H}\over d\ln Q^2})_{M^2}$ in gluon sector at 
${\cal O}(\alpha^2_s(Q^2))$), 
$C^{VF,2}_{2,Hg}$\hfil\cr
&\bosglufusi\ & $C^{VF,2}_{2,Hq}$\hfil\cr
&\bosglufusii\ &  $C^{VF,NS,2}_{2,qq}$\hfil\cr
&\bosglufusiii\ &  $C^{VF,2}_{2,qg}$\hfil\cr
&\bosglufusiv\ &  $C^{VF,PS,2}_{2,qq}$\hfil\cr
\noalign{\smallskip}
$\alpha_s^3(Q^2)$&\bosglufus\ & $C^{VF,S,2}_{2,HH}$ (by continuity of 
$({dF_{2,H}\over d\ln Q^2})_{M^2}$ in gluon sector at 
${\cal O}(\alpha^3_s(Q^2))$), 
$C^{VF,3}_{2,Hg}$\hfil\cr
&\bosglufusi\ & $C^{VF,3}_{2,Hq}$\hfil\cr
&\bosglufusii\ &  $C^{VF,3}_{2,qq}$\hfil\cr
&\bosglufusiii\ & $C^{VF,2}_{2,qH}$ (by continuity of 
$({dF_{2,i}\over d\ln Q^2})_{M^2}$ in gluon sector at ${\cal O}
(\alpha^3_s(Q^2))$), 
$C^{VF,3}_{2,qg}$\hfil\cr
&\bosglufusiv\ &  $C^{VF,PS,3}_{2,qq}$\hfil\cr
\noalign{\smallskip}
$\cdots$\hfil&$\cdots$\hfil&$\cdots$\hfil\cr
\noalign{\smallskip}
$\alpha_s^n(Q^2)$&\bosglufus\ & $C^{VF,S,n-1}_{2,HH}$ (by continuity of 
$({dF_{2,H}\over d\ln Q^2})_{M^2}$ in gluon sector at 
${\cal O}(\alpha^n_s(Q^2))$), 
$C^{VF,n}_{2,Hg}$\hfil\cr
&\bosglufusi\ & $C^{VF,n}_{2,Hq}$\hfil\cr
&\bosglufusii\ &  $C^{VF,n}_{2,qq}$\hfil\cr
&\bosglufusiii\ & $C^{VF,n-1}_{2,qH}$ (by continuity of 
$({dF_{2,i}\over d\ln Q^2})_{M^2}$ in gluon sector at 
${\cal O}(\alpha^n_s(Q^2))$), 
$C^{VF,n}_{2,qg}$\hfil\cr
&\bosglufusiv\ &  $C^{VF,PS,n}_{2,qq}$\hfil\cr}

\vfil
\eject

\noindent {\bf Table 2.}\hfil\break

\noindent Comparison of quality of fits for a wide variety of structure 
function data \hone\zeus\nmc\ref\bcdms{BCDMS collaboration: A.C. Benvenuti
{\it et al}., \PL \vyp{B223}{1989}{485}.}\ref\slac{L.W. Whitlow {\it et al}.,
\PL \vyp{B282}{1992}{475}.}
\ref\esixsix{E665 collaboration: M.R. Adams {\it 
et al}., \PR \vyp{D54}{1996}{3006}.} using our prescription for heavy 
quarks at NLO (TR) and the NLO fits MRRS and MRS(R$_2)$. We do not include 
the small--$x$, low--$Q^2$ data in the second of \hone\ in our fit in order
to make a direct comparison with the previous fits.  
  
\medskip

\hfil\vtop{{\offinterlineskip
\halign{ \strut\tabskip=0.6pc
\vrule#&  #\hfil&  \vrule#&  \hfil#& \vrule#& \hfil#& \vrule#& \hfil#&
\vrule#& \hfil#& \vrule#\tabskip=0pt\cr
\noalign{\hrule}
& Experiment && data && \omit &\omit&$\chi^2$&\omit& \omit &\cr
&\omit&& points && TR &\omit& MRRS &\omit& MRS(R$_2$)&\cr
\noalign{\hrule}
& H1 $F^{ep}_2$ && 193 && 135 && 133 && 149 &\cr
& ZEUS $F^{ep}_2$ && 204 && 274 && 290 && 308 &\cr
\noalign{\hrule}
& BCDMS $F^{\mu p}_2$ && 174 && 262 && 271 && 320 &\cr
& NMC $F^{\mu p}_2$ && 130 && 144 && 145 && 135 &\cr
& NMC $F^{\mu d}_2$ && 130 && 112 && 119 && 99 &\cr
& E665 $F^{\mu p}_2$ && 53 && 61 && 60 && 62 &\cr
& E665 $F^{\mu d}_2$ && 53 && 53 && 54 && 60 &\cr
& SLAC $F^{ep}_2$ && 70 && 98 && 96 && 95 &\cr
\noalign{\hrule}}}}\hfil

\vfil\eject

\noindent{\bf Table 3.}

\bigskip

Prescription for the order by order in $\alpha_s(Q^2)$ determination of the 
VFNS coefficient functions for $F_L(x,Q^2)$.

\medskip

\halign{ #\hfil & \quad\hfil#& \qquad #\hfil \cr
{\bf Order of} & {\bf Eqn.} & {\bf Coefficient functions determined}\cr
{\bf equality}\hfil & &\cr
\noalign{\medskip}
$\alpha_s(Q^2)$ & \bosglufus\ & $C^{VF,1}_{L,Hg}$\hfil\cr
&\bosglufusii\ & $C^{VF,NS,1}_{L,qq}$\hfil\cr
&\bosglufusiii\ &$C^{VF,1}_{L,qg}$\hfil\cr
\noalign{\smallskip}
$\alpha_s^2(Q^2)$&\bosglufus\ & $C^{VF,S,1}_{L,HH}$ (by continuity of 
$({dF_{L,H}\over d\ln Q^2})_{M^2}$ in gluon sector at 
${\cal O}(\alpha^2_s(Q^2))$), 
$C^{VF,2}_{L,Hg}$\hfil\cr
&\bosglufusi\ & $C^{VF,2}_{L,Hq}$\hfil\cr
&\bosglufusii\ &  $C^{VF,NS,2}_{L,qq}$\hfil\cr
&\bosglufusiii\ &  $C^{VF,2}_{L,qg}$\hfil\cr
&\bosglufusiv\ &  $C^{VF,PS,2}_{L,qq}$\hfil\cr
\noalign{\smallskip}
$\alpha_s^3(Q^2)$&\bosglufus\ & $C^{VF,S,2}_{L,HH}$ (by continuity of 
$({dF_{L,H}\over d\ln Q^2})_{M^2}$ in gluon sector at 
${\cal O}(\alpha^3_s(Q^2))$), 
$C^{VF,3}_{L,Hg}$\hfil\cr
&\bosglufusi\ & $C^{VF,3}_{L,Hq}$\hfil\cr
&\bosglufusii\ &  $C^{VF,3}_{L,qq}$\hfil\cr
&\bosglufusiii\ & $C^{VF,2}_{L,qH}$ (by continuity of 
$({dF_{L,i}\over d\ln Q^2})_{M^2}$ in gluon sector at ${\cal O}
(\alpha^3_s(Q^2))$), 
$C^{VF,3}_{L,qg}$\hfil\cr
&\bosglufusiv\ &  $C^{VF,PS,3}_{L,qq}$\hfil\cr
\noalign{\smallskip}
$\cdots$\hfil&$\cdots$\hfil&$\cdots$\hfil\cr
\noalign{\smallskip}
$\alpha_s^n(Q^2)$&\bosglufus\ & $C^{VF,S,n-1}_{L,HH}$ (by continuity of 
$({dF_{L,H}\over d\ln Q^2})_{M^2}$ in gluon sector at 
${\cal O}(\alpha^n_s(Q^2))$), 
$C^{VF,n}_{L,Hg}$\hfil\cr
&\bosglufusi\ & $C^{VF,n}_{L,Hq}$\hfil\cr
&\bosglufusii\ &  $C^{VF,n}_{L,qq}$\hfil\cr
&\bosglufusiii\ & $C^{VF,n-1}_{L,qH}$ (by continuity of 
$({dF_{L,i}\over d\ln Q^2})_{M^2}$ in gluon sector at 
${\cal O}(\alpha^n_s(Q^2))$), 
$C^{VF,n}_{L,qg}$\hfil\cr
&\bosglufusiv\ &  $C^{VF,PS,n}_{L,qq}$\hfil\cr}

\bigskip

In each case $C^{VF,n}_{L,aH}(z,M^2/Q^2)$ is determined by introducing
the asymptotic form multiplied by $(1-M^2/Q^2)v$ and multiplying the 
terms determined by continuity by $M^2/Q^2$.

\vfil\eject

\footatend\vfill\supereject\immediate\closeout\rfile\writestoppt
\baselineskip=14pt\centerline{{\bf References}}\bigskip{\frenchspacing%
\parindent=20pt\escapechar=` \input refs.tmp\vfill\eject}\nonfrenchspacing 
\vfill\eject\immediate\closeout\ffile{\parindent40pt
\baselineskip14pt\centerline{{\bf Figure Captions}}\nobreak\medskip
\escapechar=` \input figs.tmp\vfill\eject}

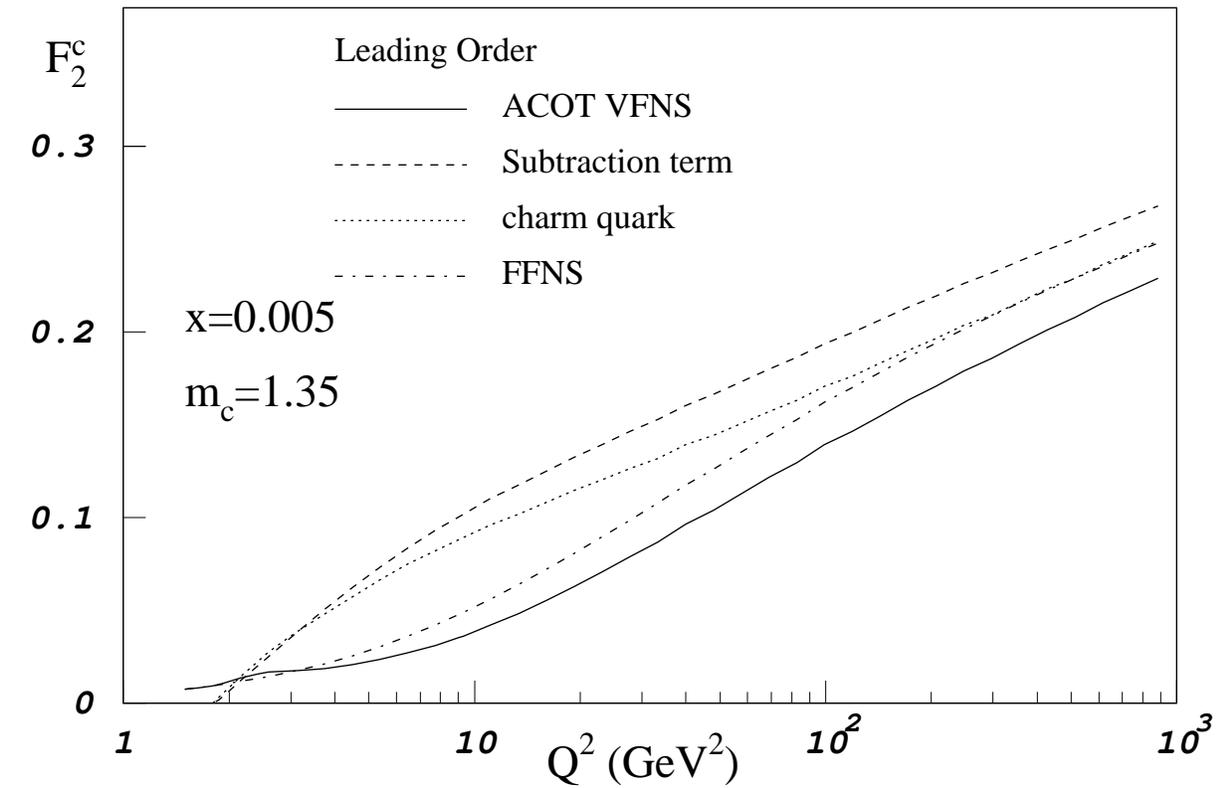
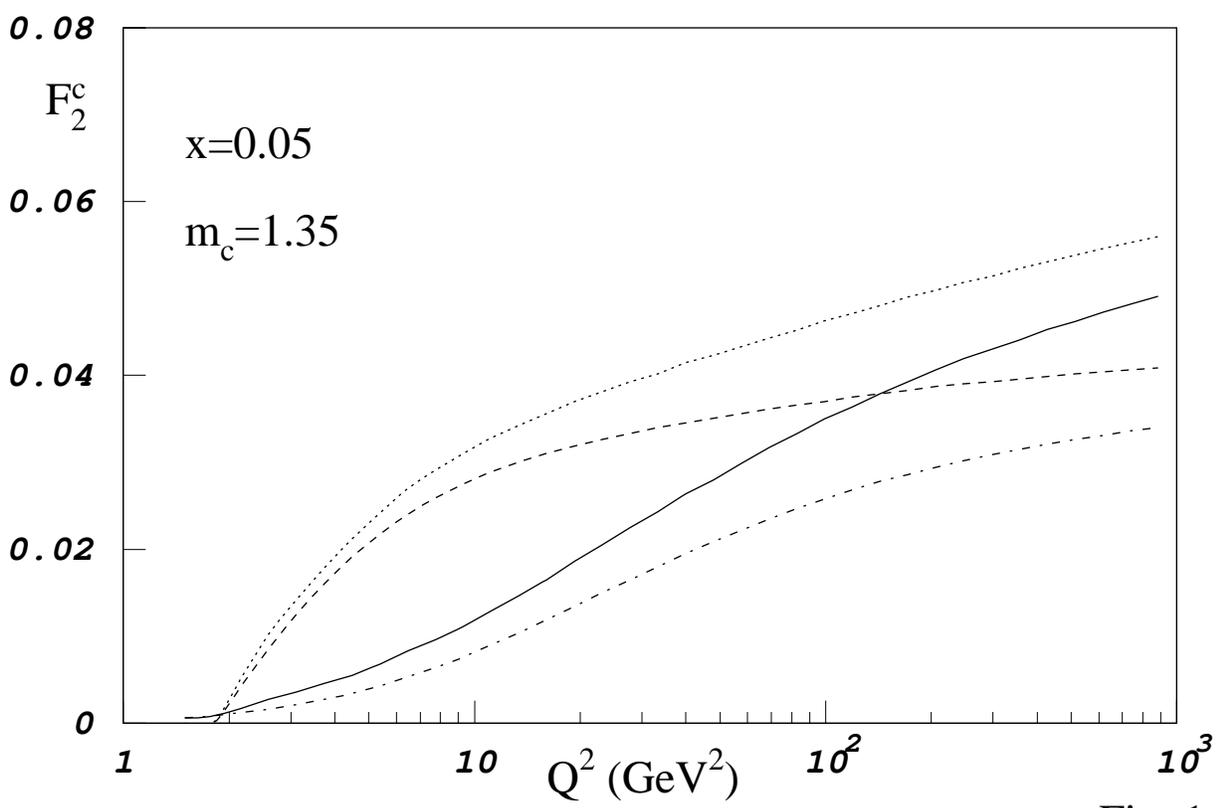
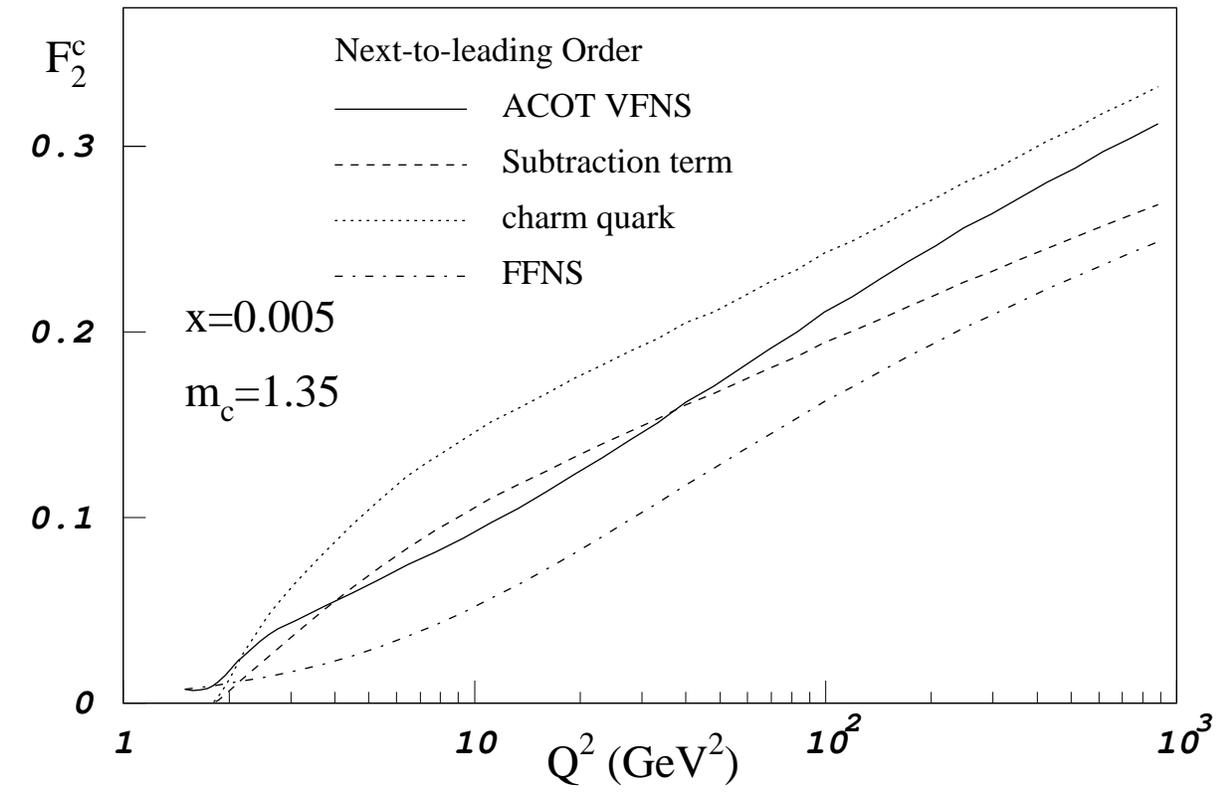
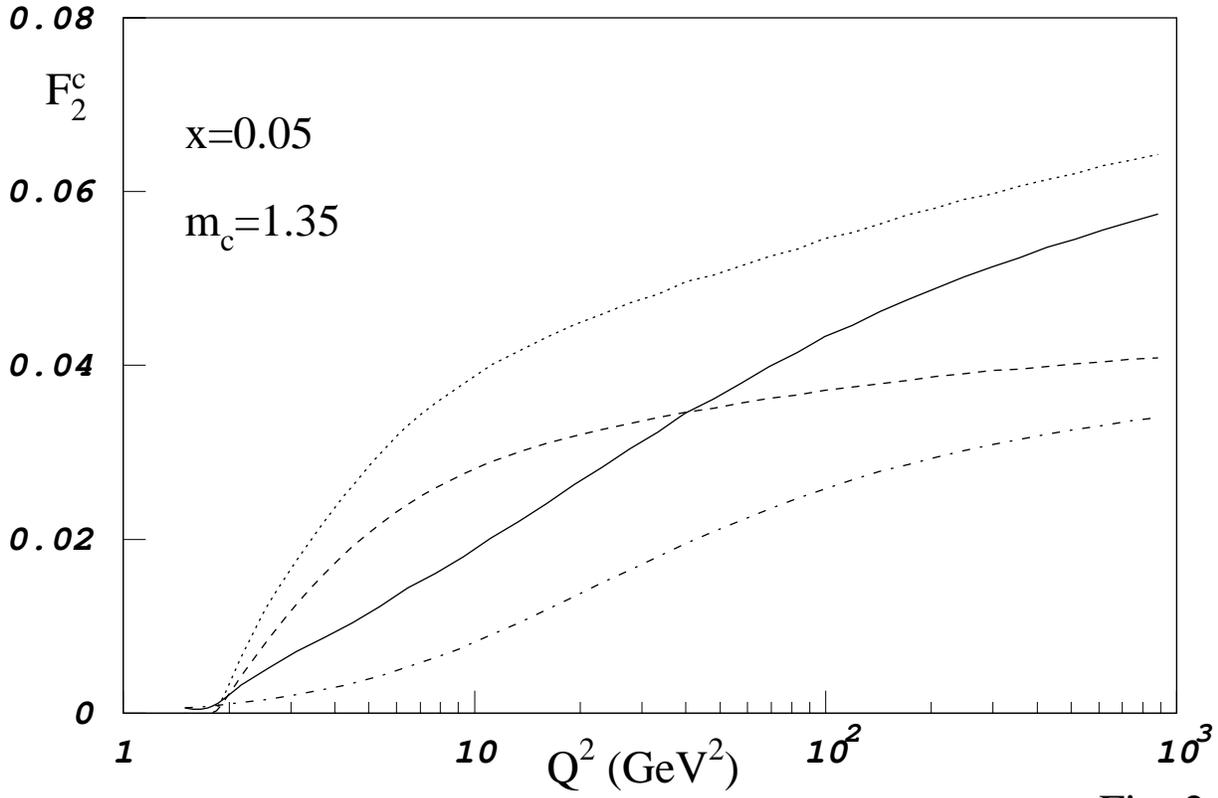
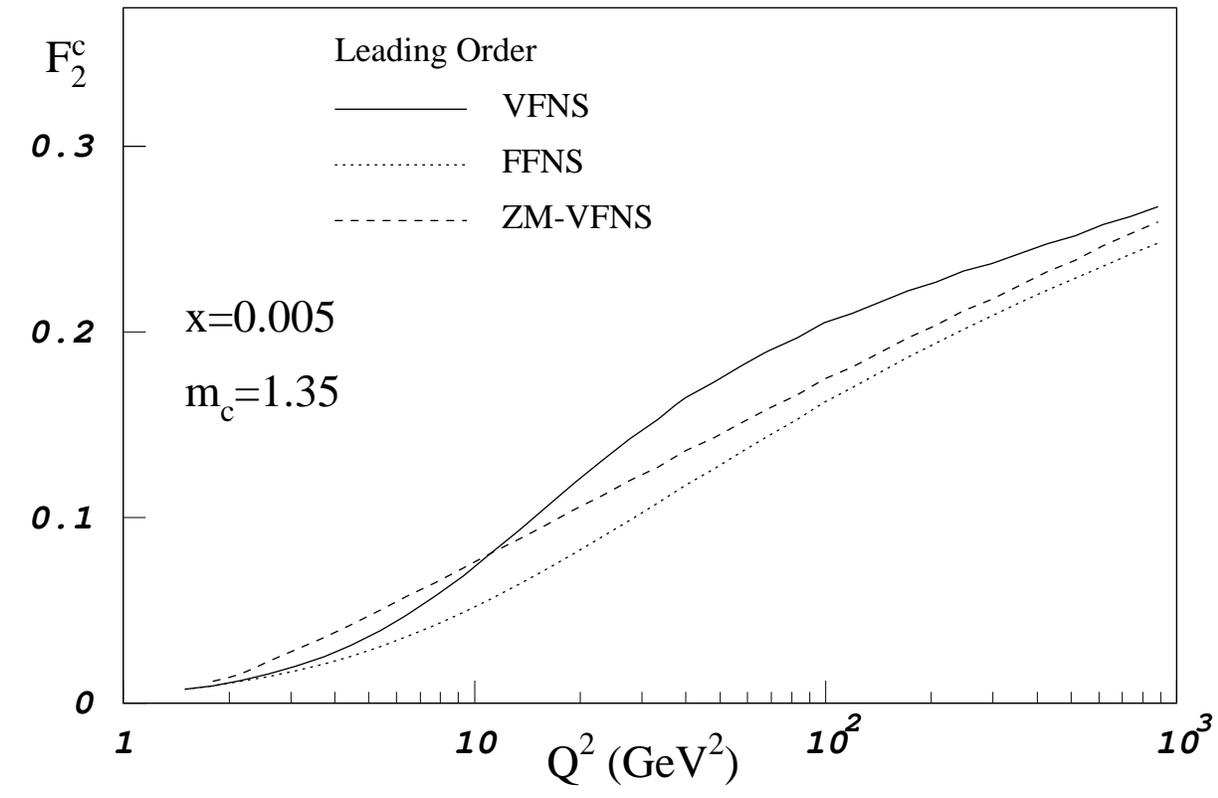
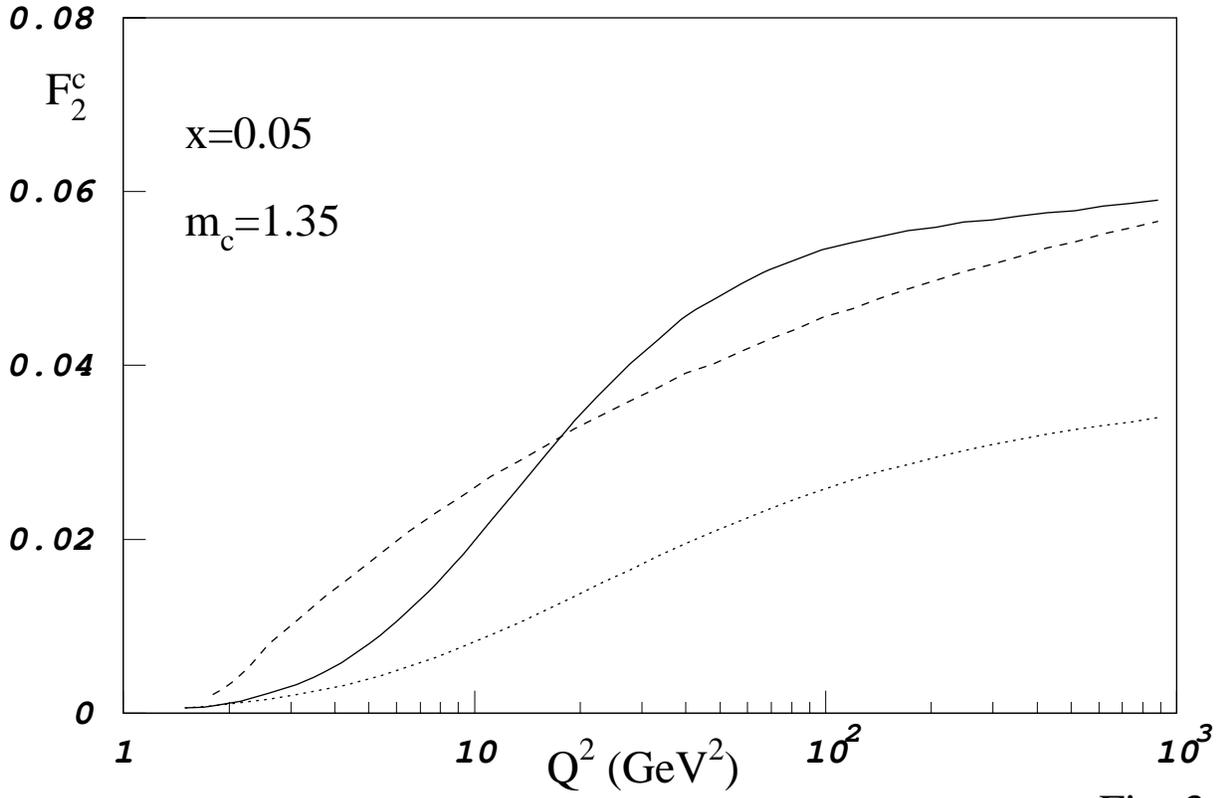
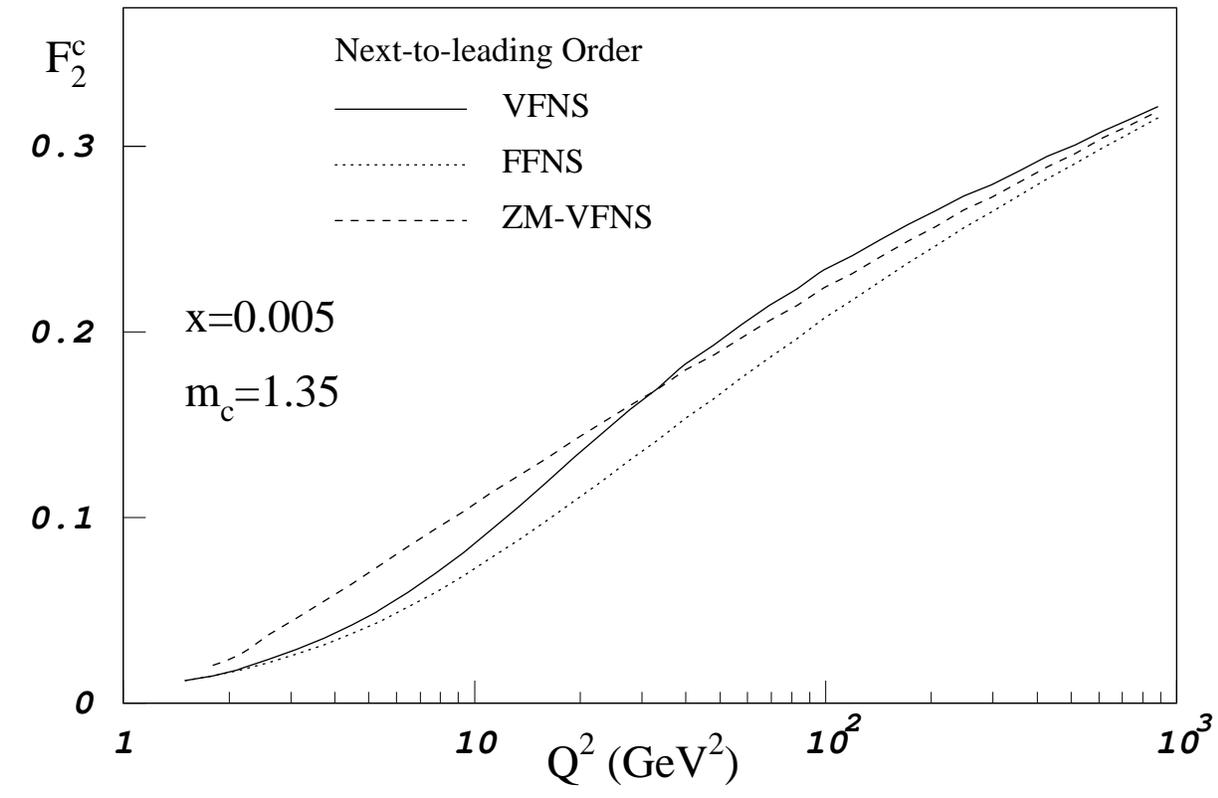
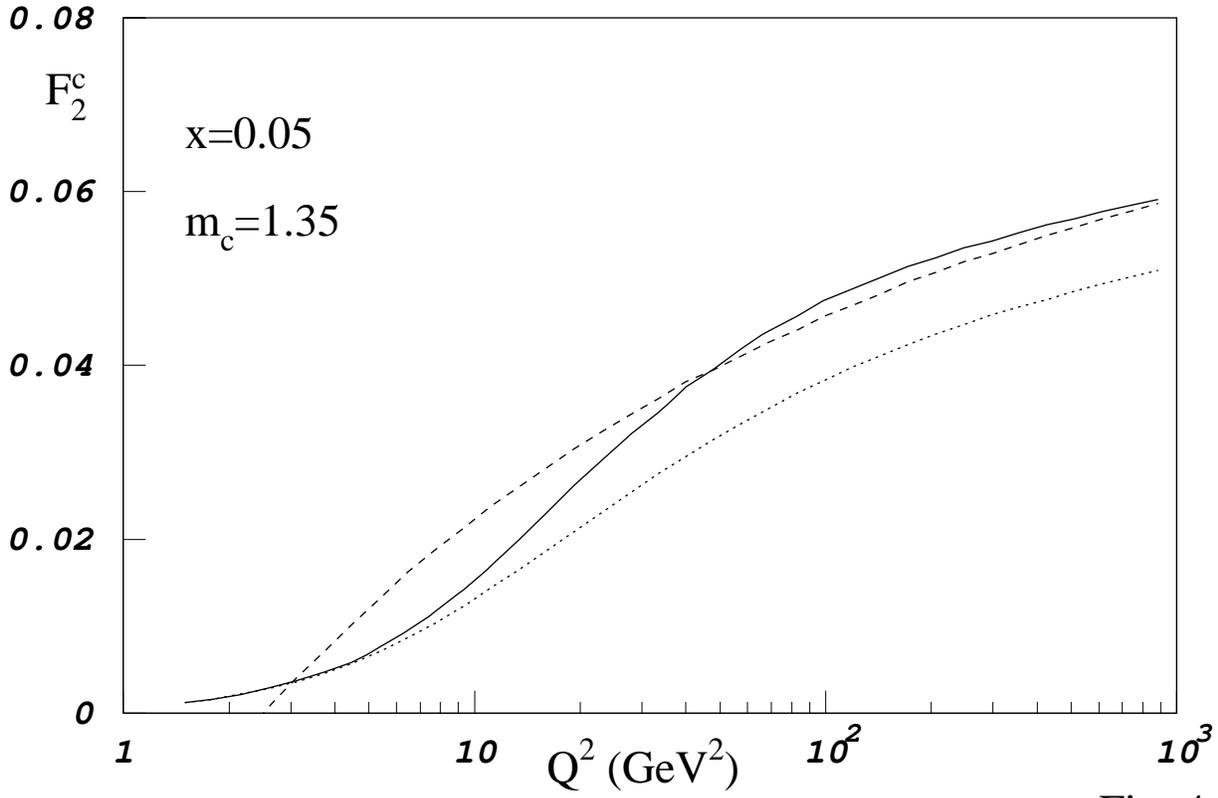
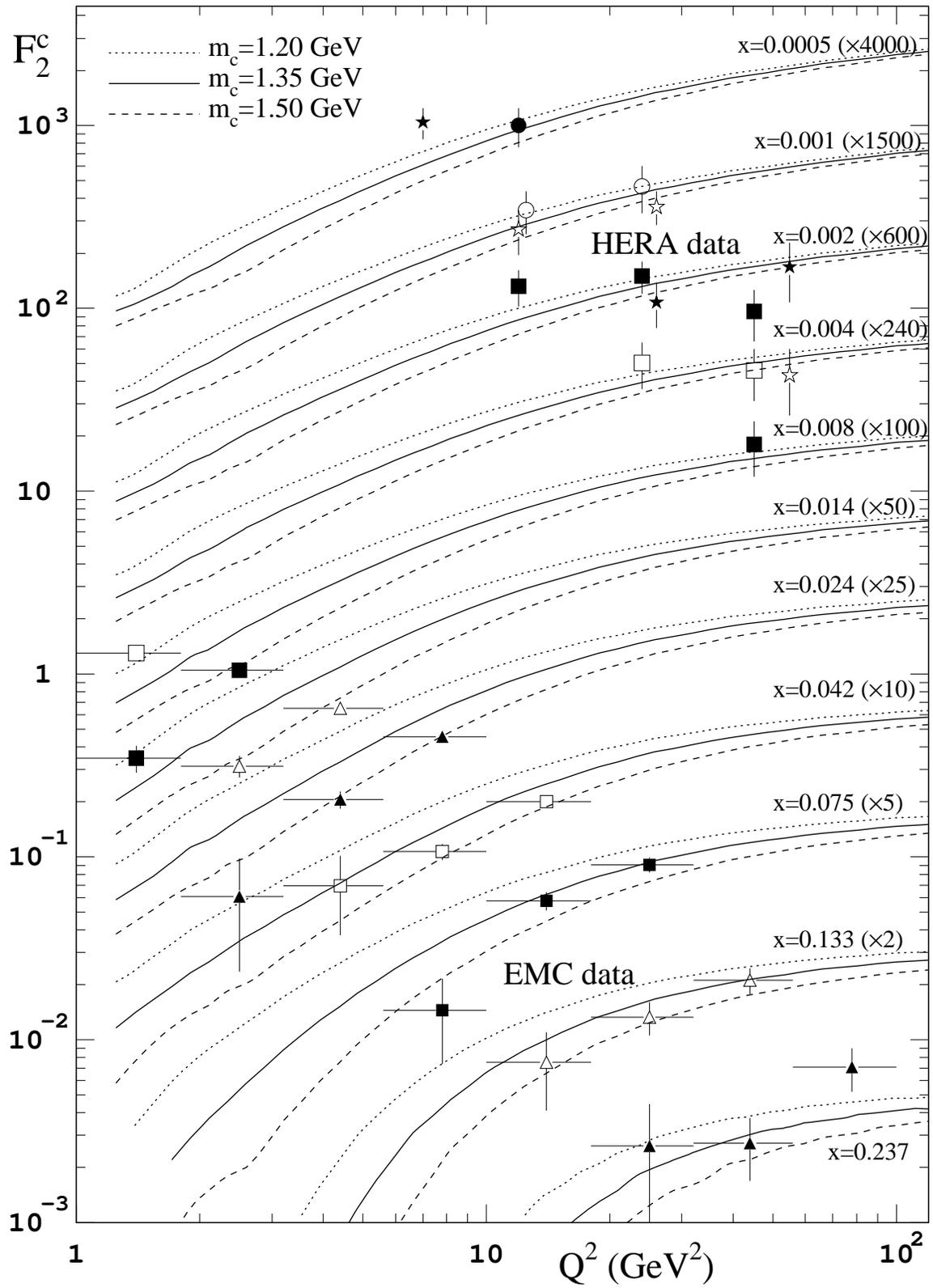
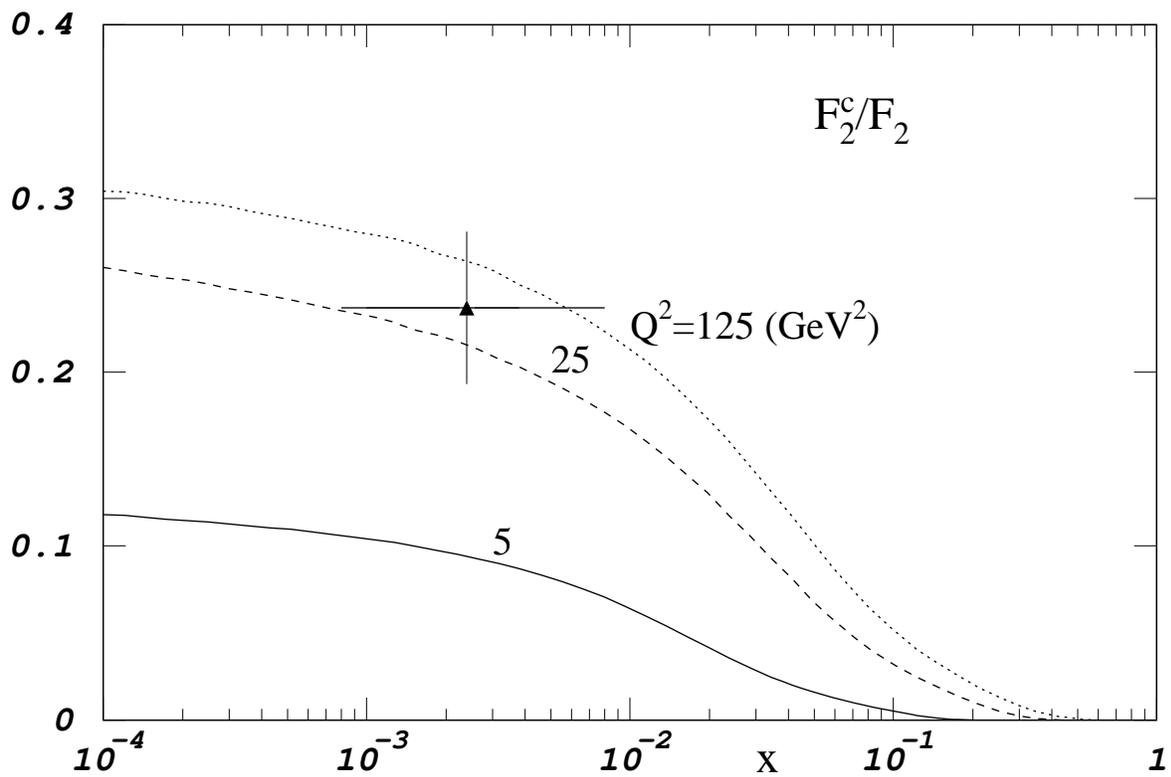
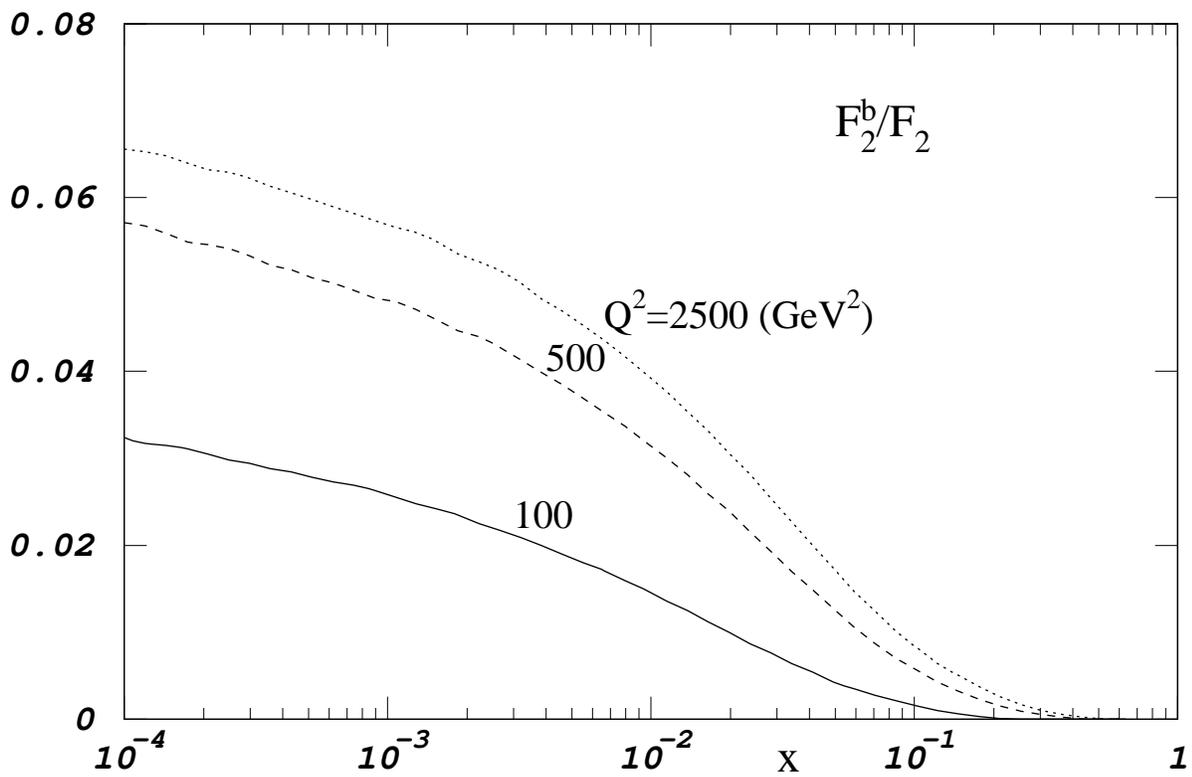
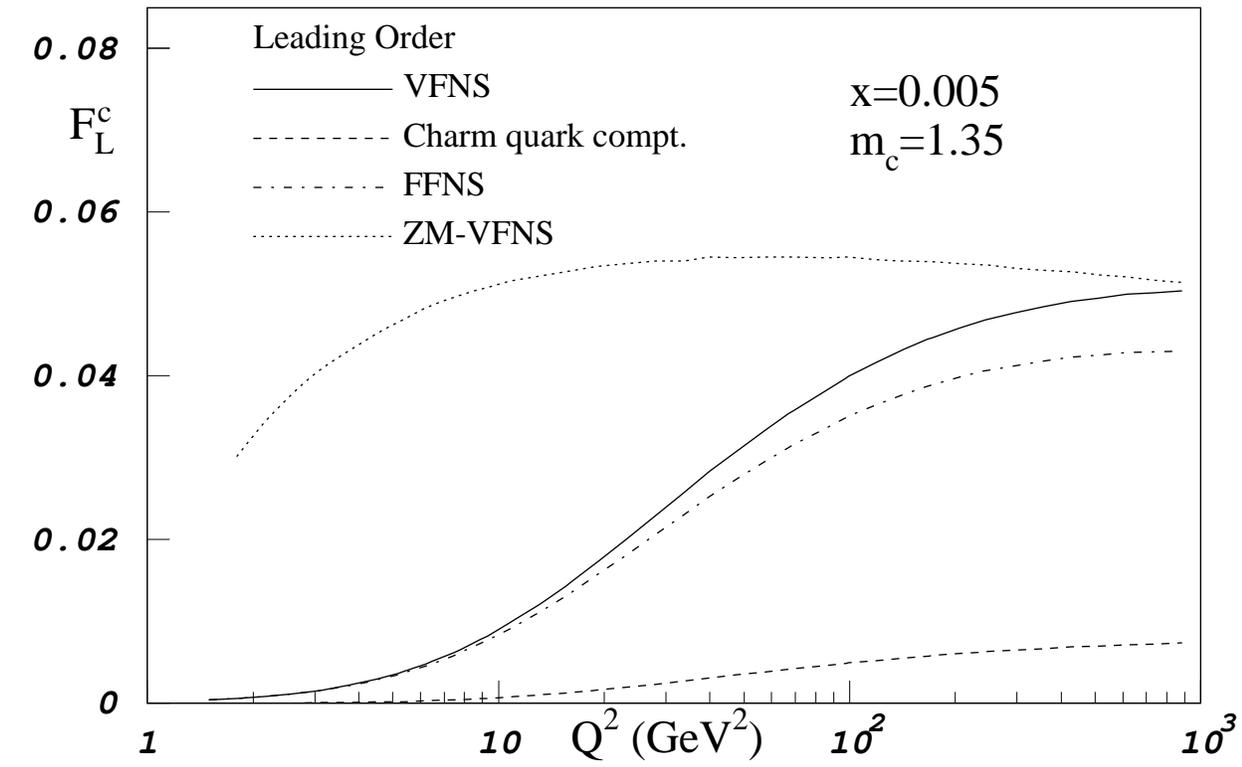
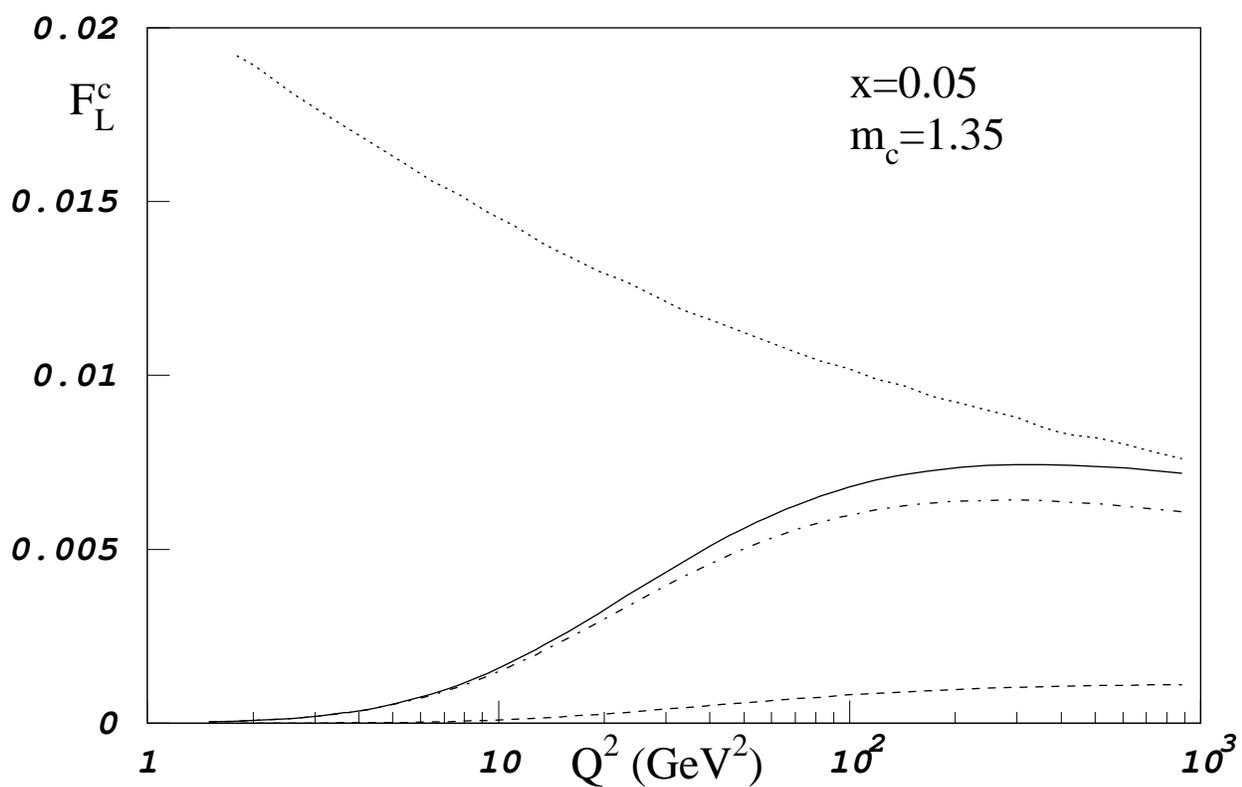
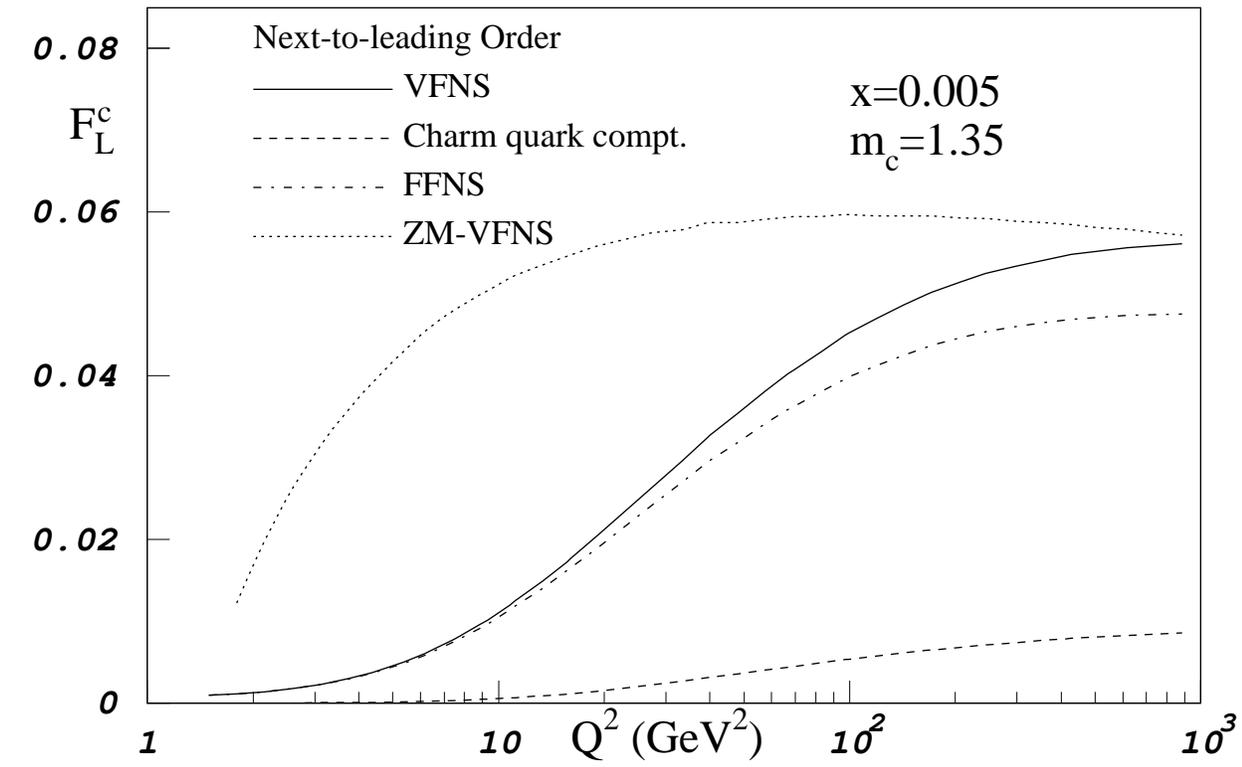
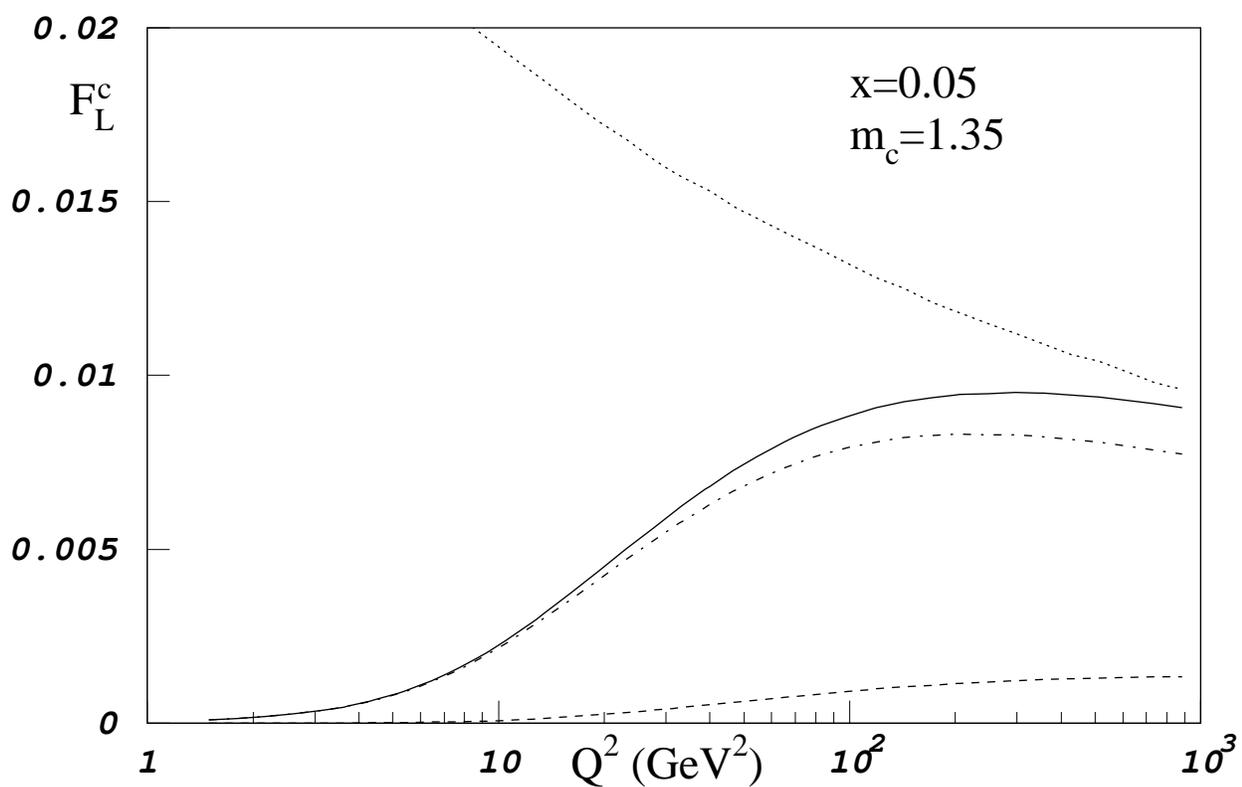
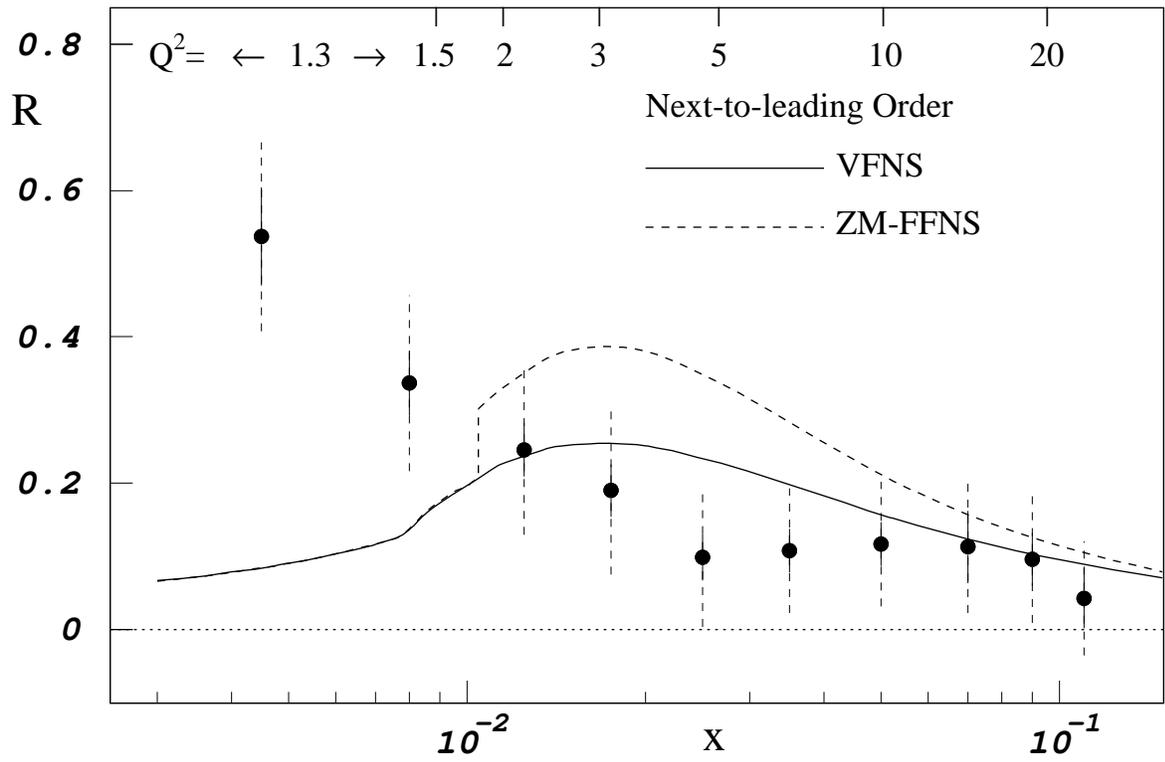
\end